\documentclass[aip, reprint, twocolumn,jcp,english]{revtex4-1}
\usepackage[T1]{fontenc}
\usepackage[latin9]{inputenc}
\usepackage{array}
\usepackage{verbatim}
\usepackage{amsmath}
\usepackage{amssymb}
\usepackage{graphicx}
\usepackage{hyperref}
\usepackage{comment}
\makeatletter
\usepackage{tikz}
\usetikzlibrary{calc}
\usetikzlibrary{arrows}
\usetikzlibrary{patterns}
\usetikzlibrary{shapes.geometric}
\tikzset{
    vertex/.style={circle,draw,minimum size=1.5em},
    edge/.style={->,> = latex'}
}

\usetikzlibrary{shapes.misc, positioning}
\usetikzlibrary{decorations.markings}

\providecommand{\tabularnewline}{\\}
\global\long\def\orthrep#1{\widetilde{#1}{\!\;\!}}%
\global\long\def\dotorthrep#1{\dot{\widetilde{#1}}{\!\;\!}}%
\makeatother
\usepackage{babel}
\begin{document}
\title{Time Evolution of ML-MCTDH Wavefunctions II: Application of the Projector
Splitting Integrator}
\begin{abstract}
 The multi-layer multiconfiguration time-dependent Hartree (ML-MCTDH)
approach suffers from numerical instabilities whenever the wavefunction
is weakly entangled. These instabilities arise from singularities
in the equations of motion (EOMs) and necessitate
the use of a regularization parameter. The Projector
Splitting Integrator (PSI) has previously been presented as an approach
for evolving ML-MCTDH wavefunctions that is free of singularities.
Here we will discuss the implementation of the multi-layer PSI with
a particular focus on how the steps required relate to those required to 
implement standard ML-MCTDH. We demonstrate the efficiency
and stability of the PSI for large ML-MCTDH wavefunctions containing
up to hundreds of thousands of nodes by considering a series of spin-boson
models with up to $10^6$ bath modes, and find that for these problems
the PSI requires roughly 3-4 orders of
magnitude fewer Hamiltonian evaluations and 2-3 orders of magnitude
fewer Hamiltonian applications than standard ML-MCTDH, and 2-3/1-2
orders of magnitude fewer evaluations/applications than approaches
that use improved regularization schemes. Finally, we consider a series
of significantly more challenging multi-spin-boson models that require
much larger numbers of single-particle functions with wavefunctions containing up to $\sim 1.3\times 10^9$ parameters to obtain accurate dynamics.
\end{abstract}
\author{Lachlan P Lindoy}
\email{ll3427@columbia.edu}
\author{Benedikt Kloss}
\author{David R Reichman}
\affiliation{Department of Chemistry, Columbia University, 3000 Broadway, New York, New York 10027, USA}
\maketitle

\section{Introduction}
In recent years considerable effort has been dedicated to addressing
the ill-conditioning of the equations of motion (EOMs) of the single
and multi-layer multi-configuration time-dependent Hartree methods.\citep{Manthe2015b,H.D.2018,Wang2018,Weike2021}  
The standard approach, in which the mean-field density matrix is regularized,
works well in many situations, has been successfully applied to obtain the dynamics of a wide range of problems,\citep{MANTHE1992,MANTHE199212,MANTHE19937,HAMMERICH1994,RAAB1999,VENDRELL20071,VENDRELL20072,VENDRELL20073,VENDRELL20074,WORTH2008,MENG2013,CRAIG2007,WANG2007,KONDOV2007,CRAIG2011,BORRELLI2012,RABANI2015,SCHULZE2016,SHIBL2017,MENDIVE2018} and a number of general purpose software packages have been developed that are capable of performing such calculations.\citep{Heidelberg,Quantics} However, as the types of problems considered
grow larger in size and complexity, it has been found that there are
regimes in which this strategy is not sufficient. These regimes include
those with very large numbers of modes,\citep{Wang2021} and those in which the Hamiltonian
directly couples all modes together in a multiplicative fashion, as is the case in Fermionic
MCTDH simulations\citep{Hinz2016} and in simulations of the polaron-transformed
spin-boson model.\citep{Wang2018,H.D.2018,MendiveTapia2020}
For these situations, the regularized ML-MCTDH dynamics can depend strongly
on components of the ML-MCTDH wavefunction that do not initially contribute
to the expanded wavefunction,\citep{Manthe2015} can show ``pseudoconvergence'' (with respect
to the regularization parameter) to incorrect dynamics,\citep{Weike2021} and can become
extremely difficult to evolve numerically which in some extreme cases
results in failure of the integrator.\citep{H.D.2018,Wang2018} 
Recently, Wang and Meyer have
presented an alternative strategy for regularizing the EOMs.\citep{Wang2018,Wang2021} Rather than regularize the mean-field
density matrices, this approach regularizes the coefficient tensors.
This approach has been applied to a wide range of spin-boson models
and has been found to perform significantly better than the standard
strategy. It has been shown to be capable of treating very large systems
and the polaron-transformed spin-boson model.\citep{Wang2018,Wang2021} 
While this approach has shown significant stability improvements over the standard 
regularization scheme, it still requires the solution EOMs that can be poorly conditioned depending on the value of the regularization parameter required.  It is possible that 
there are regimes in which convergence with respect to the regularization
parameter cannot be obtained before numerical issues arise in the solution of the EOMs.\citep{lindoy_thesis} 

 An alternative strategy for propagating ML-MCTDH wavefunctions
is the projector splitting integrator (PSI) approach,\citep{Schroeder2016,Bauernfeind2020,Kloss2020a}
originally presented by Lubich for single-layer MCTDH.\citep{Lubich2014a,Kloss2017a}
This approach employs an alternative representation of the wavefunction
when evolving the coefficient tensors, rendering the equations of
motion singularity-free. Once this evolution has been performed, it
is necessary to reconstruct the original representation, and in cases
where the ML-MCTDH EOMs are singular this reconstruction
is not fully specified. This is a fundamental issue for all ML-MCTDH
approaches that are based on the linear variational principle and
can only truly be resolved by going beyond linear variations.\citep{Manthe2015b}
However, in contrast to the standard ML-MCTDH EOMs,
where this results in significant numerical instabilities, this process
of reconstructing the original representation can be made numerically
stable, albeit still partially arbitrary. It has been suggested
that this arbitrariness may lead to problems with
this approach.  However, in principle, this approach could naturally
be paired with the optimal unoccupied SPFs approach of Manthe\citep{Manthe2015b} which
defines an optimal reconstruction of the original representation.
In practice, as we will demonstrate in this work, the arbitrary nature of the PSI does 
not significantly
influence numerical performance, and issues arising from it can
be considered to be convergence issues.

 In what follows we will discuss the implementation of the
multi-layer PSI, with a focus on how the
steps required relate to those required to implement standard ML-MCTDH. 
Following this
we will demonstrate the numerical performance of this approach by
considering a range of challenging model problems in Sec.\,\ref{Sec:Results}. 
These problems
include those that have previously been treated using an improved
regularization scheme\citep{Wang2018,Wang2021} and represent ones for which
standard ML-MCTDH can run into serious numerical issues, including the
polaron-transformed spin-boson model and the standard spin-boson model 
with a large number of bath modes. Finally, we consider an application
of the PSI to a series of multi-spin-boson models. These models require
considerably larger coefficient tensors at each node of the ML-MCTDH
wavefunction than is commonly treated using standard ML-MCTDH and
represent considerably harder challenges than any previously treated spin-boson
model.  In addition to these, further results are presented in the Supplementary Information with the aim of addressing additional aspects of the convergence of the PSI.

\section{Theory}

 Before we discuss the implementation of the PSI, we will first briefly review the ML-MCTDH wavefunction and EOMs.  In particular, we will discuss the recursive expressions for evaluating the terms present in the EOMs.  By doing so we can directly relate the evaluation of analogous quantities required by the PSI to the standard ML-MCTDH quantities.

\subsection{The ML-MCTDH Wavefunction Expansion and Trees}
\begin{figure}
\centering
\includegraphics{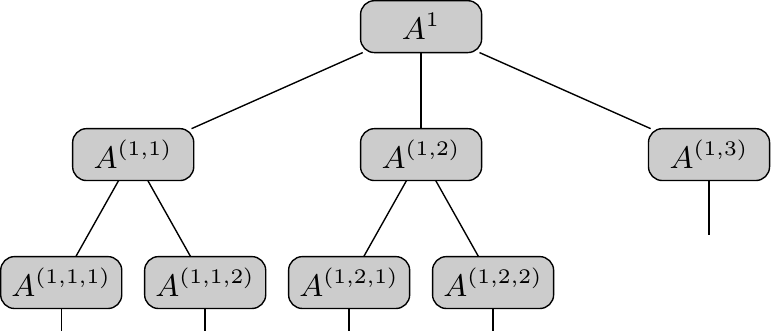}
\caption{\label{fig:mctdh_wavefunction}Representation of a tree tensor network for a five-dimensional wavefunction. Each node of the tree corresponds
to a coefficient tensor, and the lines connecting nodes represent
the contraction over indices common to the two nodes.}
\end{figure}

Instead of representing the wavefunction in the direct product basis
of all the bases of the problem's degrees of freedom which requires
an exponentially large number of wavefunction coefficients, ML-MCTDH
expands the wavefunction in a hierarchy of time-dependent bases. The
expansion coefficients of the latter are stored in a set of arrays,
or tensors, which contain auxiliary indices in addition to the indices
of the physical degrees of freedom. All tensors are connected to at
least one other tensor through a shared index, i.e. the value of the
index on both tensors is constrained to be the same in the expansion
of the wavefunction (see Fig.\,\ref{fig:mctdh_wavefunction}). This defines a general
loop-free network, and is referred to as a tree tensor network as its connectivity is
described by a tree. Any node $z_{l}=(k_1=1,k_{2},\dots,k_{l})$,
with $l$ denoting the distance from the root node and $k_{j}$ denoting
the index of child node at the $j$-th layer of the tree, stands for a wavefunction
coefficient tensor $A_{i_{1}\dots i_{d^{z_l}}i_0}^{z_{l}}$, and we
can interpret these objects as expansion coefficients in the
basis defined by the remainder of the tree. Here $i_0$ represents the index of the
array which connects this node to its parent node in the tree structure, 
that is it points towards the root node, while the remaining $d^{z_l}$ 
indices $i_{1},\dots,i_{d^{z_l}}$ point towards the leaf nodes, i.e. the nodes at the bottom of the tree that are connected to the physical degrees of freedom.
A subtree can be understood
as containing a set of (time-dependent) basis functions, referred to
as ``single-particle functions'' (SPFs), defined
on the physical degrees of freedom that are in the subtree. In the
ML-MCTDH context, it is usually understood that the physical degrees
of freedom are located on the leaf nodes, however this restriction
is not necessary. In analogy to a subtree encoding SPFs, we can think
of the complement of the subtree as encoding basis functions for
the physical degrees of freedom that are not in the subtree, the so
called ``single-hole functions'' (SHFs). 
%We may choose to define a root
%node of the tree, i.e. the highest layer of the hierarchical expansion.

The expression of a wavefunction in a ML-MCTDH format is non-unique
since the product of a matrix $M$ and its inverse $M^{-1}$ can be inserted
between any two coefficient tensor sharing a common index without changing the value of the wavefunction. However,
we can impose gauge conditions on the tensors to remove this ambiguity.
For example, we can, as is done in ML-MCTDH, require all non-root
coefficient tensors satisfy the following orthonormality condition
\begin{equation}
A_{I^{z_l}i_0}^{*z_{l}}A_{I^{z_l}j_0}^{z_{l}}=\delta_{i_0j_0}\ \ \forall\ z_{l} \neq1,\label{eq:orthonormal_cond_MLMCTDH}
\end{equation}
where we have introduced the notation $I^{z_l} = (i_{1},\dots, i_{d^{z_l}})$ for the combined set of indices pointing towards the leaf nodes. We also note the use of Einstein summation convention, implying summation over repeated indices, in Eq.\,\ref{eq:orthonormal_cond_MLMCTDH} and hereafter.
This orthonormality condition translates to $A^{z_{l}}\thinspace\forall\thinspace z_{l}\neq1$
being expansion coefficients for orthonormal SPFs in the direct product
basis of orthonormal SPFs (or primitive basis functions for the leaf nodes) of the child nodes of $z_{l}$. The associated SHFs
are not orthonormal, i.e. they have a non-unit overlap matrix, referred
to as the mean-field density matrix. The mean-field density matrix at node $z_l$ can be defined in terms of contractions between the SHFs of node $z_{l}$ with their complex conjugate. This leads to a recursive definition of the $\rho^{(z_l,k)}$, the mean-field density matrix of the $k$-th child of node $z_l$ (node $(z_l, k)$),  in terms of the mean-field density matrix $\rho^{z_{l}}$ and coefficient tensor $A^{z_{l}}$ of node $z_{l}$
\begin{equation}
\rho_{ab}^{(z_{l},k)}=A_{I^{z_l}_{k; a}i}^{*z_l}\rho_{ij}^{z_{l}}A_{I^{z_l}_{k; b}j}^{z_l},\label{eq:rho_recursive}
\end{equation}
where $I^{z_l}_{k;a} = (i_{1},\dots ,i_{k-1}, a ,i_{k+1},\dots ,i_{d^{z_l}})$ is a combined index obtained when the $k$th index in $I^{z_l}$ is replaced with the value $a$.
This expression makes use of Eq.\,\ref{eq:orthonormal_cond_MLMCTDH} for
all nodes that are not on the path between the root node and $z_{l}$.
%For the latter, Eq.\,\ref{eq:orthonormal_cond_MLMCTDH} does not lead to
%simplification because the contraction is over different sets of indices.
For the root node $z_{l}=1$, Eq.\,\ref{eq:orthonormal_cond_MLMCTDH} translates to $A^1$ being expansion coefficients for the full wavefunction in the orthonormal basis defined by the remainder
of the tree, and it is not subject to any constraint besides normalization.
To denote this special property, we will decorate its coefficient
tensor with a \textit{tilde}-symbol, namely $\orthrep{A^{1}}$.

Having obtained an unambigous ML-MCTDH representation of the wavefunction
through the constraint in Eq.\,\ref{eq:orthonormal_cond_MLMCTDH}, we can
%consider the expansion coefficients in the coefficient tensors as
treat the coefficient tensors as
variational parameters and obtain a solution to the time-dependent
Schr\"{o}dinger equation using the time-dependent
variational principle. To do this we need to further specify a dynamic gauge
condition for the overlap of coefficient tensors with their time-derivative.
Here, for simplicity, we will make use of the standard choice, namely that these 
two matrices are orthogonal for all nodes except the root node.
This leads to the following EOMs for the coefficient tensor at the root node
\begin{align}
\dotorthrep{A}_{I^1}^{1} & =-\frac{{i}}{\hbar}\sum_{r}h_{r;I^1 J^1 }^{1}\orthrep{A}_{J^1}^{1},\label{eq:ml-mctdh-eom-root}
\end{align}
and at all other nodes
\begin{equation}
\dot{A}_{I^{z_l}i_0}^{z_l}\!=\!-\frac{{i}}{\hbar}\!\sum_{r}\!Q_{I^{z_l}\!K^{z_l}}^{z_l}h_{r;K^{z_l}\!J^{z_l}}^{z_{l}}A_{J^{z_l}\!j}^{z_{l}}H_{r;jk}^{z_{l}}[(\rho^{z_{l}})^{\text{-}1}]_{ki_0}.\label{eq:mlmctdh-eom-gen}
\end{equation}
%You can always write a Hamiltonian in the sum-of-product form, it just might not be efficient
Here, we have written the Hamiltonian in a sum-of-product
(SOP) form, $\hat{H}=\sum_{r}\bigotimes_{a\in L}\hat{h}_{r}^{a}$, with
$L$ denoting the set of leaf nodes and $\hat{h}_{r}^{a}$ is the Hamiltonian
operator acting on the physical degree of freedom associated with leaf-node
$a$. The Hamiltonian at each node can be expressed in terms of the matrices
\begin{equation}
\begin{aligned}
h_{r;I^{z_l}J^{z_l}}^{z_{l}}=&\prod_{k=1}^{d^{z_{l}}}A_{I^{(z_l,k)}i_k}^{*(z_{l},k)}h_{r;I^{(z_l,k)}J^{(z_l,k)}}^{(z_{l},k)}A_{J^{(z_l,k)}
j_k}^{(z_{l},k)}\\
=&\prod_{k=1}^{d^{z_{l}}}M_{r;i_k j_k}^{(z_{l},k)},
\end{aligned}\label{eq:ham_spf}
\end{equation}
where $I^{(z_l,k)}=(i_1, \dots, i_d^{(z_l,k)})$ is the combined index for all indices associated with  all children of node $(z_l, k)$,
and
\begin{equation}
H_{r;ab}^{(z_{l},k)}=A_{I^{z_l}_{k;a} i}^{*z_{l}}\left(\prod_{\substack{k^{\prime}=1\\k^{\prime}\neq k}}^{d^{z_{l}}}M_{r;i_{k^\prime} j_{k^\prime}}^{(z_{l},k^\prime)}\right)H_{r;i j}^{z_{l}}A_{I^{z_l}_{k; b} j}^{z_{l}}.\label{eq:ham-mf}
\end{equation}
The Hamiltonians in Eqs.\,\ref{eq:ham_spf} and \ref{eq:ham-mf} are referred to as the SPF and mean-field Hamiltonian matrices, respectively. 
They can be considered as matrix elements of operators present in the SOP expansion evaluated using the SPFs and SHFs, respectively.
%They can be considered as evaluations of terms in the SOP expansion of
%the Hamiltonian on the subtree and its complement, respectively. 
Additionally,
the projector onto the space spanned by the SPFs of node $z_l$
appears in Eq.\,\ref{eq:mlmctdh-eom-gen}, and is defined as
\begin{equation}
Q_{I^{z_l}J^{z_l}}^{z_l}=\delta_{I^{z_l}J^{z_l}}-A_{I^{z_l}i}^{z_{l}}A_{J^{z_l}i}^{*z_{l}}.\label{eq:proj}
\end{equation}
The presence of this projector makes Eq.\,\ref{eq:mlmctdh-eom-gen}
non-linear. Furthermore, Eq.\,\ref{eq:mlmctdh-eom-gen} contains the
inverse of the mean-field density matrix, $(\rho^{z_{l}})^{-1}$.
Since the density matrix is not guaranteed to be of full rank and
is often almost rank-deficient, Eq.\,\ref{eq:mlmctdh-eom-gen}
can be singular and requires regularization of the matrix inverse
in order to be numerically solvable.

\subsection{Singularity-Free Versions of ML-MCTDH}
 In order to circumvent the presence of an ill-defined inverse in the
ML-MCTDH EOMs, we can make use of the gauge freedom in the ML-MCTDH 
representation of the wave function. In
particular, we can choose a representation in which both SHFs and
SPFs for a given node $z_{l}$
are orthonormal by imposing the following
orthonormality condition
\begin{equation}
A_{I^{z_l^\prime}_{p(z_l); a}i}^{*z_{l}^{\prime}}A_{I^{z_l^\prime}_{p(z_l); b}i}^{z_{l}^{\prime}}=\delta_{ab}\ \forall \ z_{l}^{'}\neq z_{l},\label{eq:orthonormality_cond_PSI}
\end{equation}
where $p(z_{l})$ indicates the index of the array which points towards
node $z_{l}$, and $I^{z_l^\prime}_{p(z_l);a}$ is a combined index of the form used previously.  %$=(i_1,\dots,i_{p(z_l)-1},a,i_{p(z_l)+1},\dots, i_{d^{z_l^\prime}})$ is a combined index obtained when the $p^{z_l}$th index is replaced with the value $a$. 
For all nodes that are not node $z_l$ or ancestors of node $z_l$ (namely they are not its parent or parent's parent, and so on up to the root node), the index pointing to the $p^{z_l}$ node is the same as the index pointing towards the root node ($p(z_l) = p(1)$). 
For these nodes Eq.\,\ref{eq:orthonormality_cond_PSI} is equivalent to Eq. 
\,\ref{eq:orthonormal_cond_MLMCTDH}.  For the nodes that are ancestors of node $z_l$,
 this corresponds to a different orthonormality condition.  Imposing this condition
 alters the value of the coefficient tensors for these nodes.
To  make this explicit, we will choose a different label, $U^{z_l}$, 
for the coefficient tensors of nodes that satisfy 
Eq.\,\ref{eq:orthonormality_cond_PSI} and are ancestors of node $z_l$.  
 This orthogonality condition leads to orthonormal SPFs for all
nodes which are not ancestors of node $z_{l}$,
and orthonormal SHFs for node $z_l$ and all of its ancestors. 
The latter property can be straightforwardly
verified by comparison with Eq.\,\ref{eq:rho_recursive}. Note that Eq.
\ref{eq:orthonormal_cond_MLMCTDH} is obtained as a special case ($z_{l}=1$)
of Eq.\,\ref{eq:orthonormality_cond_PSI}. 

Singularity-free approaches
to propagating the ML-MCTDH wavefunction make use of this representation,
in which the inverse of the density matrix $\rho^{z_{l}}$ is trivially
the identity matrix. In the following, we will discuss a particular
singularity-free method to propagate ML-MCTDH wavefunctions, the projector splitting
integrator (PSI). This algorithm can be considered as a linearized
integrator for singularity-free ML-MCTDH EOMs, similar to the commonly
used constant-mean-field integrator for standard ML-MCTDH EOMs. Instead
of solving Eq.\,\ref{eq:mlmctdh-eom-gen} for the non-root node coefficient
tensors under the orthonormality constraint Eq.\,\ref{eq:orthonormal_cond_MLMCTDH},
the PSI involves solving two differential equations for every node
$z_{l}$ serially, that is node-by-node. First, we propagate the coefficient
tensor $\orthrep{A}^{z_{l}}$ forward in time, where the \emph{tilde}-symbol
indicates that we enforce orthonormality condition in Eq.\,\ref{eq:orthonormality_cond_PSI}
for node $z_{l}$ according to
\begin{equation}
\dotorthrep{A}_{I^{z_l}i_{0}}^{z_l}(t)=-\frac{{i}}{\hbar}\sum_{r}h_{r;I^{z_l}J^{z_l}}^{z_{l}}\orthrep{A}_{J^{z_l}j}^{z_{l}}(t)\orthrep{H}_{r_{ji_{0}}}^{z_{l}}.\label{eq:Atilde-PSI-eom}
\end{equation}
Here, we explicitly denote time-dependence, and the other quantities
are taken to be time-independent, which renders Eq.\,\ref{eq:R-PSI-eom}
linear. 
\emph{Tilde}-symbols on the SHF Hamiltonian matrices are used
to indicate that these quantities are evaluated using orthonormal SHFs. %the expansion
%of the wavefunction with the current orthonormality condition. 
The
SPF Hamiltonian matrix elements are unchanged from those in Eq.\,\ref{eq:mlmctdh-eom-gen}.
%always evaluated using orthonormal functions%basis
%and thus do not require this notation to distinguish them from the
%quantities in Eq.\,\ref{eq:mlmctdh-eom-gen}. 
Note that the structure
of Eq.\,\ref{eq:Atilde-PSI-eom} is similar to Eq.\,\ref{eq:mlmctdh-eom-gen},
with the exception of the omission of the mean-field density matrix inverse, which is trivially the identity matrix due to the chosen orthonormality condition
and the second term in projector $Q^{z_{l}}$ (see Eq.\,\ref{eq:proj}).  Second, we use an orthogonal decomposition
\begin{equation}
\orthrep{A}_{I^{z_l}i_{0}}^{z_{l}}=A_{I^{z_l}j_{0}}^{z_{l}}R_{j_{0}i_{0}}^{z_{l}}\quad|\quad A_{I^{z_l}j_{0}}^{*z_{l}}A_{I^{z_l}j_{0}^{\prime}}^{z_l}=\delta_{j_{0},j_{0}^{\prime}},\label{eq:AR-decomp}
\end{equation}
which gives a different orthonormality condition,
\begin{equation}
\begin{aligned}
U_{I^{z_l}_{p(z_l); a}i}^{*z_{l}^{\prime}}U_{I^{z_l}_{p(z_l) b}i}^{z_{l}^{\prime}}&=\delta_{ab}\ \forall\  z_{l}^{'}\in \mathrm{ancestors}(z_{l}), \\
A_{I^{z_l}i}^{*z_{l}^\prime}A_{I^{z_l}j}^{z_{l}^\prime} & =\delta_{ij}\ \forall\ z_l^{\prime} \not\in \mathrm{ancestors}(z_l),
\end{aligned}\label{eq:bond-orthonormality-cond}
\end{equation}
where $\mathrm{ancestors}(z_l)$ denotes the set of nodes that are ancestors of node $z_l$.
We can interpret $R^{z_l}$ as an additional, temporary
node in the tree. %, and here it is decorated with a \emph{tilde} symbol to
%indicate it as the coefficient tensor for which the remainder of the
%tree defines an orthonormal basis. 
We propagate $R^{z_{l}}$
backwards in time according to 
\begin{equation}
\dot{R}_{ij}^{z_l}(t)=-\frac{{i}}{\hbar}\sum_{r}A_{K^{z_l}i}^{*z_{l}}h_{r;K^{z_l}J^{z_l}}^{z_{l}}A_{J^{z_l}k}^{z_{l}}{R}_{kl}^{z_{l}}(t)\orthrep{H}_{r;lj}^{z_{l}}.\label{eq:R-PSI-eom}
\end{equation}
The structure of Eq.\,\ref{eq:R-PSI-eom} is again similar to that of Eq.\,\ref{eq:mlmctdh-eom-gen},
with the exception that Eq.\,\ref{eq:R-PSI-eom} only accounts for the
nontrivial term in projector $Q^{z_{l}}$ and the lack of the inverse
of the density matrix due to the chosen orthonormality condition.
The fact that this equation is solved backwards in time stems from
the negative sign in front of the second term in projector $Q^{z_{l}}$.
The requirement to solve Eqs.\,\ref{eq:Atilde-PSI-eom} and \ref{eq:R-PSI-eom}
sequentially, one node at a time, arises from the change of orthonormality
conditions associated with propagating different nodes. Note that
seriality is not inherently necessary for singularity-free approaches,
as discussed in the companion paper and in Refs. \onlinecite{Weike2021} and \onlinecite{Ceruti2021a}.

The PSI involves the changing of orthonormality
conditions between different nodes interspersed with the solution of linear
differential equations. In the following, we will introduce how
to efficiently convert between ML-MCTDH wavefunctions with different
orthonormality conditions before discussing the algorithm in detail.

\subsection{Conversion Between Different Orthonormality Conditions and Update
of Hamiltonian Matrix Elements}
\begin{figure}
\centering
\includegraphics{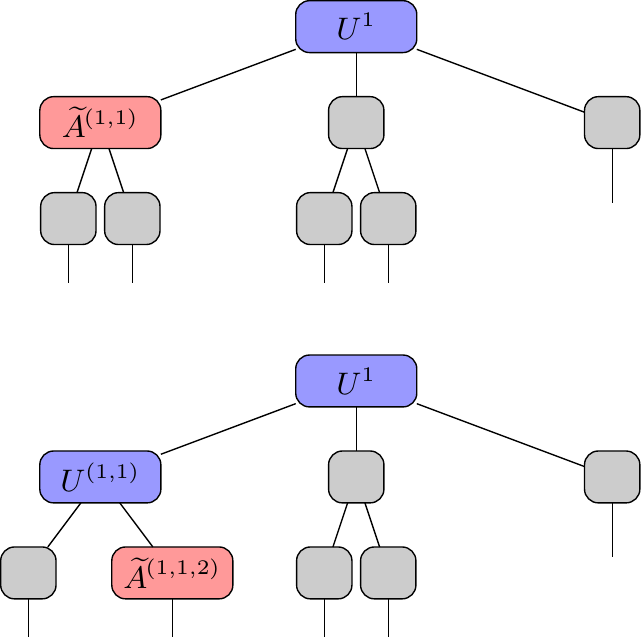}
\caption{\label{fig:mctdh_wavefunction_psi_rep} Diagrammatic representations of the ML-MCTDH wavefunction from Fig.\,\ref{fig:mctdh_wavefunction} transformed into the alternative representation where both the SPF and SHFs of node $z_l$ are orthonormal.  For the top tree $z_l = (1,1)$, while for the bottom tree $z_l=(1,1,2)$.  The coefficient tensors of the unlabelled (grey) nodes satisfy the orthonormality condition given in Eq.\,\ref{eq:orthonormal_cond_MLMCTDH}, and are left unchanged by this transformation.  The coefficient tensors of the nodes labelled with $U^{z_l}$ (blue) satisfy the orthonormality condition given in Eq.\,\ref{eq:orthonormality_cond_PSI}, and so will generally differ from the standard coefficient tensors.  Finally, the coefficient tensors labelled with $\orthrep{A}^{z_l}$ (red) are the transformed coefficient tensors given by Eq.\,\ref{eq:AR-decomp} that evolves according to the EOM given in Eq.\,\ref{eq:Atilde-PSI-eom}.}
\end{figure}

The orthonormality condition used in ML-MCTDH (Eq.\,\ref{eq:orthonormal_cond_MLMCTDH})
is just the special case of the orthonormality condition given in Eq.
\ref{eq:orthonormality_cond_PSI} when $z_l=1$. In order to be able to convert between Eq.\,\ref{eq:orthonormal_cond_MLMCTDH} and 
Eq.\,\ref{eq:orthonormality_cond_PSI} for any arbitrary $z_l$ and back, there are 
two operations we need to to perform:
\begin{itemize}
    \item Given that Eq.\,\ref{eq:orthonormality_cond_PSI} holds for node $z_l$ we need to be able to convert to a representation where it holds for the child node $(z_l, k)$ for any value of $k$.
    \item Given that Eq.\,\ref{eq:orthonormality_cond_PSI} holds for the node $(z_l,k)$ we need to able to convert to a representation where it holds for its parent node $z_l$.
\end{itemize}
We will begin by assuming that we have a representation in which Eq.\, \ref{eq:orthonormality_cond_PSI} holds for node $z_l$, 
and present how to transform the ML-MCTDH wavefunction so that this condition is 
enforced for a child node $(z_l,k)$.  In order to do this we need to construct a new coefficient tensor $U^{z_l}$
for node $z_l$ so that the orthogonality condition in
Eq.\,\ref{eq:orthonormality_cond_PSI}
is satisfied for node $(z_l,k)$.  This tensor is related to the coefficient tensors $\orthrep{A}^{z_l}$ by multiplication with a matrix ${R}^{(z_l,k)}$,
\begin{equation}
\orthrep{A}_{I^{z_l}_{k; b}i}^{z_{l}}=U_{I^{z_l}_{k; a}i}^{z_l}{R}_{ba}^{(z_l,k)}\quad|\quad U_{I^{z_l}_{k; a}i}^{*z_{l}}U_{I^{z_l}_{k; b}i}^{z_{l}}=\delta_{ab}.\label{eq:decomposition-descending}
\end{equation}
%For concreteness we will consider converting a representation with
%orthonormal SPFs (Eq. \ref{eq:orthonormal_cond_MLMCTDH}) to an expansion
%with orthonormal SHFs for a children node of the root node, $(0,k)$.\_We
%define a new coefficient tensor $A^{0}$, satisfying \ref{eq:orthonormality_cond_PSI}
%for node $(0,k)$, that is related to the original coefficient tensor
%$\widetilde{A}^{0}$ by multiplication with a matrix $\widetilde{R}^{0,k}$,
%\begin{equation}
%\widetilde{A}_{I_{/k}i_{k}}^{0}=B_{I_{/k}j_{k}}^{0}\widetilde{R}_{j_{k}i_{k}}^{0,k}\quad|\quad B_{I_{/k}j_{k}}^{*0}B_{I_{/k}j_{k}^{\prime}}^{0}=\delta_{j_{k},j_{k}^{\prime}}.\label{eq:orthonormal_decomp}
%\end{equation}
Note that the decomposition given in Eq.\,\ref{eq:decomposition-descending} is not unique,
which can be seen by inserting $I=U^{-1}U$ with a unitary $U$
between ${R}^{(z_l,k)}$ and $\orthrep{U}^{z_l}$. 
There are arbitrarily many matrices that provide equally valid decompositions.
%e are free to use any approach to obtain this decomposition.
For instance, one could apply a QR matrix decomposition on $\orthrep{A}^{z_l}$,
reshaped as a matrix of size
\begin{equation}
    \left(\displaystyle\prod_{\substack{j=0\\j\neq k}}^{d^{z_l}}n^{z_l}_{j}\right)\times n^{z_l}_{k},\label{eq:mat_dimens}
\end{equation}
with $n^{z_l}_{0}=N^{z_l}$, which returns a unique decomposition
with ${R}^{(z_l,k)}$ an upper triangular matrix if $\orthrep{A}^{z_l}$
does not have linearly dependent columns. If $\orthrep{A}^{z_l}$
does have linearly dependent columns, the corresponding entries in
$U^{z_l}$ can be chosen freely as long as they constitute orthonormal
columns. Contracting the matrix ${R}^{(z_l,k)}$ with the coefficient
tensor at node $(z_l,k)$,
\begin{equation}
    A_{I^{(z_l,k)}j_0}^{(z_l,k)}{R}_{j_{0}i_{0}}^{(z_l,k)}=\orthrep{A}_{I^{(z_l,k)}i_0}^{(z_l,k)}, \label{eq:contraction-descending}
\end{equation}
leaves us with an expansion of the wavefunction in terms of orthonormal
SPFs and SHFs for node $(z_l,k)$. 
Since we have modified the coefficient tensor of the node $z_l$, we need to update the mean-field Hamiltonian matrix elements for node $(z_l,k)$,
%\begin{equation}
%\begin{aligned}
%\widetilde{H}_{r_{i_{0},j_{0}}}^{0,k}&=B_{i_{0}^{\prime},i_{1}^{\prime}\dots %i^{\prime}{}_{k-1}i_{0}i_{k+1}\dots i_{d(z_{l})}}^{*0}\\ %&\left(\bigotimes_{k^{\prime}=1,k^{\prime}\neq %k}^{d(z_{l})}A_{i_{k^{\prime}}^{\prime},I_{/0}^{\prime}}^{*(z_{l},k^{\prime})}h_{r_{I_{/%0}^{\prime}J_{/0}^{\prime}}}^{(z_{l},k^{\prime})}A_{j_{k^{\prime}}^{\prime},J_{/0}^{\pri%me}}^{(z_{l},k^{\prime})}\right)\\&B_{j_{0}^{\prime},j_{1}^{\prime}\dots %j^{\prime}{}_{k-1}j_{0}j_{k+1}\dots j_{d(z_{l})}.}^{0} 
%\end{aligned}\label{eq:ham-mf-orthonormal}
%\end{equation}
\begin{equation}
\orthrep{H}_{r;ab}^{(z_{l},k)}=U_{I^{z_l}_{k;a} i}^{*z_{l}}\left(\prod_{\substack{k^{\prime}=1\\k^{\prime}\neq k}}^{d^{z_{l}}}M_{r;i_{k^\prime} j_{k^\prime}}^{(z_{l},k^\prime)}\right)\orthrep{H}_{r;i j}^{z_{l}}U_{I^{z_l}_{k; b} j}^{z_{l}},\label{eq:mf-update-descending}
\end{equation}
where $M_{r;i_{k^\prime} j_{k^\prime}}^{(z_{l},k^\prime)}$ is defined in Eq.\,\ref{eq:ham_spf}, and in order to ensure we can apply this equation for all nodes in the tree, we will 
define $\orthrep{H}^1_{r;ij}=1$ for $i=j=1$.
Note that this equation is very closely related to the standard
ML-MCTDH expression for the mean-field matrix $H_r^{(z_l,k)}$ for a child of node $z_l$, Eq.\,\ref{eq:ham-mf}.  
It is the same expression but with $A^{z_l}$ replaced with $U^{z_l}$.
As such, the process for evaluating these matrices does not change significantly when compared
to standard ML-MCTDH, other than that some additional optimizations become possible due to 
the orthonormality of these SHFs (see Sec. IV of the Supplementary Information). %Appendix\,\ref{appendix:a}).
Note that when transferring the  orthogonality condition to a child node, the SPF Hamiltonian matrices for the child node
remain unchanged and do not need to be updated.%, while the SPF Hamiltonian
%matrices for the parent node do not need to be recomputed because
%they are not used in the algorithm.

The reverse operation, converting the orthogonality condition from
node $(z_l,k)$ to its parent node $z_l$ proceeds as follows. First,
a decomposition of the form given in Eq.\,\ref{eq:AR-decomp} is obtained, 
\begin{equation}
\orthrep{A}_{I^{(z_l,k)}i_{0}}^{(z_{l},k)}=A_{I^{(z_l,k)}j_{0}}^{(z_{l},k)}{R}_{j_{0}i_{0}}^{(z_{l},k)}\quad|\quad A_{I^{(z_l,k)}j_{0}}^{(z_{l},k)*}A_{I^{(z_l,k)}j_{0}^{\prime}}^{(z_l,k)}=\delta_{j_{0}j_{0}^{\prime}}.\label{eq:decomposition-ascending}
\end{equation}
Then the SPF Hamiltonian matrices for node $z_l$ are updated according
to 
\begin{equation}
\begin{aligned}
h_{r;I^{z_l}J^{z_l}}^{z_{l}}=&\prod_{k=1}^{d^{z_{l}}}A_{I^{(z_l,k)}i_k}^{*(z_{l},k)}h_{r;I^{(z_l,k)}J^{(z_l,k)}}^{(z_{l},k)}A_{J^{(z_l,k)}
j_k}^{(z_{l},k)},%\\
%=&\prod_{k=1}^{d^{z_{l}}}M_{r;i_k j_k}^{(z_{l},k)},
\end{aligned}\label{eq:spf-update-ascending}
\end{equation}
which is exactly the same expression as in the standard ML-MCTDH case.
Finally we contract ${R}^{(z_l,k)}$ with $U^{z_l}$,
\begin{equation}
\orthrep{A}_{I^{z_l}_{k;b}i}^{z_l}={R}_{ba}^{(z_l,k)}U_{I^{z_l}_{k;a}i}^{z_l}.\label{eq:contraction-ascending}
\end{equation}
When transferring the orthogonality condition to the parent of node $z_l$,
the SHF Hamiltonian matrices for the parent remain unchanged and do
not need to be updated, while the SHF Hamiltonian matrices for node $z_l$ do not appear in the algorithm's working equations and
thus do not need to recomputed.

In calculations it is not necessary to form the
SPF Hamiltonian matrices $h_{r}^{z_{l}}$, all required operations can be expressed in terms of the matrices $M_r^{(z_{l},k)}$ associated with 
the child nodes of node $z_l$.  The SPF Hamiltonian matrix is
the Kronecker product of each of these matrices.  This representation
leads to an efficient scheme for applying the SPF Hamiltonian matrix to
the coefficient tensors in terms of a series of matrix-tensor contractions.
Additionally, it is useful to make use of an alternative SOP form as discussed in Sec. IV of the Supplementary Information.%Appendix \ref{appendix:a}.

We can convert a ML-MCTDH wavefunction for which Eq.\,\ref{eq:orthonormality_cond_PSI}
is satisfied for node $z_l$ to one for which it is satisfied for node
$z_l^\prime$ by performing the above operations sequentially along the
path through the tree that connects nodes $z_l$ and $z_l^\prime$.% in the tree associated with the
%ML-MCTDH expansion.

\subsection{Integrating the Singularity-Free Equations of Motion: the Projector Splitting
Integrator}

\begin{figure}[ht]
\begin{centering}
\includegraphics{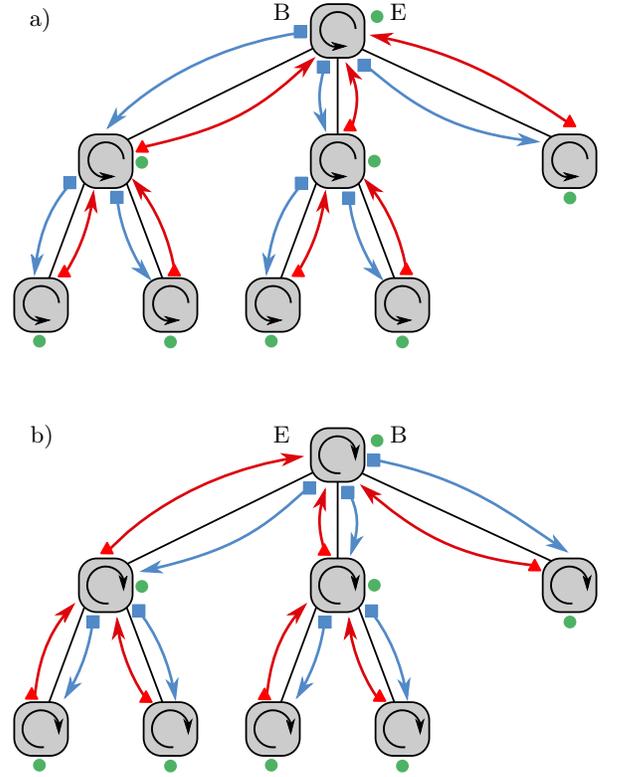}
\par\end{centering}
\caption{\label{fig:psi_tree_traversal} The two node traversal schemes used
in the implementation of the PSI described in this work. a) Represents the forward 
walk step. b) Representation of the backwards walk step.
The blue squares correspond to the construction of the orthonormal SHFs
followed by the updating of mean-field matrices. The green circles
correspond to evolution of the transformed coefficient tensors. The
red triangles correspond to reformation of the standard coefficient
tensors. The arrows connecting nodes correspond to transfers of the
non-orthogonal $\boldsymbol{{R}}^{z}(t)$ matrix between nodes of
the tree, and the black arrows in each node indicate the order in
which these operations occur, e.g. operations on a node are applied
counter-clockwise in the forward step and clockwise in the backwards
step. For both walks the beginning (B) and end (E) points are
shown.}
\end{figure}

The PSI propagates the coefficient tensor
at node $z_{l}$, $A^{z_{l}}$, for a fixed time step $d t$ using
a wavefunction expansion with orthonormal SPFs and SHFs for node $z_{l}$,
while keeping the coefficient tensors at other nodes constant during
the update at node $z_{l}$. Neglecting the time dependence of quantities
involving other coefficient tensors is similar in spirit to the well-established
constant mean-field (CMF) integration scheme for the standard ML-MCTDH
EOMs, with the exception that the PSI updates
the coefficients at the nodes sequentially. The order in which nodes
are propagated is in principle arbitrary. Here we will discuss an
efficient and easily illustrated order, namely an Euler tour traversal,\citep{DSAGoodrich} 
starting from the root node. 
For the single-layer case this traversal reduces to the previously published 
version for MCTDH wavefunctions.\citep{Kloss2017a} 
%I'm a bit hesitant to call it tree-traversal.  Tree-traversals only access each 
%node once.
%however an efficient and
%easily illustrated order is obtained by using tree-traversals. 
%The
%starting point of this traversal can be chosen at any node of the
%tree. Here, we choose the root node as a reference, in which case the
%ML-MCTDH PSI reduces to its previously published single-layer version.
% previously published single-layer version
In analogy with the single-layer version, a symmetric, second-order
integrator is obtained by combining a propagation step with its adjoint.
Specifically, this means that a single time-step consists of two substeps: a forward
and a backward walk through the tree structure, each of which propagates all
coefficient tensors by $dt/2$.
%Specifically, this means that a single time-step consists of the following
%three substeps: 
%\begin{itemize}
%    \item A forward-walk that propagates all non-root coefficient
%tensors by $dt/2$,
%\item Propagation of the root node coefficient
%tensor by $dt$, and
%\item  A backward-walk that propagates all
%non-root coefficient tensors by $dt/2$.
%\end{itemize}
We restate the equations
of motion to be solved for the reader's convenience expressed using the matrix
notation employed in the companion paper:
\begin{equation}
\begin{aligned}\dotorthrep{{\boldsymbol{{A}}}}^{\boldsymbol{{z}}_{l}}(t)=-\frac{{i}}{\hbar}\sum_{r} & \boldsymbol{{h}}_{r}^{\boldsymbol{{z}}_{l}}\orthrep{{\boldsymbol{{A}}}}^{\boldsymbol{{z}}_{l}}(t)\orthrep{{\boldsymbol{{H}}}}_{r}^{\boldsymbol{{z}}_{l}},\end{aligned}
\label{eq:sf_eom_1-1}
\end{equation}
and
\begin{equation}
\begin{aligned}\dot{{\boldsymbol{{R}}}}{}^{\boldsymbol{{z}}_{l}}(t)= & -\frac{{i}}{\hbar}\sum_{r}\boldsymbol{{A}}^{\boldsymbol{{z}}_{l}\dagger}\boldsymbol{{h}}_{r}^{\boldsymbol{{z}}_{l}}\boldsymbol{{A}}^{\boldsymbol{{z}}_{l}}\boldsymbol{{R}}^{\boldsymbol{{z}}_{l}}(t)\orthrep{{\boldsymbol{{H}}}}_{r}^{\boldsymbol{{z}}_{l}},\end{aligned}
\label{eq:sf_eom_2-1}
\end{equation}
for $z_{l}\neq1$, and 
\begin{equation}
\dotorthrep{\boldsymbol{A}}^{1}(t)=-\frac{i}{\hbar}\boldsymbol{{h}}{}^{1}\orthrep{\boldsymbol{A}}^{1}(t)\label{eq:sf_root_node_eom}
\end{equation}
for the root node coefficient tensor. The integrator is described
schematically in Fig.\,\ref{fig:psi_tree_traversal} and verbally
in the following paragraphs.

\paragraph*{Forward walk:}

The forward walk starts and ends with a wavefunction that fulfills
Eq.\,\ref{eq:orthonormality_cond_PSI} for the root node $z_{l}=1$. 
The walk on the tree is counter-clockwise, namely the next
node will always be the closest to the last visited node in a
counter-clockwise direction. This ensures that every link on the 
tree is traversed exactly
twice, once while descending and once while ascending. The first node visited after the root  
node is its leftmost child node. During the
forward walk, descending the tree only involves transferring the orthonormality condition given by Eq.\,\ref{eq:orthonormality_cond_PSI} from node $z_l$ to the next node (one of its children), according to the procedure
outlined above. Furthermore, we update the mean-field Hamiltonian
matrix elements, $\orthrep{{\boldsymbol{{H}}}}_{r}^{\boldsymbol{{z}}_{l}}$,
to be consistent with the current expansion of the wavefunction (see
Eq.\,\ref{eq:decomposition-descending}). When
the walk ascends from node $(z_l,k)$ to node $z_l$, we solve the EOMs
Eqs.\,\ref{eq:sf_eom_1-1} and \ref{eq:sf_eom_2-1}. Specifically,
we propagate $\orthrep{\boldsymbol{A}}^{(z_l,k)}(t_{0})$ to
$\orthrep{\boldsymbol{A}}^{(z_l,k)}(t_{0}+dt/2)$ using
Eq.\,\ref{eq:sf_eom_1-1} and for all non-root nodes use the decomposition $\orthrep{\boldsymbol{A}}^{(z_l,k)}(t_{0}+dt/2)={\boldsymbol{{A}}}^{(z_l,k)}(t_{0}+dt/2)\boldsymbol{R}^{(z_l,k)}(t_{0}+dt/2)$
(see Eq.\,\ref{eq:decomposition-ascending}). After updating SPF
Hamiltonian matrix elements of node $z_l$ using ${\boldsymbol{{A}}}^{(z_l,k)}(t_{0}+dt/2)$
(see Eq.\,\ref{eq:spf-update-ascending}), we propagate $\boldsymbol{R}^{(z_l,k)}(t_{0}+dt/2)$
backwards in time to $\boldsymbol{R}^{(z_l,k)}(t_{0})$ using Eq.\,\ref{eq:sf_eom_2-1}.
Contracting $\boldsymbol{R}^{(z_l,k)}(t_{0})$ with the coefficient tensor
of node $z_l$ establishes the orthogonality condition Eq.\,\ref{eq:orthonormality_cond_PSI}
for node $z_l$ and completes the step from node $(z_l,k)$ to $z_l$.

%\paragraph*{root node update:}

%The root node coefficient tensor is propagated from $\orthrep{\boldsymbol{A}}^{1}(t_{0})$
%to $\orthrep{\boldsymbol{A}}^{1}(t_{0}+dt)$ according to Eq.
%\ref{eq:sf_root_node_eom}, which is identical to the EOM for the
%root node tensor in standard ML-MCTDH.

\paragraph*{Backward walk:}

The backward walk starts and ends with a wavefunction that fulfills
Eq.\,\ref{eq:orthonormality_cond_PSI} for $z_{l}=1$, i.e. at the
root node. The walk on the tree is clockwise, i.e. the next node is
the closest to the last visited node in a clockwise direction. During the walk,
every link on the tree is traversed exactly twice, once while descending
and once while ascending. The first node visited after the root node is its rightmost
child node. During the backward walk, ascending the tree only
involves transferring the orthonormality condition given by Eq.\,\ref{eq:orthonormality_cond_PSI} from a node $(z_l, k)$ to its parent node $z_l$. Furthermore, we will
update the SPF Hamiltonian matrix elements, $\boldsymbol{h}_{r}^{z{l}}$,
to be consistent with the current expansion of the wavefunction (see
Eq.\,\ref{eq:spf-update-ascending}). When the walk descends
from node $z_l$ to node $(z_l,k)$, we solve the EOMs Eqs.\,\ref{eq:sf_eom_1-1}
and \ref{eq:sf_eom_2-1}. Specifically, we obtain $\boldsymbol{R}^{(z_l,k)}$
by decomposing $\orthrep{\boldsymbol{A}}^{z_l}$ according  to Eq.\,\ref{eq:decomposition-descending}, update the mean-field
matrix elements for node $(z_l,k)$ (see Eq.\,\ref{eq:mf-update-descending})
and propagate $\boldsymbol{R}^{(z_l,k)}$ backwards in time
from $\boldsymbol{R}^{(z_l,k)}(t_0+dt)$ to $\boldsymbol{R}^{(z_l,k)}(t+dt/2)$
using Eq.\,\ref{eq:sf_eom_2-1}. $\boldsymbol{R}^{(z_l,k)}(t+dt/2)$
is contracted with ${\boldsymbol{{A}}}^{(z_l,k)}(t_{0}+dt/2$)
(see Eq.\,\ref{eq:contraction-descending}), yielding $\orthrep{\boldsymbol{A}}^{(z_l,k)}(t_{0}+dt/2)$
and restoring the orthogonality condition Eq.\,\ref{eq:orthonormality_cond_PSI}
for node $(z_l,k)$. $\orthrep{\boldsymbol{A}}^{(z_l,k)}(t_{0}+dt/2)$
is propagated forward in time according to Eq.\,\ref{eq:sf_eom_1-1},
which completes the step from node $z_l$ to $(z_l,k)$.

%The evolution of the root node coefficient tensor is the last step of the forward loop and the
%first step of the backward loop.  As such, the two evolutions each through the time steps $dt/2$ can be combined into a single step

\subsection{Discussion of the PSI}
The PSI can be readily implemented within existing ML-MCTDH implementations.
Since the working equations are similar to those employed in standard
ML-MCTDH, the only components which may not already be part of the
code are the Euler tour traversal and the orthonormal decomposition of coefficient
tensors in Eq.\,\ref{eq:decomposition-descending}. Algorithms for the former
are readily available,\citep{DSAGoodrich} 
%probably add another reference. This book only discusses the traversal for binary trees.
and efficient linear algebra routines to perform
an orthogonal matrix decomposition%, for example a QR or singular value decomposition, 
are part of most linear algebra packages. Employing
the latter requires reshaping of the coefficient tensor into a matrix
with size given by Eq.\,\ref{eq:mat_dimens}, with $k$ chosen as required.
%I don't think this point is necessary, yes it is true but it is an unimportant 
%implementation detail, and it is one that I think would lead to considerably 
%poorer performance.  The reshaping is essentially free compared (\mathcal{O}(N^2)) 
%to the decomposition (\mathcal{O}(N^3)).  
%The larger number of non-contiguous memory accesses required in doing this likely 
%removes any benefits.
%Note that, in principle, the
%decomposition can be implemented manually without recourse to matrix
%reshaping.

A potential drawback of the PSI presented here is its inherent
seriality. CMF schemes for ML-MCTDH on the other hand allow for straightforward
parallel integration. Parallel versions of the PSI should be possible
to construct, with a version for single-layer MCTDH having already
been proposed \citep{Ceruti2021a} and a related scheme having been
discussed for ML-MCTDH.\citep{Weike2021} At present, such
approaches tend to require much smaller time steps than the PSI 
(see Sec. II of the Supplementary Information), which
largely negates the speedup obtained through parallelization in many
cases.

The PSI has been shown to be stable with respect to almost or fully
rank deficient wavefunction parametrizations in its MCTDH\citep{Kloss2017a}
and ML-MCTDH formulations,\citep{Bauernfeind2020,Kloss2020a,Ceruti2021, lindoy_thesis}
including the important subcategory of matrix product states (maximally
layered trees with physical degrees of freedom at each layer).\citep{Schroeder2016,Kloss2017,Paeckel2019}
The stability of the PSI with respect to nearly rank-deficient wavefunction
parametrizations can be attributed to the fact that the EOMs solved
are always well-conditioned. In ML-MCTDH, the regularization parameter
controls how ill-conditioned the equations are and how rapidly the variational
parameters with vanishing weight in the wavefunction expansion evolve.
In addition, it influences the weight above which the evolution of these parameters
becomes accurate.
The PSI on the other hand has no adjustable parameter except for
time-step size, and rank-deficiency leads to non-uniqueness of the
orthogonal decomposition but does not increase the stiffness of the
underlying EOMs. In exceptional cases, any ML-MCTDH approach can suffer
from inaccurate results for certain choices of initial unoccupied SPFs\citep{Manthe2018} and the PSI could additionally show dependence
on the choice of redundant parameters in the orthogonal decompositions.
We will provide numerical evidence in the following that this effect
can be suppressed by simply increasing the number of SPFs.

%The PSI generally allows for favorably large time steps, although existing
%benchmarks focus on other types of problems than the ones typically
%considered when testing ML-MCTDH. 
The following section will address
the lack of benchmarks and demonstrate the stability and performance
of the PSI in a set of typical applications of ML-MCTDH. 
Before we
move on, we would like to point to developments towards a dynamical
adaption of the number of SPFs at each node of the ML-MCTDH expansion
during the propagation.\citep{Larsson2017,Yang2020,MendiveTapia2020,Ceruti2021b} These approaches
can speed up challenging calculations, but more importantly algorithms that make
use of higher powers of the Hamiltonian\citep{Yang2020, MendiveTapia2020} address,
at least partially, the fundamental shortcoming of the time-dependent
variational principle.  They provide information about the optimal evolution 
of redundant variational parameters, however at the cost of requiring the 
evaluation of the action of higher powers of the Hamiltonian on the wavefunction.
While some of these approaches have been formulated
for matrix product states, they are equally applicable to generic
ML-MCTDH wavefunctions.

\section{Results \label{Sec:Results}}

In order to demonstrate the capabilities of the PSI, we will consider
a series of 3 models. The first, a simple two-dimensional bi-linearly
and bi-quadratically coupled harmonic oscillator model, will be used
to explore the extent to which the choice of unoccupied SPFs influences
the dynamics obtained. The second set of models will be a series of
spin-boson models with strong system-bath coupling and varying numbers
of bath modes. This set of models will be used to demonstrate the
effect that increasing the bath size has on the convergence properties
of the PSI and how well the PSI can treat Hamiltonians with terms that
couple all modes to all other modes. The final set of models
that will be considered are a series of multi-spin-boson models with
varying numbers of two-level systems (TLSs). These models will serve
as a test of the PSI when larger number of SPFs are required at each
node. 

\subsection{Two-Dimensional, Bi-Linearly and Bi-Quadratically Coupled Harmonic
Oscillator Model }

 We start by considering a two-dimensional oscillator model
that is described by the Hamiltonian\citep{Manthe2015b} 
\begin{equation}
\begin{aligned}\hat{{H}}= & \frac{{1}}{2}\left(-\frac{{\partial^{2}}}{\partial x^{2}}+x^{2}\right)+\frac{{1}}{2}\left(-\frac{{\partial^{2}}}{\partial y^{2}}+x^{2}\right)\\
 & +\frac{{1}}{4}xy+\frac{{1}}{2}\left(x^{2}-\frac{{1}}{2}\right)\left(y^{2}-\frac{{1}}{2}\right).
\end{aligned}
\end{equation}
This model
is particularly useful when assessing the behavior of unoccupied
SPFs. When the initial wavefunction is taken to be the non-interacting
ground state, ($\Psi_{0}(x,y)=\chi_{0}(x)\chi_{0}(y)$, where $\chi_{n}$
is the n-th eigenfunction of the unitless one-dimensional harmonic
oscillator), it is possible to evaluate the optimal unoccupied SPFs
analytically, and their short time evolution, giving\citep{Manthe2015b}
\begin{align}
\phi_{opt}^{(1)} & (z)=\chi_{2}(z),\\
\phi_{opt}^{(2)} & (z)=\chi_{1}(z),\ \ \ z=x,y.\label{eq:non_dominant_spf}
\end{align}

For short times, the first of these function will provide
the dominant correction to the evolution of the wavefunction and
will be referred to as the dominant unoccupied SPF. In Ref. \onlinecite{Manthe2015b}, 
it was demonstrated that when two SPFs are used for each mode,
the dynamics obtained by conventional MCTDH are sensitive to the choice
of the initially unoccupied SPF. The dynamics obtained using the PSI
has also been shown to exhibit a dependence on the initially unoccupied
SPFs.\citep{Kloss2017a} When the unoccupied SPF initially has considerable overlap
with the non-dominant unoccupied SPF, Eq.\,\ref{eq:non_dominant_spf}, it may not capture the
dominant contribution to the dynamics. 
It should be noted that in these tests, only a very
specific set of initial conditions are considered.

 Here we consider a series of different initial configurations
in an attempt to further explore the dependence on initial conditions.
We will employ a $N_{hilb}$-dimensional Hilbert space of harmonic
oscillator basis functions for each mode and uniformly sample normalized
unoccupied SPFs within the space that is orthogonal to the harmonic
oscillator ground state. In Table\,\ref{tab:The-number-of},
we present the percentage of wavepacket evolutions (out of a total of 1000) in which the unoccupied SPF did not evolve following the dominant unoccupied SPFs. 
\begin{table}[ht]
\begin{centering}
\begin{tabular}{>{\centering}p{1cm}|>{\centering}p{1cm}>{\centering}p{1cm}>{\centering}p{1cm}>{\centering}p{1cm}}
$N_{SPF}$ & \multicolumn{4}{c}{$2$}\tabularnewline
\hline 
$N_{hilb}$ & $10$ & $10^{2}$ & $10^{3}$ & $10^{4}$\tabularnewline
$\%_{f}$ & 0 & 0 & 1.2 & 10.8\tabularnewline
\hline 
\multicolumn{1}{>{\centering}p{1cm}}{} &  &  &  & \tabularnewline
$N_{SPF}$ & \multicolumn{4}{c}{$3$}\tabularnewline
\hline 
$N_{hilb}$ & $10$ & $10^{2}$ & $10^{3}$ & $10^{4}$\tabularnewline
$\%_{f}$ & 0 & 0 & 0 & 0\tabularnewline
\hline 
\end{tabular}
\par\end{centering}
\caption{\label{tab:The-number-of}The percentage of wavepacket evolutions
($\%_{f}$) where the unoccupied SPFs failed to evolve following the
dominant unoccupied SPF for different local Hilbert space dimensions
($N_{hilb}$). 1000 wavepacket evolutions were performed for each
case with the initial states of the unoccupied SPFs being uniformly
sampled from the space orthogonal to the harmonic oscillator ground
state. Here we consider cases where we have 2 and 3 SPFs per mode.}
\end{table}
When using two SPFs per mode we observe that none of the 1000 trajectories that were run fail
to evolve according to the dominant optimal unoccupied SPF when using primitive Hilbert
space dimension of up to 100.  It is
only for large primitive Hilbert space dimensions (10000), that an appreciable
number of trajectories ($\sim11\%)$ follow the non-dominant solution. The
average overlap of a randomly sampled vector in a $D$-dimensional
space with an arbitrary vector in this space scales as $D^{\text{\text{{-}}}1}$.
As such, for small local Hilbert space dimensions a randomly sampled
vector will, on average, have a relatively large overlap with the dominant
solution, and at least for this model will track the dominant solution. 

As has previously been observed for the PSI,\citep{Kloss2017a} 
when 3 SPFs are used per mode the dynamics becomes independent
of the states of the initially unoccupied SPFs, and all trajectories
track the two optimal unoccupied SPFs. This suggests that, at least
for this model, the issues associated with the unoccupied SPFs are
a result of insufficient convergence with respect to the number of
SPFs. While one of the unoccupied SPF provides the larger contribution
to the dynamics initially, both contribute considerably to the dynamics,
and as soon as there are enough SPFs to capture both important SPFs,
the dependence on initial conditions entirely vanishes. As the number
of SPFs increases towards a full Hilbert space calculation, the PSI is 
guaranteed to produce the correct result regardless
of the initially unoccupied SPFs. Additionally, even when there are
insufficiently many SPFs, this problem is unlikely to manifest
unless specific poor choices of the initially unoccupied SPF are selected.
Whether this holds true generally is outside of the scope of this
work. However we do not observe issues related to the initially unoccupied
SPFs in the following applications to models that are more representative
of those that would be treated using ML-MCTDH based approaches.

\subsection{Spin-Boson Model \label{Sec:Spin_boson} }
The two-mode model treated in the previous section is not
particularly representative of the types of models conventionally
treated using ML-MCTDH. We now consider a number of larger problems,
starting with a series of spin-boson models in which a single two-level
system (TLS) is linearly coupled to a harmonic bath.\citep{garg1985,Leggett1987,nitzan2006,Weiss2012} The spin-boson
Hamiltonian may be written as
\begin{equation}
\hat{{H}}=\varepsilon\hat{{\sigma}}_{z}+\Delta\hat{{\sigma}}_{x}+\hat{{\sigma}}_{z}\sum_{k}g_{k}(\hat{{a}}_{k}^{\dagger}+\hat{{a}}_{k})+\sum_{k}\omega_{k}\hat{{a}}_{k}^{\dagger}\hat{{a}},
\end{equation}
 where $\hat{{a}}_{k}^{\dagger}$ and $\hat{{a}_{k}}$ are
the bosonic creation and annihilation operators associated with the
$k$th bath mode, and 
\begin{equation}
\hat{{\sigma}}_{x}=\left|0\right\rangle \left\langle 1\right|+\left|1\right\rangle \left\langle 0\right|
\end{equation}
 and
\begin{equation}
\hat{{\sigma}}_{z}=\left|0\right\rangle \left\langle 0\right|-\left|1\right\rangle \left\langle 1\right|
\end{equation}
are Pauli matrices. The influence of the system on the bath
is fully characterised by the bath spectral density\citep{Leggett1987,Weiss2012}
\begin{equation}
\begin{aligned}J(\omega) & =\frac{{\pi}}{2}\sum_{k}g_{k}^{2}\delta(\omega-\omega_{k})\end{aligned}
\end{equation}
 from which the system-bath coupling constants, $g_{k}$,
and bath frequencies, $\omega_{k}$, can be obtained. We will
consider an Ohmic spectral density with an exponential cutoff
\begin{equation}
J(\omega)=\frac{{\pi}}{2}\alpha\omega e^{-\omega/\omega_{c}},
\end{equation}
 where $\alpha$ is the dimensionless Kondo parameter and
$\omega_{c}$ is the cutoff frequency of the bath. 

%I rearranged this section as I realised that the way it was structured before was a bit confusing.
An alternative representation of the spin-boson Hamiltonian can be obtained by applying the polaron transform \citep{Weiss2012}
\begin{equation}
\hat{{T}}=\exp\left[-\hat{{\sigma}}_{z}\sum_{k}\frac{{g_{k}}}{\omega_{k}}\left(\hat{{a}}_{k}^{\dagger}-\hat{{a}}_{k}\right)\right],
\end{equation}
which displaces the bath modes depending on the states of the TLS, to this Hamiltonian.
The polaron-transformed spin-boson Hamiltonian can be written as
\begin{equation}
\begin{aligned}\hat{{H}}_{pol}=\  & \hat{{T}}^{\dagger}\hat{{H}}\hat{{T}}\\
=\  & \varepsilon\hat{{\sigma}}_{z}+\sum_{k}\omega_{k}\hat{{a}}_{k}^{\dagger}\hat{{a}}\\
 + & \left[\hat{{\sigma}}_{+}\exp\left[-2\sum_{k}\frac{{g_{k}}}{\omega_{k}}\left(\hat{{a}}_{k}^{\dagger}-\hat{{a}}_{k}\right)\right]+\mathrm{{h.c.}}\right],
\end{aligned}
\label{eq:polaron_transformed_hamiltonian}
\end{equation}
in which the final term directly couples all bath and system modes in a multiplicative fashion.
It has previously been show that the polaron-transformed spin-boson model 
provides a considerably harder challenge for ML-MCTDH than the standard representation.\citep{H.D.2018,Wang2018}

 Before we can apply the PSI to
simulating the unitary dynamics arising from either of these Hamiltonians, it
is necessary to discretize the bath and consequently the spectral
density. Following standard approaches,\citep{Wang2001c,Wang2018} we will consider a discrete bath of $N$ modes with system-bath coupling constant given by 
\begin{equation}
g_{k}^{2}=\frac{{2}}{\pi}\frac{{J(\omega_{k})}}{\rho(\omega_{k})}.
\end{equation}
Here, $\rho(\omega)$ is the density of frequencies, which
is taken to be
\begin{equation}
\rho(\omega)=\gamma e^{-\omega/\omega_{c}},\label{eq:exponential_frequency_dist}
\end{equation}
with $\gamma$ chosen to ensure correct normalization and the frequencies are given by
\begin{equation}
\begin{split}\int_{0}^{\omega_{k}}\rho(\omega)\mathrm{{d}\omega}=k, & \ \ \ \ k=1,2,3,\dots,N.\end{split}
\label{eq:discrete_frequencies}
\end{equation}

 In the following discussion, we will consider the time-dependent
population difference between the states of the two-level system as
our observable of interest. This population difference is given by
\begin{equation}
P(t)=\left\langle \Psi(t)\right|\hat{{\sigma}}_{z}\left|\Psi(t)\right\rangle,
\end{equation}
 where the initial wavefunction is taken as
\begin{equation}
\left|\Psi(0)\right\rangle =\left|0\right\rangle \otimes\left|vac\right\rangle ,
\end{equation}
with $\left|vac\right\rangle $ denoting the vacuum state of the bath in the standard representation.  Note here we only work at $T=0$.
Following Refs. \onlinecite{Wang2018} and \onlinecite{Wang2021}, we will quantify the convergence of
the population dynamics by using the relative cumulative deviation
\begin{equation}
\Delta P=\frac{{1}}{\left(P_{max}-P_{min}\right)}\frac{{1}}{\tau}\int_{0}^{\tau}|P(t)-P_{ref}(t)|\mathrm{{d}}t
\end{equation}
 from the population difference obtained from some reference
calculation, $P_{ref}(t)$, obtained over a time period $\tau$. Here,
$P_{min}$ and $P_{max}$ are the minimum and maximum population differences
throughout this time-period. 

 We will consider a relatively strongly coupled, unbiased ($\epsilon=0$)
spin-boson model with $\alpha=2$ and a cutoff frequency of $\omega_{c}=25\Delta$.
For all calculations here and in the Supplementary
Information, the tree structure is constructed as follows. On the
top layer, we consider $N_{b}+1$ groups of SPFs, 1 of which accounts
for the system degrees of freedom and contains a complete basis for
the system (here this is 2 functions). The remaining $N_{b}$ groups
of SPFs treat the bath degrees of freedom, with the bath modes being
partitioned between the groups so that they contain modes of similar
frequencies. The bath SPFs in each of the $N_{b}$ groups are then
represented using as close to a balanced tree as possible with each node having up to
$N_{lower}$ SPF groups. Mode combination and adiabatic contraction are used for
the bottom layer, with modes being grouped up to some maximum local
Hilbert space.   %Maybe add a figure?

\subsubsection*{Standard Representation}
 We will begin by considering a bath of $N=500$ oscillators
with $N_{b}=2$ and $N_{lower}=2$, and we will use 16 SPFs at each
node in the bath subtrees. For each bath mode we will use a basis
of 15 harmonic oscillator eigenfunctions and will allow for a maximum
combined Hilbert space dimension of 3000.  This will typically allow
for 3 bath modes to be combined. We use an adaptive order, adaptive
time step Lanczos integration scheme\citep{Park1986,Sidje1998} for integrating the
linear systems of equations associated with each node, and in all
calculations that follow we used an integration tolerance of $10^{\text{{-}12}}$.
In Table III of the Supplementary Information we consider the effect
of varying this tolerance on the accuracy and efficiency of the simulations,
and find that such a tolerance is more than sufficient to provide
converged dynamics. 
\begin{figure}[ht]
\begin{centering}
\includegraphics{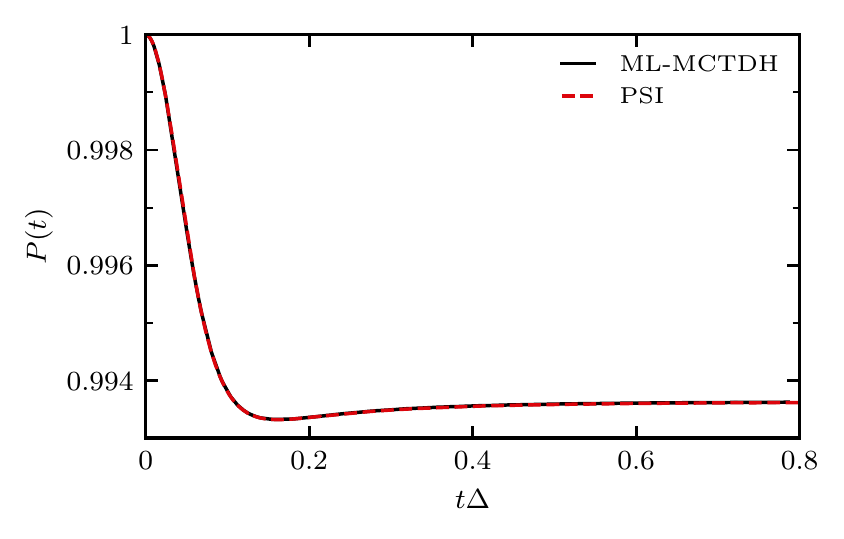}
\par\end{centering}
\caption{\label{fig:system_2_comparison} The time-dependent expectation value
$P(t)$ obtained for a spin-boson model with $\varepsilon/\Delta=0$,
$\alpha=2$, $\omega_{c}=25\Delta$, and with $N=500$ bath modes.
Here the results obtained using the PSI are compared against the results
obtained using the standard ML-MCTDH approach in Ref. \onlinecite{Wang2018}.}
\end{figure}

 The dynamics of $P(t)$ for this model has previously
been obtained using standard ML-MCTDH and with
the alternative regularization scheme mentioned in the introduction.\citep{Wang2018} In Fig.
\ref{fig:system_2_comparison}, we compare the results obtained using
the PSI to their results in an effort to
demonstrate the validity of our implementation of the PSI. Here we see population difference dynamics that agree
well over the entire time period considered. Having validated our implementation
of the multi-layer version of the PSI, we
are now in a position to look at some more challenging cases.

\subsubsection*{Polaron-Transformed Representation}
\begin{figure}[ht]
\begin{centering}
\includegraphics{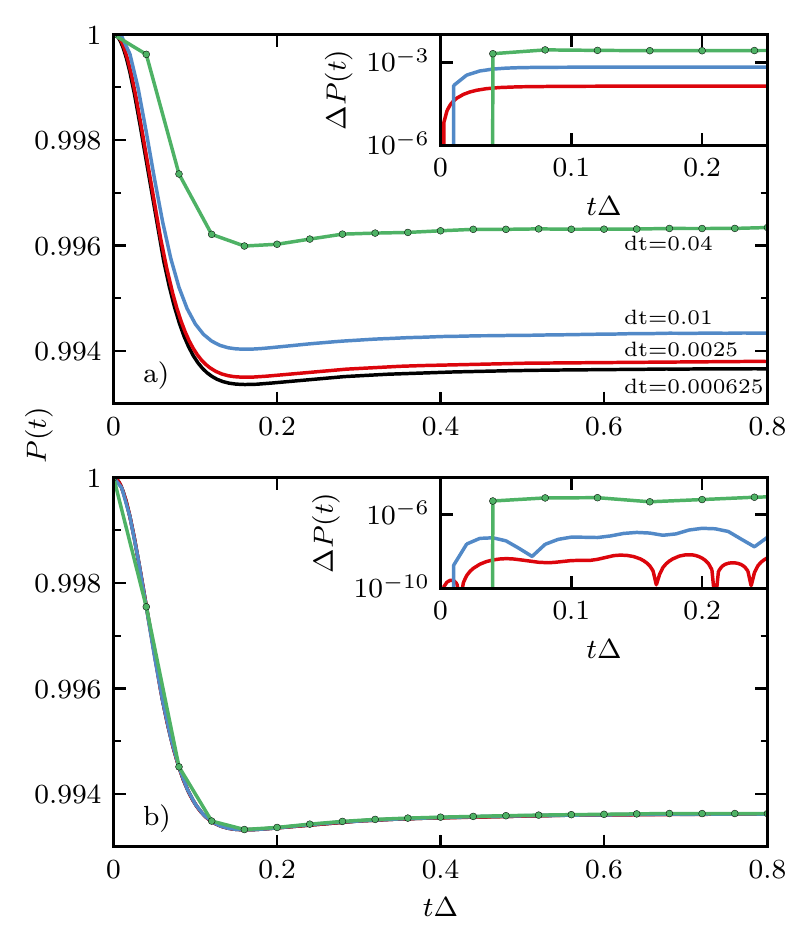}
\par\end{centering}
\caption{\label{fig:system_2_timestep_convergence}The effect of time step on
the convergence of $P(t)$ for a polaron-transformed spin-boson model with $\varepsilon/\Delta=0$,
$\alpha=2$, $\omega_{c}=25\Delta$, and with $N=500$ bath modes.
a) Results obtained using a fixed time step integrator
(with time steps shown). b) Results alternative
scheme that makes use of a series of smaller time steps to integrate form $t=0$ to $t=dt$ but uses the same
fixed step as in panel a) for all times $t>dt$. The insets in each panel
show the absolute difference of the results obtained with a given
time step and the reference calculation obtained with $dt\Delta=0.000625$. }
\end{figure}

It has previously been shown that the standard ML-MCTDH approach fails
to capture the dynamics of the polaron-transformed representation of the spin-boson model.\citep{H.D.2018,Wang2018}  This is due to the
inability of the regularized ML-MCTDH EOMs to accurately capture the rapid
evolution of unoccupied SPFs required for this
representation, and an alternative regularization scheme is required
to obtain accurate dynamics.\citep{H.D.2018,Wang2018} 
%This Hamiltonian is obtained by applying the polaron transform \citep{Weiss2012}
%\begin{equation}
%\hat{{T}}=\exp\left[-\hat{{\sigma}}_{z}\sum_{k}\frac{{g_{k}}}{\omega_{k}}\left(\hat{{a}}_{k}^{\dagger}-\hat{{a}}_{k}\right)\right]
%\end{equation}
%which displaces the bath modes depending on the states of the TLS.
%The polaron-transformed Hamiltonian can be written as
%\begin{equation}
%\begin{aligned}\hat{{H}}_{pol}=\  & \hat{{T}}^{\dagger}\hat{{H}}\hat{{T}}\\
%=\  & \varepsilon\hat{{\sigma}}_{z}+\sum_{k}\omega_{k}\hat{{a}}_{k}^{\dagger}\hat{{a}}\\
% + & \left[\hat{{\sigma}}_{+}\exp\left[-2\sum_{k}\frac{{g_{k}}}{\omega_{k}}\left(\hat{{a}}%_{k}^{\dagger}-\hat{{a}}_{k}\right)\right]+\mathrm{{h.c.}}\right],
%\end{aligned}
%\label{eq:polaron_transformed_hamiltonian}
%\end{equation}
%in which the final term directly couples all bath and system modes in a multiplicative fashion.
%It has previously been shown that the standard ML-MCTDH approach fails
%to capture the dynamics in this representation due to the
%inability of the regularized equations to accurately capture the rapid
%evolution of unoccupied SPFs required for this
%Hamiltonian, and an alternative regularization scheme is required
%to obtain accurate dynamics.\citep{H.D.2018,Wang2018} 
In panel a) of Fig.\,\ref{fig:system_2_timestep_convergence}, we present the
dynamics of the polaron-transformed representation of the spin-boson model
considered in Fig.\,\ref{fig:system_2_comparison} obtained using the PSI with 
a range of constant time steps.

The population dynamics obtained for the polaron-transformed Hamiltonian
converges considerably slower than the population dynamics obtained
when using the standard representation (shown in Fig.\,\ref{fig:system_2_comparison}).
By considering the time-dependent deviation from the reference calculation
shown in the inset of panel a) of Fig.\,\ref{fig:system_2_timestep_convergence},
it can be seen that the significant deviation observed is dominated
by errors made in the first time step, following which the deviation
from the reference calculations reaches a plateau. In order to accurately
describe the dynamics of this Hamiltonian, and in particular the contribution
arising from the final term in Eq.\,\ref{eq:polaron_transformed_hamiltonian}
that couples all modes to all other modes in a multiplicative manner, the SPFs need to undergo
rapid evolution. Within the PSI, the coefficients
for each node are evolved over each time period with the SPFs held
constant. When the SPFs evolve rapidly, this provides a poor approximation
to the true dynamics, as the full evolution operator quickly leaves the
region of the Hilbert space described by the constant SPF
and SHF basis. As such, in contrast to the standard
ML-MCTDH approach where numerical issues arise from difficulties in
describing this rapid evolution of SPFs, the deviations observed for
the PSI can be attributed to errors associated
with the constant mean-field approximation.

 In order to demonstrate this point, we consider an alternative
integration scheme in which the initial evolution through the interval $dt$
is split into a series of smaller time steps, with each time step being
an order of magnitude larger than the previous until we reach time
$t=dt$ (i.e. if 5 steps are used the first uses a time step
of $10^{\text{-}5}dt$). The results obtained using this
scheme are shown in panel b) of Fig.\,\ref{fig:system_2_timestep_convergence}.
With this modification to the integration protocol, we see a three orders
of magnitude reduction in the deviation of the population dynamics
even for a time step of $dt\Delta=0.04$, and as a consequence
the population dynamics converges more rapidly in this case than
for the standard spin-boson model. 

These results suggest that the PSI is readily able to accurately capture 
rapid evolution of the initially unoccupied
SPFs.  In addition, this approach can benefit from an adaptive time step
integration scheme. The error associated with the PSI
can vary significantly throughout a simulation depending on how rapidly the SPF and mean-field Hamiltonian matrices evolve. In principle,
an adaptive time step integration scheme would resolve this issue.

We have applied the adaptive integration approach described in Ref.
\onlinecite{Kloss2017a} for the MCTDH case to the multi-layer case. This scheme involves propagating two
approximations to the ML-MCTDH wavefunction obtained using time steps
that differ by a factor of two, and as such allows for an estimate of the
error of the integration to be obtained.  However, we find that due
to a build-up of floating point errors in the evaluation of the Hilbert
angle (or overlap) between the two approximations of the wavefunction used 
in this approach, the error estimate
itself has limited accuracy for deep tree structures ($\sim\!10^{\text{{-}6}}$). 
This approach provides a reasonable adaptive integration scheme when moderate integration
errors are allowed ($\sim\!10^{\text{{-}5}}$), however, for problems
requiring stricter tolerances we find that such an approach often
becomes impractical as it is unable to accurately estimate the error.

\subsubsection*{Dependence on Initially Unoccupied SPFs }
 We next consider the dependence of the dynamics obtained
for the  model with different choices for the initially unoccupied
SPFs. To do this we perform a series of simulations with different
initially unoccupied SPFs. For each node in the tree, we sample the
coefficients of the unoccupied SPF uniformly within the local Hilbert
space such that they are orthogonal to our initially occupied state.
We run a series of 20 different simulations for both the standard
and polaron-transformed spin-boson model with $N=500$ bath modes,
and consider the average relative cumulative deviation over the 20
trajectories taking the mean of all trajectories as the reference.
The resultant mean deviations are given in Table\,\ref{tab:System2_unoccupied_function_variance}.
\begin{table}[ht]
\begin{centering}
\begin{tabular}{r|cc}
$dt\Delta$ & $\left\langle \Delta P\right\rangle $ : Standard & $\left\langle \Delta P\right\rangle $ : Polaron\tabularnewline
\hline 
$4\times10^{\text{-2}}$ & $1.9\times10^{\text{-3}}$ & $3.0\times10^{\text{-4}}$\tabularnewline
$2\times10^{\text{-2}}$ & $9.7\times10^{\text{-5}}$ & $4.7\times10^{\text{-5}}$\tabularnewline
$1\times10^{\text{-2}}$ & $1.0\times10^{\text{-5}}$ & $4.9\times10^{\text{-6}}$\tabularnewline
$5\times10^{\text{-3}}$ & $3.8\times10^{\text{-6}}$ & $1.2\times10^{\text{-6}}$\tabularnewline
$2.5\times10^{\text{-3}}$ & $1.5\times10^{\text{-6}}$ & $1.6\times10^{\text{-6}}$\tabularnewline
$1.25\times10^{\text{-3}}$ & $2.5\times10^{\text{-7}}$ & $1.3\times10^{\text{-6}}$\tabularnewline
$6.25\times10^{\text{-4}}$ & $8.5\times10^{\text{-8}}$ & $1.7\times10^{\text{-6}}$\tabularnewline
\hline 
\end{tabular}
\par\end{centering}
\caption{\label{tab:System2_unoccupied_function_variance}Average cumulative
deviation of 20 simulations from their mean obtained with different
initially unoccupied SPFs. }
\end{table}
Some deviations in the population dynamics obtained with
different initially unoccupied SPFs  are observed. For larger
time steps, these relative deviations can be rather large, on the order of $10^{\text{{-}3}}$. As the time step is decreased, the
mean deviations likewise decrease, and converge towards the results shown in
Fig.\,\ref{fig:system_2_comparison}. For this problem, the choice of the redundant initial conditions does not significantly impacts the dynamics obtained provided they are converged with respect to the integration time step. %dependence on initial
%conditions is associated with a failure to converge the dynamics with
%respect to the time step of integration. 

\subsubsection*{Larger Bath Sizes}
Finally, for this model, we explore the effect that increasing the number of bath
modes has on the accuracy and efficiency of the dynamics obtained
using the PSI. We consider a series of
six bath sizes ranging from 500 to $10^6$ modes and treat the dynamics using the standard representation for the Hamiltonian. We once again
use the relative deviation with respect to a reference calculation
for each bath size as a measure of the accuracy, and use the total
number of Hamiltonian evaluations, $N_{MF}$, and average number of
Hamiltonian applications per node, $\langle N_{H}\rangle$, as measures
of efficiency. For variable mean-field based approaches, such as those
considered in Refs. \onlinecite{Wang2018} and \onlinecite{Wang2021}, these two quantities are the same
where as for the PSI they can differ. Depending on
the specific application, either of these two steps can dominate the
numerical cost associated with the calculation. For all calculations
we used the alternative integration scheme introduced in the previous
section, and the approach outlined above for constructing the tree
topology. For all trees we use 16 SPFs per node
for the first nine layers of nodes (or as many layers as the tree
has), and four SPFs per node for all other nodes.
This corresponds to a similar tree structure to that used for this
model in Ref. \onlinecite{Wang2021}, and is sufficient to converge the dynamics
with respect to the number of SPFs. 
\begin{table*}
\begin{centering}
\begin{tabular}{r|c>{\centering}p{2cm}>{\centering}p{2cm}>{\centering}p{2cm}>{\centering}p{2cm}>{\centering}p{2cm}>{\centering}p{2cm}}
$N$ & $ $ & \multicolumn{2}{c}{500} & \multicolumn{2}{c}{2,000} & \multicolumn{2}{c}{5,000}\tabularnewline
\hline 
\hline 
$dt\Delta$  & $N_{MF}$ & $\Delta P$ & $\langle N_{H}\rangle$ & $\Delta P$ & $\langle N_{H}\rangle$ & $\Delta P$ & $\langle N_{H}\rangle$\tabularnewline
\hline 
$4\times10^{\text{-2}}$ & 50 & $1.6\times10^{\text{-}2}$ & $9.9\times10^{2}$ & $9.4\times10^{\text{-}3}$ & $8.6\times10^{2}$ & $1.3\times10^{\text{-}2}$ & $8.0\times10^{2}$\tabularnewline
$2\times10^{\text{-2}}$ & 90 & $1.6\times10^{\text{-}4}$ & $1.4\times10^{3}$ & $1.3\times10^{\text{-}4}$ & $1.3\times10^{3}$ & $2.2\times10^{\text{-}4}$ & $1.1\times10^{3}$\tabularnewline
$1\times10^{\text{-2}}$ & 170 & $1.2\times10^{\text{-5}}$ & $2.1\times10^{3}$ & $2.3\times10^{\text{-5}}$ & $1.9\times10^{3}$ & $1.6\times10^{\text{-5}}$ & $1.7\times10^{3}$\tabularnewline
$5\times10^{\text{-3}}$ & 330 & $5.0\times10^{\text{-6}}$ & $3.2\times10^{3}$ & $2.3\times10^{\text{-6}}$ & $2.9\times10^{3}$ & $3.2\times10^{\text{-6}}$ & $2.7\times10^{3}$\tabularnewline
$2.5\times10^{\text{-3}}$ & 650 & $3.4\times10^{\text{-6}}$ & $5.3\times10^{3}$ & $8.1\times10^{\text{-7}}$ & $4.9\times10^{3}$ & $7.1\times10^{\text{-7}}$ & $4.6\times10^{3}$\tabularnewline
$1.25\times10^{\text{-3}}$ & 1290 & $1.8\times10^{\text{-7}}$ & $9.3\times10^{3}$ & $2.5\times10^{\text{-7}}$ & $8.8\times10^{3}$ & $2.1\times10^{\text{-7}}$ & $8.3\times10^{3}$\tabularnewline
$6.25\times10^{\text{-4}}$ & 2570 & Reference & $1.7\times10^{4}$ & Reference & $1.6\times10^{4}$ & Reference & $1.5\times10^{4}$\tabularnewline
\hline 
\multicolumn{1}{r}{} &  &  &  &  &  &  & \tabularnewline
$N$ & $ $ & \multicolumn{2}{c}{$10^4$} & \multicolumn{2}{c}{$10^5$} & \multicolumn{2}{c}{$10^6$}\tabularnewline
\hline 
\hline 
$dt\Delta$ & $N_{MF}$ & $\Delta P$ & $\langle N_{H}\rangle$ & $\Delta P$ & $\langle N_{H}\rangle$ & $\Delta P$ & $\langle N_{H}\rangle$\tabularnewline
\hline 
$4\times10^{\text{-2}}$ & 50 & $1.6\times10^{\text{-}2}$ & $7.5\times10^{2}$ & $1.2\times10^{\text{-}2}$ & $6.7\times10^{2}$ & $1.4\times10^{\text{-}2}$ & $5.3\times10^{2}$\tabularnewline
$2\times10^{\text{-2}}$ & 90 & $5.1\times10^{\text{-}4}$ & $1.0\times10^{3}$ & $4.1\times10^{\text{-}4}$ & $9.5\times10^{2}$ & $3.0\times10^{\text{-}4}$ & $8.0\times10^{2}$\tabularnewline
$1\times10^{\text{-2}}$ & 170 & $1.5\times10^{\text{-5}}$ & $1.6\times10^{3}$ & $1.6\times10^{\text{-5}}$ & $1.4\times10^{3}$ & $1.7\times10^{\text{-5}}$ & $1.3\times10^{3}$\tabularnewline
$5\times10^{\text{-3}}$ & 330 & $2.6\times10^{\text{-6}}$ & $2.5\times10^{3}$ & $2.3\times10^{\text{-6}}$ & $2.4\times10^{3}$ & $4.2\times10^{\text{-6}}$ & $2.2\times10^{3}$\tabularnewline
$2.5\times10^{\text{-3}}$ & 650 & $3.9\times10^{\text{-7}}$ & $4.4\times10^{3}$ & $5.8\times10^{\text{-7}}$ & $4.1\times10^{3}$ & $1.3\times10^{\text{-6}}$ & $4.0\times10^{3}$\tabularnewline
$1.25\times10^{\text{-3}}$ & 1290 & $1.0\times10^{\text{-7}}$ & $8.1\times10^{3}$ & $2.7\times10^{\text{-7}}$ & $7.7\times10^{3}$ & $8.6\times10^{\text{-7}}$ & $7.4\times10^{3}$\tabularnewline
$6.25\times10^{\text{-4}}$ & 2570 & Reference & $1.5\times10^{4}$ & Reference & $1.4\times10^{4}$ & Reference & $1.3\times10^{4}$\tabularnewline
\hline 
\end{tabular}
\par\end{centering}
\caption{\label{tab:bath_size_convergence_model_2}The convergence with respect
to time step ($dt$) of the deviation in the population
dynamics of a spin-boson model obtained using the PSI. The spin-boson model considered has $\varepsilon/\Delta=0$,
$\alpha=2$, $\omega_{c}=25\Delta$, and with varying numbers of bath
modes (specified in the tables). In all cases, the dynamics was obtained
using the alternative integration scheme described above and the relative
cumulative deviation (from the indicated reference) was evaluated for dynamics obtained to times
$t\Delta=0.8$. }
\end{table*}
In Table\,\ref{tab:bath_size_convergence_model_2}, the convergence
behavior of the PSI is compared for baths
with 500, 2,000, 5,000, $10^4$, $10^5$, and $10^6$ modes. For
all bath sizes, convergence of the population dynamics (here taken
to be a deviation of $\sim\!10^{\text{{-}4}}$) is observed for time steps
of $dt\Delta=1\times10^{\text{-2}}$ or smaller. As the
number of bath modes increases, the average number of Hamiltonian applications
required decreases slightly. However, this decrease is not particularly
significant and can likely be attributed to the tree structure used.
For problems with large numbers of bath modes we have a considerably
larger proportion of nodes that use four SPFs rather than sixteen
SPFs. For these nodes the solution of Eq.\,\ref{eq:Atilde-PSI-eom}
using a Krylov subspace integration scheme requires slightly fewer
function applications although this difference decreases as the time step
decreases. For all bath sizes this corresponds to a total of 170
Hamiltonian evaluations and with $\lesssim2\times10^{3}$ average Hamiltonian
applications per node. In comparison, standard ML-MCTDH calculations
require roughly three to four orders of magnitude more mean-field
evaluations and two to three orders of magnitude more Hamiltonian
evaluations to obtain the same error for baths of up to $10^4$ modes.\citep{Wang2018,Wang2021} The difference in efficiency is
less extreme when compared to the improved regularization scheme 
presented in Ref. \onlinecite{Wang2018}, 
however even in this case the PSI requires roughly 30-80 times fewer Hamiltonian evaluation
and 3-8 times fewer Hamiltonian applications for baths of up to $10^4$
modes. This difference becomes more pronounced when considering the
$10^5$ mode case. For the PSI, the number of Hamiltonian evaluations
and applications required for convergence is essentially independent
of the number of bath modes. For the improved regularization scheme results presented in Ref. \onlinecite{Wang2021}, a factor of 10 increase in the number of 
evaluations required is observed when moving from $10^4$ to $10^5$ modes. 
For reference, the PSI calculations for the $10^5$ mode model using a
time step of $dt\Delta=1\times10^{\text{-2}}$ required less than three
hours on a single core of an Intel i5-8250U CPU. 

These results suggests that
for the PSI the number of Hamiltonian evaluations and applications
at a given time step is limited by the physics of the problem, and
as the number of bath modes increases, little change in the number of these operations occurs.
For ML-MCTDH approaches that employ regularized EOMs,
the required regularization parameter depends on the number of bath
modes and as a consequence considerably more integration steps are
required to obtain converged results for larger baths. 

 All calculations presented in this section have used a relatively
deep tree structure. In contrast, the results obtained in Refs.
\onlinecite{Wang2018} and \onlinecite{Wang2021} for this model with bath sizes up to $N=10^4$
were obtained using wider trees. In Table V of the Supplementary
Information, we present results obtained using wider tree structures 
generated with $N_{b}=4$ or $5$, and $N_{lower}=4$. We find that the conclusions reached
in this section do not change when using this alternative tree topology, and thus the accuracy
and efficiency when measured in terms of the number Hamiltonian applications
and Hamiltonian evaluations of the PSI
does not depend significantly on the tree structure. In addition to
these results, we have considered the full set of spin-boson models
considered in Ref. \onlinecite{Wang2018}. For completeness these results are
presented in Sec. VII  of the Supplementary Information, and once again the PSI can 
obtain converged results with roughly 1-2
orders of magnitude fewer Hamiltonian evaluations and comparable to
an order of magnitude fewer Hamiltonian applications than the improved
regularization approach. 

For all of the calculations presented so far, relatively few SPFs
are required to obtain converged results, and as such do not represent
particularly challenging problems. We will now consider a significantly more 
challenging physical system for which considerably more SPFs are required in order to obtain
accurate dynamics. 

\subsection{Multi-Spin-Boson Model }
 As a final application of the PSI
approach we consider a generalization of the spin-boson
model, in which a set of $M$ TLSs are each linearly coupled
to a \textit{common} harmonic oscillator bath. The Hamiltonian for this system
may be written as
\begin{equation}
\begin{aligned}\hat{{H}}= & \sum_{i=1}^{M}\left(\varepsilon\hat{{\sigma}}_{i,z}+\Delta\hat{{\sigma}}_{i,x}\right)+\sum_{k}\omega_{k}\hat{{a}}_{k}^{\dagger}\hat{{a}}\\
 & +\sum_{i=1}^{M}\hat{{\sigma}}_{i,z}\sum_{k}g_{ik}(\hat{{a}}_{k}^{\dagger}+\hat{{a}}_{k}), \label{eq:ham_msb}
\end{aligned}
\end{equation}
where $\hat{{a}}_{k}^{\dagger}$ and $\hat{{a}_{k}}$ are
the bosonic creation and annihilation operators associated with the
$k$th bath mode, and $\hat{{\sigma}}_{i,x}$ and $\hat{{\sigma}}_{i,z}$
are the Pauli matrices associated with spin $i$. Eq.\,\ref{eq:ham_msb} provides an interesting model
in which to explore the interplay between coherent system interactions and the effects
of dissipation arising from the bath.\citep{ORTH2010, MCCUTCHEON2010, WINTER2014, DEFILIPPIS2021}  For Ohmic and sub-Ohmic baths, the presence of the additional TLSs have been shown to enhance localization, reducing the system-bath coupling strength at which the delocalization to localization transition occurs.\citep{ORTH2010, MCCUTCHEON2010, WINTER2014, DEFILIPPIS2021} Additionally, the model with a super-Ohmic bath arises in the description of low temperature glasses,\citep{JOFFRIN1975,Kassner1989,BROWN1998,SCARDICCHIO2021} where it has been suggested that signatures of many-body localization may be experimentally observable.\citep{SCARDICCHIO2021} In what follows, we will restrict ourselves to the case of an Ohmic bath.
In contrast to the standard spin-boson model, where the influence of the bath on the
system is entirely encoded within the spectral density, the MSB
model involves an $M\times M$ matrix of spectral densities. The elements
of this matrix of spectral densities, are given by
\begin{equation}
J_{ij}(\omega)=\frac{{\pi}}{2}\sum_{k}g_{ik}g_{jk}\delta(\omega-\omega_{k})
\end{equation}
and can be used to specify the system-bath coupling constants,
$g_{ik}$, between spin $i$ and mode $k$, and the frequency of
mode $k$, $\omega_{k}$. We will consider diagonal spectral densities
that are Ohmic with an exponential cutoff
\begin{equation}
J_{ii}(\omega)=\frac{{\pi}}{2}\alpha\omega e^{-\omega/\omega_{c}},
\end{equation}
and off-diagonal spectral densities of the form\citep{WINTER2014}
\begin{equation}
J_{ij}(\omega)=\frac{{\pi}}{2}\alpha\omega e^{-\omega/\omega_{c}}\cos\left(\omega r_{ij}/\nu\right),
\end{equation}
where $\alpha$ is the dimensionless Kondo parameter, $\omega_c$ is the cutoff
frequency of the bath, $r_{ij}=|x_{i}-x_{j}|$ is the distance between the
TLSs $i$ and $j$, and $\nu$ is the speed of sound in the bath.
As a result of this cosine term, the peak in the off-diagonal bath
correlation functions is delayed compared to the diagonal correlation
functions to time $t=\tau_{ij}=r_{ij}\nu$. This gives rise to inherently
non-Markovian phonon-mediated interactions between TLSs. This form
for the spectral density matrix arises when considering a set of TLSs coupled
to a common one-dimensional bath of phonons.\citep{WINTER2014} 

 In order to perform simulations of the unitary dynamics
of this system, it is necessary to discretize the bath. As in the
single spin-boson case we have used an exponential density of bath
frequencies (Eq.\,\ref{eq:exponential_frequency_dist}), and have
discretized the bath into a set of $N$ frequencies according to Eq.
\ref{eq:discrete_frequencies}. For each of mode, there is a set of
$M$ coupling constants that need to be determined. These coupling
constants need to satisfy the set of $M\times M$ equations for the
spectral densities. If all TLSs are not at the same position in space,
it is not possible to satisfy all equations with a single set of $M$
coupling constants. Rather we consider a set of $M$ modes, each with
the same frequency, that each couple to all the TLSs. This gives
us an $M\times M$ matrix of coupling constants, $\boldsymbol{{g}}_{k}$,
for each frequency that needs to satisfy the equation
\begin{equation}
\boldsymbol{{g}}_{k}^{T}\boldsymbol{{g}}_{k}=\frac{{2}}{\pi}\frac{{\boldsymbol{{J}}(\omega_{k})}}{\rho(\omega_{k})},
\end{equation}
where $\boldsymbol{{J}}(\omega)$ is the matrix of spectral
densities. Here we construct a symmetric matrices of coupling constants
$\boldsymbol{{g}}_{k}$ by taking the principal square root of the right hand side.
Applying this process, we arrive at a bath of $N\times M$ modes, with $M$
distinct frequencies. 

 We will consider a series of MSB models with
up to six TLSs, with randomly sampled positions, all coupled to the
same bath. We choose parameters such that the timescales associated
with the bath dynamics are comparable to the timescales of the TLS
dynamics, the physical parameters used are shown in Table\,\ref{tab:Multi-spin-boson-parameters}.
\begin{table}[ht]
\begin{centering}
\begin{tabular}{cccccc}
 & $\epsilon_{i}/\Delta$ & $\Delta_{i}/\Delta$ & $\alpha$ & $\omega_{c}/\Delta$ & \tabularnewline
\cline{2-5} \cline{3-5} \cline{4-5} \cline{5-5} 
 & 0 & 1 & 2 & $2.5$ & \tabularnewline
 &  &  &  &  & \tabularnewline
$x_{1}\Delta/\nu$ & $x_{2}\Delta/\nu$ & $x_{3}\Delta/\nu$ & $x_{4}\Delta/\nu$ & $x_{5}\Delta/\nu$ & $x_{6}\Delta/\nu$\tabularnewline
\hline 
2.08463 & 4.59884 & 4.73564 & 4.62870 & 2.39182 & 0\tabularnewline
%5.92845 & 8.44266 & 8.57946 & 8.47252 & 6.23564 & 3.84382\tabularnewline
\end{tabular}
\par\end{centering}
\caption{\label{tab:Multi-spin-boson-parameters}MSB model parameters
used in the calculations performed in this section. All parameters
are defined with respect to the Tunnelling amplitude, $\Delta$, which
specifies the energy scales involved.  As only the relative positions are important, the positions of the TLS, $x_i$, have been shifted from their randomly sampled values.}
\end{table}

 In the following calculations we will evaluate the dynamics
of the time-dependent population difference for each TLS,
\begin{equation}
P_{i}(t)=\left\langle \Psi(t)\right|\hat{{\sigma}}_{i,z}\left|\Psi(t)\right\rangle ,
\end{equation}
where the initial wavefunction is taken to be
\begin{equation}
\left|\Psi(0)\right\rangle =\bigotimes_{i=1}^{M}\left|0\right\rangle _{i}\otimes\left|vac\right\rangle .
\end{equation}
We will measure the convergence of this dynamics by looking at the
average relative deviation obtained for the population dynamics of
each of the TLSs
\begin{equation}
\langle\Delta P\rangle\!=\!\frac{{1}}{M}\!\sum_{i=1}^{M}\frac{{1}}{\left(P_{i,max}\!-\!P_{i,min}\right)}\frac{{1}}{\tau}\!\int_{0}^{\tau}\!\!\!|P_{i}(t)\!-\!P_{i,ref}(t)|dt,
\end{equation}
where $P_{i,ref}(t)$ is a reference result for the population dynamics
of TLS $i$, and $P_{i,min}$ and $P_{i,max}$ are the minimum and maximum
values of population difference. 

\subsubsection*{Convergence with Respect to $N_{SPF}$}
\begin{figure*}[t]
\begin{centering}
\includegraphics{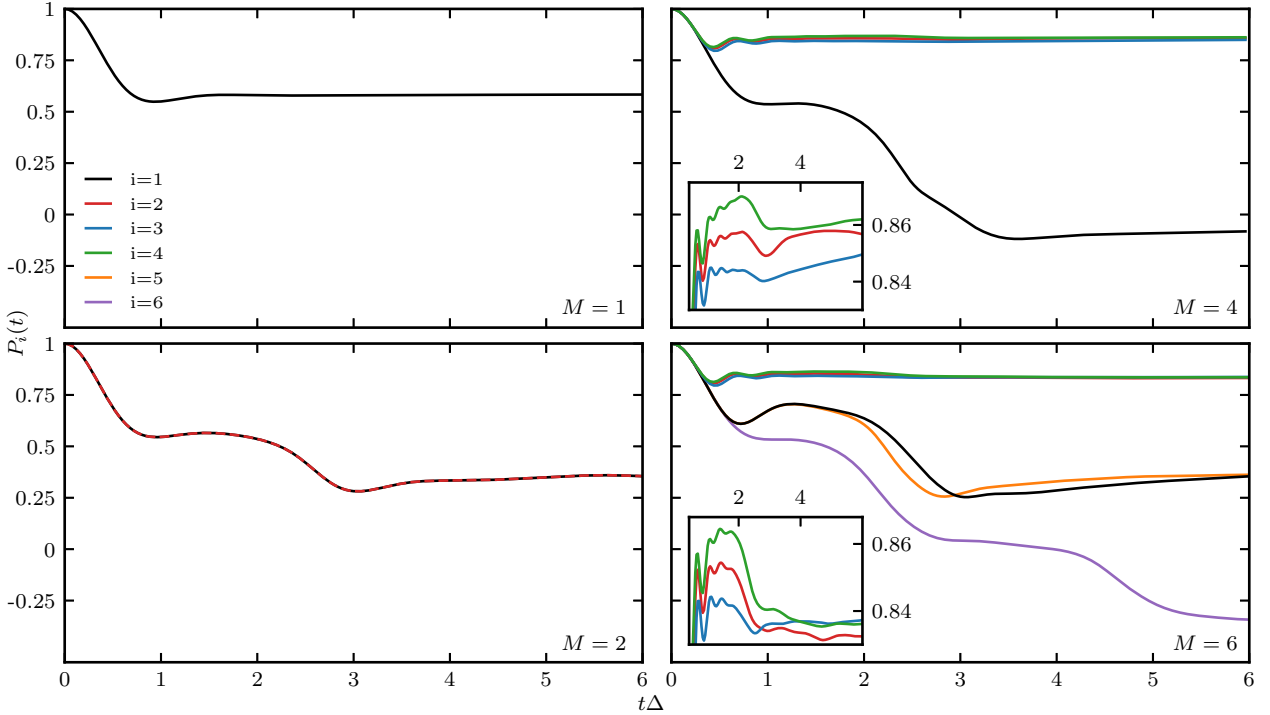}
\par\end{centering}
\caption{\label{fig:system_3_results}Population dynamics for each of the TLSs
in a series of MSB models with varying numbers of TLSs
coupled to a common heat bath. A bath containing 512 distinct frequencies
is used in each case which correspond to baths containing 512, 1024,
2048, and 3072 bath modes for 1,2,4, and 6 TLSs, respectively. The
inlays for the 4 and 6 spin cases, provide a closeup of the population
dynamics for TLSs 2,3, and 4 all of which are located in a similar
point in space. }
\end{figure*}

 We have considered the dynamics of four different MSB
models with $M=$1, 2, 4, and 6 TLSs. The bath is discretized using 512
distinct frequencies, and as a consequence baths of 512, 1024, 3048,
and 3072 modes are used for the 1, 2, 4, and 6 TLSs systems, respectively.
The ML-MCTDH wavefunctions used in these calculations employed three
groups of SPFs at the top level, one of which accounts for the degrees of freedom of the TLSs
 and includes as many SPFs as there are functions
in the multi-TLS Hilbert space ($2^{M}$). The remaining two groups
accounted for the bath degrees of freedom. The partitioning of bath
modes uses the same procedure as used above for the spin-boson model,
and employed two SPF groups for all subsequent layers. Simulations were
performed with a range of different numbers of SPFs for treating the
bath degrees of freedom, $N_{spf}$, however, in all cases twice as
many functions ($2N_{spf})$ were used for the root node. The largest
calculations considered here used $N_{spf}=100$. For the system with six TLSs
this corresponds to expanding the wavefunction in terms of a basis
of $2^{6}\times200^{2}=2,560,000$ states at the top layer, and $100^{3}=1,000,000$
basis states for all other non-leaf layers. A total of $\sim1.3\times 10^9$ variational parameters were used to parameterize the wavefunction.
A primitive basis of 30 harmonic oscillator
basis functions is used for each leaf node with frequency less than
the bath cutoff frequency and 10 basis functions for all higher frequency
modes. Mode combination was used with multiple physical modes being
combined together up until a maximum Hilbert space dimension of 3000 was reached. All calculations
presented in this section were performed with a time step of $dt\Delta=0.04$,
unless otherwise specified, which was found sufficient to provide results 
converged to a relative cumulative deviation of $\sim\!10^{\text{{-}5}}$.

 The results used as the reference calculations for each of the
 different MSB models are shown in Fig.
\ref{fig:system_3_results}. These calculations used $N_{spf}=48$,
80, 100, and 100 for $M=1$, 2, 4, and 6 TLSs, respectively. The dynamics
obtained for the one TLS case are not particularly interesting; we
see rapid initial decay of the population difference following which
it plateaus to a finite value as the system is at sufficiently strong
coupling that the dynamics are within the localised phase. For the
two TLSs case, the population dynamics obtained for each TLS is identical
due to symmetry. For short times the dynamics closely resembles that
of the single TLS case, however, starting at around $t\Delta\sim2.5$
the population difference begins to decay further before reaching
a lower plateau at later times. This corresponds to the time required
for propagation of phonons between the two TLSs, and so this further
decay arises due to phonon mediated interactions between the TLSs.
For the four and six TLSs cases, the population dynamics of TLSs 2,
3, and 4 are very similar (as shown in the insets). This arises due
to the close proximity of these TLSs and thus the timescale on which the bath
induces coupling between these TLSs is considerably faster than the
timescale of the bath dynamics. In this regime, the bath-mediated
interactions can be well described by a relatively strong, ferromagnetic,
$\hat{{\sigma}}{}_{i,z}\hat{{\sigma}}{}_{j,z}$ coupling between the
TLSs that results in strong correlations between the dynamics of these
spins. At later times, these TLSs interact with the other TLSs present,
and can result in significant changes in the population dynamics of
the other spins due to the cumulative effect of the three spins. 
\begin{table*}[t]
\begin{centering}
\begin{tabular}{r|>{\centering}p{2cm}>{\centering}p{2cm}>{\centering}p{2cm}>{\centering}p{2cm}>{\centering}p{2cm}>{\centering}p{2cm}>{\centering}p{2cm}>{\centering}p{2cm}}
$M$ & \multicolumn{2}{c}{1} & \multicolumn{2}{c}{2} & \multicolumn{2}{c}{4} & \multicolumn{2}{c}{6}\tabularnewline
\hline 
\hline 
$N_{spf}$ & $\langle\Delta P\rangle$ & $\langle N_{H}\rangle$ & $\langle\Delta P\rangle$ & $\langle N_{H}\rangle$ & $\langle\Delta P\rangle$ & $\langle N_{H}\rangle$ & $\langle\Delta P\rangle$ & $\langle N_{H}\rangle$\tabularnewline
\hline 
4 & $3.5\times10^{\text{-}3}$ & $2.2\times10^{3}$ & $1.7\times10^{\text{-}1}$ & $2.1\times10^{3}$ & $1.2\times10^{\text{-}1}$ & $2.4\times10^{3}$ & $1.4\times10^{\text{-}1}$ & $2.4\times10^{3}$\tabularnewline
8 & $3.8\times10^{\text{-}4}$ & $2.4\times10^{3}$ & $8.7\times10^{\text{-}3}$ & $2.3\times10^{3}$ & $5.5\times10^{\text{-}2}$ & $2.5\times10^{3}$ & $5.5\times10^{\text{-}2}$ & $2.5\times10^{3}$\tabularnewline
12 & $4.0\times10^{\text{-}5}$ & $2.5\times10^{3}$ & $3.7\times10^{\text{-}3}$ & $2.3\times10^{3}$ & $3.0\times10^{\text{-}2}$ & $2.5\times10^{3}$ & $7.7\times10^{\text{-}2}$ & $2.5\times10^{3}$\tabularnewline
16 & $8.1\times10^{\text{-}6}$ & $2.5\times10^{3}$ & $3.7\times10^{\text{-}3}$ & $2.4\times10^{3}$ & $2.4\times10^{\text{-}2}$ & $2.6\times10^{3}$ & $3.9\times10^{\text{-}2}$ & $2.5\times10^{3}$\tabularnewline
24 & $2.7\times10^{\text{-}6}$ & $2.6\times10^{3}$ & $4.8\times10^{\text{-}4}$ & $2.6\times10^{3}$ & $1.0\times10^{\text{-}2}$ & $2.6\times10^{3}$ & $2.6\times10^{\text{-}2}$ & $2.7\times10^{3}$\tabularnewline
32 & $1.3\times10^{\text{-}6}$ & $2.7\times10^{3}$ & $3.0\times10^{\text{-}4}$ & $2.6\times10^{3}$ & $6.0\times10^{\text{-}3}$ & $2.7\times10^{3}$ & $1.7\times10^{\text{-}2}$ & $2.6\times10^{3}$\tabularnewline
48 & Reference & $3.8\times10^{3}$ & $4.9\times10^{\text{-}5}$ & $2.7\times10^{3}$ & $2.1\times10^{\text{-}3}$ & $3.2\times10^{3}$ & $1.1\times10^{\text{-}2}$ & $2.6\times10^{3}$\tabularnewline
64 &  &  & $2.4\times10^{\text{-}5}$ & $2.8\times10^{3}$ & $1.5\times10^{\text{-}3}$ & $3.3\times10^{3}$ & $7.0\times10^{\text{-}3}$ & $2.7\times10^{3}$\tabularnewline
80 &  &  & Reference & $2.9\times10^{3}$ & $7.1\times10^{\text{-}4}$ & $2.7\times10^{3}$ & $3.3\times10^{\text{-}3}$ & $2.7\times10^{3}$\tabularnewline
100 &  &  &  &  & Reference & $3.3\times10^{3}$ & Reference & $2.7\times10^{3}$\tabularnewline
\hline 
\end{tabular}
\par\end{centering}
\caption{\label{tab:system_3_spf_convergence}Convergence of the population
dynamics obtained for the MSB models with varying numbers
of TLSs, $M$, as a function of the number of SPFs, $N_{spf}$. Convergence
is measured through the average relative deviation of the population
dynamics from reference calculations indicated in the table. 
We also present the average number of Hamiltonian evaluations
per node required to obtain this dynamics out to a time $t\Delta=6$.}
\end{table*}

 In order to accurately capture the dynamics of these systems,
it is necessary to have enough SPFs that the correlations between
individual TLSs and bath as well as the bath-mediated correlations
between the TLSs can be well described. This becomes a more challenging
problem as the number of TLSs increase, and as a result we find that
significantly more SPFs are required to obtain accurate dynamics.
In Table\,\ref{tab:system_3_spf_convergence} we present the average
relative deviations of the population dynamics of the systems and
number of Hamiltonian applications required for varying numbers of
SPFs. The number of Hamiltonian applications required for a given
number of SPFs appears to be roughly independent of the number of
TLSs considered. However there is a trend of requiring more Hamiltonian
applications as the number of SPFs increases. This dependence is not
very strong, with at most a factor of 2 increase being observed. As
such, the increase in the number of Hamiltonian applications does
not dramatically impact the computational effort associated with the calculations
as the number of SPFs increase. It would be interesting to see whether
this holds for the standard ML-MCTDH approach and improved schemes
that employ regularization of singular mean-field density matrices.
Whether the increase in the size of these mean-field density matrices
due to an increase in the number of SPFs alters the size of the regularization
parameter required to obtain converged results is not immediately
obvious. Furthermore, whether the larger number of initially unoccupied
SPFs increases the time over which the regularization impacts the
dynamics remains to be seen. 

\begin{figure}[ht]
\begin{centering}
\includegraphics{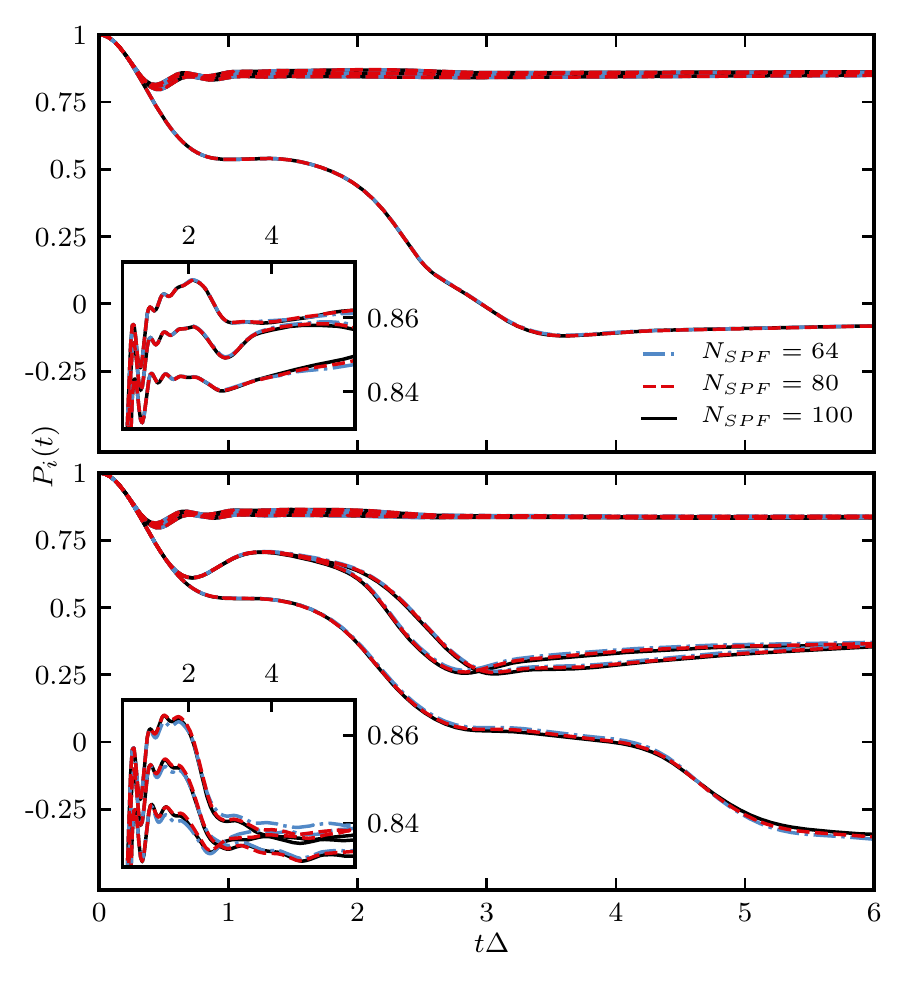}
\par\end{centering}
\caption{\label{fig:system_3_spf_convergence}Convergence of the population
dynamics with respect to the number of SPFs for $M=4$ (top) and 6
(bottom) TLSs. The population difference as a function of time for
each TLS is shown. In each case, results were obtained using $N_{spf}=64$,
80, and 100 SPFs. The insets show a zoomed in view of the population
difference dynamics for spins 2,3, and 4, which experience similar
population dynamics.}
\end{figure}

 The convergence of the population dynamics with respect
to the number of SPFs differs considerably for systems with different
numbers of TLSs. For the single TLS (spin-boson model) case, convergence
is very rapid with deviations $<10^{\text{{-}4}}$ being obtained
for $N_{spf}\geq12$. As the number of TLSs increase it becomes considerably
more challenging to converge dynamics. For the two TLSs case, a deviation
$<10^{\text{{-}4}}$ is not obtained until $N_{spf}\geq48$. Where
as for the four and six TLSs case, the deviations between the population
dynamics obtained for the largest numbers of SPFs ($N_{spf}=80$ and
100) are $7.1\times10^{\text{-}4}$ and $3.3\times10^{\text{-}3}$
, respectively. These relatively large deviations indicate that the
dynamics are not fully converged with respect to the number of SPFs
used. To give a more clear indication of the scale of these deviations,
the population dynamics obtained for the four and six TLS models using
$N_{spf}=64$, 80, and 100 are shown in Fig.\,\ref{fig:system_3_spf_convergence}.

 For the four TLSs case, the population dynamics are converged
to within the thickness of the lines shown when represented on the
larger scale, however, some small deviations are observed in the population
dynamics of TLSs 2,3, and 4 at times $t\Delta\geq4$ (shown
in the insets). For many applications, this level of convergence is
sufficient. For the six TLSs case, more significant deviations are
observed between the results with different numbers of SPFs, with
the onset of these significant deviations being at shorter times $t\Delta\geq1.5$.
The timescale of the dynamics of TLSs 1, 5, and 6 is not fully converged,
with the $N_{spf}=100$ case showing a slight shift to shorter times.
In both cases, the results obtained with $N_{spf}=80$ agree with
the results obtained using $N_{spf}=100$ to longer times than the
results obtained $N_{spf}=64$, and show smaller deviations once they
occur. In order to obtain fully converged results it would be necessary
to consider even larger number of SPFs than have been considered here. 

 The primary factor limiting the number of SPFs that can be
treated is the $\mathcal{{O}}(N_{spf}^{4})$ scaling of operations
at each node, and $\mathcal{{O}}(N_{spf}^{3})$ scaling of the memory
requirements. For reference, the six TLSs calculations with $N_{spf}=100$,
required $\sim32$ GB of memory to store the required data for the
evolution. The memory limitations can be offset significantly through
the use of out-of-core storage techniques. However, even with such
techniques the size of systems that are practical to treat are still
limited by the $\mathcal{{O}}(N_{spf}^{4})$ scaling of the number of operations
applied to each node. The serial updating of nodes within the PSI limits
the extent to which it can be parallelized. 
Only the linear algebra kernels applied in the evolution
of each node can readily be parallelized. As
a consequence, the wall time requirements can become very large for
systems with larger numbers of SPFs. For reference, the six TLSs calculations
with $N_{spf}=100$ took $\sim 480$ hours running
on 8 cores of an Intel Xeon E5-2690 v3 CPU. For this case the speed
up due to the use of parallelized linear algebra kernels was sublinear.
However, it is likely that with further optimization better parallel
performance could be obtained. This poor parallel performance arising
from the serial nature of the node updates represents one of the main
limitations of the current PSI approach. %In a future publication we
%intend to discuss possible strategies for parallelizing this approach
%and compare their numerical performance to the serial PSI.

\subsubsection*{Effect of Varying Timestep}
\begin{table}[ht]
\begin{centering}
\begin{tabular}{r|>{\centering}p{1.6cm}>{\centering}p{1.6cm}>{\centering}p{1.6cm}>{\centering}p{1.6cm}}
$N_{spf}$ & 4 & 16 & 32 & 48\tabularnewline
\hline 
\hline 
$dt\Delta$ & \multicolumn{4}{c}{$\langle\Delta P\rangle$}\tabularnewline
\hline 
$6.4\times10^{\text{-1}}$ & $3.5\times10^{\text{-}2}$ & $6.7\times10^{\text{-}3}$ & $1.1\times10^{\text{-}3}$ & $4.6\times10^{\text{-}4}$\tabularnewline
$3.2\times10^{\text{-1}}$ & $1.6\times10^{\text{-}2}$ & $1.2\times10^{\text{-}3}$ & $2.4\times10^{\text{-}4}$ & $7.1\times10^{\text{-}5}$\tabularnewline
$1.6\times10^{\text{-1}}$ & $5.0\times10^{\text{-}3}$ & $5.6\times10^{\text{-}4}$ & $7.6\times10^{\text{-}5}$ & $1.4\times10^{\text{-}5}$\tabularnewline
$8\times10^{\text{-2}}$ & $2.5\times10^{\text{-}3}$ & $4.5\times10^{\text{-}4}$ & $5.3\times10^{\text{-}5}$ & $4.9\times10^{\text{-}6}$\tabularnewline
$4\times10^{\text{-2}}$ & $5.1\times10^{\text{-}4}$ & $4.0\times10^{\text{-}4}$ & $3.1\times10^{\text{-}5}$ & $4.4\times10^{\text{-}6}$\tabularnewline
$2\times10^{\text{-2}}$ & $2.8\times10^{\text{-}4}$ & $8.5\times10^{\text{-}4}$ & $3.2\times10^{\text{-}5}$ & $4.0\times10^{\text{-}6}$\tabularnewline
$1\times10^{\text{-2}}$ & $1.3\times10^{\text{-}4}$ & $5.4\times10^{\text{-}4}$ & $1.8\times10^{\text{-}5}$ & $2.1\times10^{\text{-}6}$\tabularnewline
$5\times10^{\text{-3}}$ & Reference & Reference & Reference & Reference\tabularnewline
\hline 
\end{tabular}
\par\end{centering}
\caption{Convergence of the population dynamics of the two TLS model with respect
to time step for different numbers of SPFs. In each calculation the
model parameters are as given in the main text. For each value of
$N_{spf}$, convergence is measure with respect to a different reference,
the calculation obtained with that value of $N_{spf}$ and with a time step
of $dt\Delta=5\times10^{\text{-3}}$.}
\end{table}
As the number of SPFs increase, the deviation between results obtained
with a given time step and the reference calculation obtained with
a time step of $dt\Delta=5\times10^{\text{-3}}$ decreases
significantly. This result is readily understood when considering
the approximations involved in obtaining the PSI. The PSI uses an
expansion of the full wavefunction in terms of an orthonormal basis
at each node of the tree. The evolution of the coefficient tensor
at each node involves the construction of the propagator obtained
from the projection of the Hamiltonian onto this local basis. As the
size of this local basis increases, the projection of the Hamiltonian
onto this basis better approximates the Hamiltonian in the full Hilbert
space, and as a consequence the propagator evaluated using this projected
Hamiltonian is accurate for longer times, and fewer Hamiltonian evaluations
are required when integration the equations. In the limit of a complete
basis expansion at each node of the tree, the propagator becomes exact
and the approximation of a constant mean-field becomes exact for arbitrary
times. These results further suggest that the projector splitting
integrator could benefit significantly from an adaptive time step integration
scheme. In particular, as the number of SPFs is increased such a scheme
would be able to take larger time steps, potentially reducing the computational
effort required to obtain converged results. 

There is one additional point that we need to consider when
discussing the convergence of the dynamics of this model. In principle,
the dynamics should be converged with respect to the number of discrete
bath modes, $M$, in order to reach the continuum bath limit. The
memory and computer time requirements scale as $\mathcal{{O}}(M\log(M))$.
Here, due to the large wall time requirements necessary to perform
the calculations with large tree sizes and large number of SPFs we
have restricted the bath to only 512 frequencies for the four and
six TLSs cases. For the one and two TLSs cases, calculations using
larger baths are feasible and are shown in Sec. VII.A of the Supplementary
Information. 
Finally, as in the spin-boson model case, it is possible to perform a polaron transform
for the MSB model. We have considered this in Sec. VII.B of the Supplementary
Information for the one and two TLSs cases.  These results show that, once again, 
the PSI has no issue when treating the challenging polaron-transformed representation.
%We have found that similar numbers of SPFs are required to 
%obtain converged results using both representations of the problem.
%However, we find that more Hamiltonian applications are required for the polaron
%transformed calculations at a given $N_{spf}$. As such, we will not 
%consider the polaron transform further. 
\section{Conclusion}
In this paper we have discussed the implementation of the PSI for
ML-MCTDH wavefunctions and have presented a series of numerical applications.  Here, we have 
consider three different types of models:
\begin{itemize}
    \item A two-dimensional, bi-linearly and bi-quadratically coupled harmonic oscillator model.
    \item A series of spin-boson models using both the standard and polaron-transformed forms of the Hamiltonian.
    \item A series of multi-spin-boson models.
\end{itemize}
These three types of models have allowed us to explore the stability and efficiency of the PSI in a number of different regimes. 

The two-dimensional oscillator model serves as a useful test for how rank deficiency 
in the ML-MCTDH wavefunction influences the accuracy of the dynamics obtained by the PSI.
The dynamics of unoccupied SPFs is arbitrary within any dynamical method that is based on
linear variations, such as standard ML-MCTDH or the PSI.  As a consequence, the dynamics of 
the wavefunction can depend on the value chosen for these unoccupied SPFs that do not initially
contribute to the wavefunction.  Here we find that while the dynamics obtained for this
two-dimensional model can depend on the value of the initially unoccupied SPFs when 2 SPFs are
used, uniformly sampling the initially unoccupied SPFs typically leads to
dynamics that corresponds to the optimal choice of the initially unoccupied SPF.  Further, if we
increase the number of SPFs used to 3, then the results obtained do not depend on the initial 
choice of these functions.  It is only when we have insufficient flexibility in the wavefunction to
account for the two unoccupied SPFs that contribute significantly to the dynamics at short times
that we find this dependence on initial conditions.  While this two-dimensional model 
provides a convenient test case for the behavior of unoccupied 
SPFs it is not representative of the types of problems that ML-MCTDH would typically be 
applied to.  

We next considered a series of spin-boson models that more closely resemble the types of 
problems that are typically treated using ML-MCTDH, and allow us to observe how the
PSI performs for problems with considerably more modes.  Converged dynamics were obtained for 
systems with up to $10^6$ degrees of freedom represented using tree structures with $\mathcal{O}(10^5)$ nodes.  By comparison with results that
have previously been obtained using standard ML-MCTDH (and an improved variant) for systems with fewer modes,\citep{Wang2018, Wang2021}
we find that the PSI approach provides an efficient and accurate approach.  
When compared to standard ML-MCTDH, we find that the PSI approach required between 3-4 orders of magnitude fewer Hamiltonian evaluations and 2-3 orders of magnitude
fewer Hamiltonian applications in order to obtain well converged results.  Even when compared to improved approaches, the PSI still demonstrated considerably better performance  with up to a factor of 30 reduction in effort.  
When the polaron-transformed form of this model is considered, we find that a very small time step
is required to obtain converged results.  We show that this is due to large errors that occur during the first time step.  The polaron-transformed Hamiltonian couples all modes together in a multiplicative fashion which leads to rapid initial evolution of the SPFs, and as a consequence the linearization is only accurate for short times.  Motivated by this, we discuss a simple modification
that while likely not optimal, considerably improves the convergence of the
dynamics obtained using the PSI. This modification makes use of variable but not adaptive time steps with the goal of using smaller times steps to integrate the initial steps where large 
numbers of SPFs are unoccupied. Our results suggest that an adaptive time step integration scheme
would be beneficial for the PSI.  We have briefly discussed one such scheme that has previously been used for the MCTDH case and that we find is not generally useful for the multi-layer case.  

Finally, we have considered a series of multi-spin-boson
models that require considerably larger numbers of SPFs in order to obtain accurate results.
We have presented results obtained using ML-MCTDH wavefunctions
with up to 100 SPFs for each node and that contain a total of $1.3\times10^9$ 
variational parameters.  
These results demonstrate that the PSI provides a stable approach even for ML-MCTDH
wavefunctions that involve very large coefficient tensors. Furthermore, we find
that as the number of SPFs increases, the approximations involved in the PSI become valid 
for longer times.  As a consequence, the number of time steps required to accurately integrate
the dynamics decreases as the representation of the wavefunction becomes more accurate.  

We have identified two areas in which further development of this
approach would be beneficial: adaptive time step control and parallelization
of the algorithm. We have briefly discussed some efforts we have made towards addressing 
these issues and have pointed out the potential short comings of these approaches.
Development in these areas remains an important area for
future work.  We believe that the results presented in this
paper demonstrate that the PSI, even in its current form, is a robust and 
efficient approach for evolving ML-MCTDH wavefunctions, and that 
these results will encourage the implementation of such approaches within
pre-existing ML-MCTDH packages. To help facilitate such development,
the source code for the implementation of the PSI discussed here will be made
available upon request.  

\begin{acknowledgements}
L.P.L. and D.R.R. were supported by the Chemical Sciences, Geosciences and Biosciences Division of the Office of Basic Energy Sciences, Office of Science, U.S. Department of Energy
\end{acknowledgements}

\section*{Data Availability}
The data that support the findings of this study are available from the corresponding author upon reasonable request.
\bibliography{lib_ben}

\end{document}

\title{Time Evolution of ML-MCTDH Wavefunctions: Supplementary Information}
{
\let\clearpage\relax
\author{Lachlan P Lindoy}
\email{ll3427@columbia.edu}
\author{Benedikt Kloss}
\author{David R Reichman}
\affiliation{Department of Chemistry, Columbia University, 3000 Broadway, New York, New York 10027, USA}
\maketitle
\tableofcontents
}
\pagebreak
\section{Linearization of the EOMs}

\subsection{Constant Mean-Field and Linearization\label{subsec:Constant-Mean-Field-Schemes}}
The ML-MCTDH EOMs
\begin{equation}
\dot{{\boldsymbol{{A}}}}^{1}(t)=-i\left(\frac{1}{\hbar}\boldsymbol{{h}}{}^{1}(t)-\boldsymbol{{X}}^{1}(t)\right)\boldsymbol{{A}}^1(t),\label{eq:root-eom-mlmctdh}
\end{equation}
and
\begin{equation}
\dot{{\boldsymbol{{A}}}}^{z_{l}}(t)=  -\frac{{i}}{\hbar}\boldsymbol{{Q}}^{z_{l}}(t)\sum_{r}\boldsymbol{{h}}_{r}^{z_{l}}(t)\boldsymbol{{A}}^{z_{l}}(t)\boldsymbol{{H}}_{r}^{z_{l}}(t)\boldsymbol{{\rho}}^{z_{l}}(t)^{-1}-i\boldsymbol{{A}}^{z_{l}}(t)\boldsymbol{{x}}^{z_{l}}(t)+i\boldsymbol{{X}}^{z_{l}}(t)\boldsymbol{{A}}^{z_{l}}(t)
\label{eq:mlmctdh_eom}
\end{equation}
are a non-linear set of coupled differential equations
that define the evolution of the coefficient tensors at each node
of the ML-MCTDH wavefunction. The coupling between these equations
arises from the dependence of the Hamiltonian matrices and mean-field
density matrices on the coefficient tensors of each node. These equations
can be solved directly, using general purpose integration schemes
suitable for non-linear ODEs, and such approaches are referred to
as variable mean-field (VMF) integration schemes.\citep{BECK20001}
However, in many situations it is beneficial to use alternative schemes
developed specifically for the ML-MCTDH approach. A commonly employed
alternative is based on the use of the constant mean-field (CMF) approximation.\citep{Beck1997,BECK20001,MANTHE2006168} To obtain
such a scheme, we start by expanding the configurations and SHF coefficient
tensors present in the mean-field density matrix and SPF and mean-field
Hamiltonian matrices using a Taylor series expansion around a point
$t$ to a time $t+dt$. Performing this expansion and using the standard choice
of the constraint operator ($\boldsymbol{{x}}^{z_{l}}(t)=\boldsymbol{0}$) for simplicity,
the ML-MCTDH EOMs become
\begin{equation}
\dot{\boldsymbol{A}}^{1}(t)=-\frac{i}{\hbar}\boldsymbol{h}{}_{t}^{1}\boldsymbol{A}^1(t)+\mathcal{O}(dt),\label{eq:eom_linearised_root}
\end{equation}
and
\begin{equation}
\dot{\boldsymbol{A}}^{z_{l}}(t)=  -\frac{{i}}{\hbar}\boldsymbol{Q}^{z_{l}}(t)\sum_{r}\boldsymbol{h}_{rt}^{z_{l}}\boldsymbol{A}^{z_{l}}(t)\boldsymbol{H}_{rt}^{z_{l}}\left(\boldsymbol{\rho}_{t}^{z_{l}}\right)^{\text{-}1} +\mathcal{\mathcal{O}}(dt),
\label{eq:cmf_nonroot}
\end{equation}
where we have introduced the notation $\boldsymbol{{O}}_{t}$ to denote
the now time-independent matrix obtained when $\boldsymbol{{O}}(t)$
is evaluated using the coefficient tensors obtained at time $t$. Truncating
these equations to lowest order in $dt$, we obtain the first-order
CMF scheme. This corresponds to a forward Euler integration scheme. In practice higher order integration
schemes are used.\citep{Beck1997,BECK20001,MANTHE2006168} However,
in all CMF approaches the EOMs for each node are independent,
and as such it is possible to solve them in parallel. The EOM for
the root node is linear in the coefficient tensors and so can be formally
solved to give
\begin{equation}
\boldsymbol{A}^{1}(t+dt)=\exp\left[-\frac{i}{\hbar}\boldsymbol{h}{}_{t}^{1}dt\right]\boldsymbol{A}^1(t)+\mathcal{O}(dt^{2}).
\end{equation}

During each time step the root node coefficient tensor experiences
unitary dynamics under the constant Hamiltonian matrix that is obtained
by expressing the Hamiltonian operator in terms of an orthonormal,
incomplete basis of configurations associated with the root node.  These configurations
change at each time step but are constant through a step. The EOM
for all other nodes are non-linear due to the presence of the projector,
and typically general purpose integrators are used for treating these
equations. This, however, is not the only possible approach. We can
instead further approximate Eq.\,\ref{eq:cmf_nonroot} by inserting
a Taylor series expansion of the coefficient tensors $\boldsymbol{A}^{z_{l}}(t)$
that are present in the projector $\boldsymbol{{Q}}^{z_{l}}(t)$.
Doing this gives rise to linear EOMs for each non-root node which have
the same leading error as Eq.\,\ref{eq:cmf_nonroot}. As was the
case for the CMF integration schemes, it is possible to obtain higher
order methods through the use of more sophisticated linearization
schemes. These now linear uncoupled EOMs can be formally solved to
give
\begin{equation}
\boldsymbol{A}^{z_{l}}(t+dt)=\exp\Bigg[-\frac{{i}}{\hbar}dt\mathcal{{Q}}_{t}^{z_{l}}\mathcal{{H}}_{t}^{z_{l}}\Bigg]\boldsymbol{A}^{z_{l}}(t)+\mathcal{\mathcal{O}}(dt^{2}),
\end{equation}
Here we have introduced the operators $\mathcal{{Q}}_{t}^{z_{l}}$
and $\mathcal{{H}}_{t}^{z_{l}}$, defined by
\begin{equation}
\begin{aligned}\mathcal{{Q}}_{t}^{z_{l}}\boldsymbol{A} & =\boldsymbol{Q}_{t}^{z_{l}}\boldsymbol{A},\end{aligned}
\label{eq:Qsuperop}
\end{equation}
and
\begin{equation}
\begin{aligned}\mathcal{{H}}_{t}^{z_{l}}\boldsymbol{A} & =\sum_{r}\boldsymbol{h}_{rt}^{z_{l}}\boldsymbol{A}\boldsymbol{H}\left(\boldsymbol{\rho}_{t}^{z_{l}}\right)^{-1}.\end{aligned}
\end{equation}

During each time step the non-root node coefficient tensors
experience unitary dynamics under a projected Hamiltonian matrix.
This Hamiltonian matrix is is obtained by expressing the Hamiltonian
operator in terms of a non-orthonormal, incomplete basis that is
constructed using a direct product of the configurations and SHFs
associated with this node. This will be important in the discussion
of alternative equations of motion for evolving ML-MCTDH wavefunctions
that are free of singularities.

\subsection{Linearization and the Singularity-Free EOMs \label{subsec:Linearisation-and-Evolution}}
In order to demonstrate that the singularity-free EOMs can be used
to evolve the ML-MCTDH wavefunction, we will start by linearizing the
ML-MCTDH EOMs for the root node,
\begin{equation}
\dotorthrep{\boldsymbol{A}}^{1}(t)=\dot{\boldsymbol{A}}^{1}(t)=-i\left(\frac{1}{\hbar}\boldsymbol{{h}}{}^{1}(t)-\boldsymbol{{X}}^{1}(t)\right)\orthrep{\boldsymbol{A}}^1(t),\label{eq:alt_rep_eom_root}
\end{equation}
and non-root nodes,
\begin{equation}
\dot{\boldsymbol{A}}^{z_{l}}(t)=  -\frac{i}{\hbar}\boldsymbol{Q}^{z_{l}}(t)\sum_{r}\boldsymbol{h}_{r}^{z_{l}}(t)\orthrep{\boldsymbol{A}}^{z_{l}}(t)\orthrep{\boldsymbol{H}}_{r}^{z_{l}}(t)\left(\boldsymbol{R}^{z_{l}}(t)\right)^{\text{-}1} -i\boldsymbol{A}^{z_{l}}(t)\boldsymbol{x}^{z_{l}}(t)+i\boldsymbol{X}^{z_{l}}(t)\boldsymbol{A}^{z_{l}}(t).
\label{eq:alt_rep_eom}
\end{equation}
For the root node, we obtain
the same result as in the standard ML-MCTDH representation (Eq.\,\ref{eq:eom_linearised_root}),
while for the non-root nodes we may write
\begin{equation}
\dot{\boldsymbol{A}}^{z_{l}}(t)=-\frac{i}{\hbar}\boldsymbol{Q}_{t+dt}^{z_{l}}\sum_{r}\boldsymbol{h}_{rt}^{z_{l}}\boldsymbol{A}^{z_{l}}(t)\boldsymbol{R}_{t}^{z_{l}}\orthrep{\boldsymbol{H}}_{rt}^{z_{l}}\left(\boldsymbol{R}_{t}^{z_{l}}\right)^{-1}+\mathcal{O}(dt),
\end{equation}
where we have used the same notation as in Sec. \ref{subsec:Constant-Mean-Field-Schemes},
and we have expanded $\boldsymbol{Q}^{z_{l}}(t)$ around
the point $t+dt$. As was the case for the standard ML-MCTDH EOM,
each of these equations are independent of each other. As such, they
can each be formally solved to give
\begin{equation}
\boldsymbol{A}^{z_{l}}(t+dt)=  \exp\Bigg[-\frac{{i}}{\hbar}dt\mathcal{Q}_{t+dt}^{z_{l}}\mathcal{H}_{\boldsymbol{R}^{z_{l}}t}^{z_{l}}\Bigg]\boldsymbol{A}^{z_{l}}(t)+\mathcal{\mathcal{O}}(dt^{2}),
\label{eq:linearised_eom_new_rep}
\end{equation}
where $\mathcal{Q}_{t+dt}^{z_{l}}$ is given in Eq.\,\ref{eq:Qsuperop},
and
\begin{equation}
\mathcal{H}_{\boldsymbol{R}^{z_{l}}t}^{z_{l}}\boldsymbol{A=}\sum_{r}\boldsymbol{h}_{rt}^{z_{l}}\boldsymbol{A}\boldsymbol{R}_{t}^{z_{l}}\orthrep{\boldsymbol{H}}_{rt}^{z_{l}}\left(\boldsymbol{R}_{t}^{z_{l}}\right){}^{-1}.
\end{equation}
Whenever
the mean-field density matrix is singular, Eq.\,\ref{eq:linearised_eom_new_rep}
requires the evaluation of the exponential of an unbounded superoperator,
and as such, we are not able to evaluate this term without regularisation.
If, however, we apply the matrix $\boldsymbol{R}_{t}^{z_{l}}=\boldsymbol{R}^{z_{l}}(t)$
to both sides of this equation, we find
\begin{equation}
\boldsymbol{A}^{z_{l}}(t+dt)\boldsymbol{R}^{z_{l}}(t)=  \left(\exp\Bigg[-\frac{{i}}{\hbar}dt\mathcal{Q}_{t+dt}^{z_{l}}\mathcal{H}_{\boldsymbol{R}^{z_{l}}t}^{z_{l}}\Bigg]\boldsymbol{A}^{z_{l}}(t)\right)\boldsymbol{R}^{z_{l}}(t) +\mathcal{\mathcal{O}}(dt^{2}),
\label{eq:linearised_eom_new_rep-1}
\end{equation}
where the exponentiated superoperator acts only on
on $\boldsymbol{A}^{z_{l}}(t)$. Evaluating the action of the exponential
of the superoperator on $\boldsymbol{A}^{z_{l}}(t)$, we obtain
\begin{equation}
\begin{aligned}\boldsymbol{A}^{z_{l}}(t+dt)\boldsymbol{R}^{z_{l}}(t)= & \exp\Bigg[-\frac{{i}}{\hbar}dt\mathcal{Q}_{t+dt}^{z_{l}}\mathcal{H}_{t}^{z_{l}}\Bigg]\orthrep{\boldsymbol{A}}^{z_{l}}(t)+\mathcal{\mathcal{O}}(dt^{2}),\end{aligned}
\label{eq:linearised_eom_new_rep-1-1}
\end{equation}
where we have now introduced a new Hamiltonian superoperator
\begin{equation}
\mathcal{H}_{t}^{z_{l}}\boldsymbol{A=}\sum_{r}\boldsymbol{h}_{rt}^{z_{l}}\boldsymbol{A}\orthrep{\boldsymbol{H}}_{rt}^{z_{l}},
\end{equation}
and have made no further approximations. Eq.\,\ref{eq:linearised_eom_new_rep-1-1}
requires the evaluation of the action of the exponential of a
bounded superoperator on the transformed coefficient tensor, which no longer requires 
the use of regularisation.
In principle, this exponential could be applied directly.
However, in order to arrive at the PSI EOMs,
\begin{equation}
\begin{aligned}\dotorthrep{{\boldsymbol{{A}}}}^{z_{l}}(t)= & -\frac{{i}}{\hbar}\sum_{r}\boldsymbol{{h}}_{r}^{z_{l}}(t)\orthrep{{\boldsymbol{{A}}}}^{z_{l}}(t)\orthrep{{\boldsymbol{{H}}}}_{r}^{z_{l}}(t),\end{aligned}
\label{eq:psi_eom_1}
\end{equation}
and
\begin{equation}
\dot{{\boldsymbol{{R}}}}{}^{z_{l}}(t)= -\frac{{i}}{\hbar}\sum_{r}\boldsymbol{{A}}^{z_{l}\dagger}(t)\boldsymbol{{h}}_{r}^{z_{l}}(t)\boldsymbol{{A}}^{z_{l}}(t)\boldsymbol{{R}}^{z_{l}}(t)\orthrep{{\boldsymbol{{H}}}}_{r}^{z_{l}}(t),
\label{eq:psi_eom_2}
\end{equation} 
it is necessary to take an alternative approach.
Using the definition of the projector $\mathcal{Q}_{t+dt}^{z_{l}}=1-\mathcal{P}_{t+dt}^{z_{l}}$,
we can apply a Lie-Trotter splitting of the exponential giving
\begin{equation}
\boldsymbol{A}^{z_{l}}(t+dt)\boldsymbol{R}^{z_{l}}(t)=  \exp\Bigg[-\frac{{i}}{\hbar}(-dt)\mathcal{P}_{t+dt}^{z_{l}}\mathcal{H}_{t}^{z_{l}}\Bigg]\exp\Bigg[-\frac{{i}}{\hbar}dt\mathcal{H}_{t}^{z_{l}}\Bigg]\orthrep{\boldsymbol{A}}^{z_{l}}(t)+\mathcal{\mathcal{O}}(dt^{2}).
\label{eq:linearised_eom_new_rep-1-1-1}
\end{equation}
The term involving evolution under the unprojected Hamiltonian
arises as a linearization of Eq.\,\ref{eq:psi_eom_1}, and so it can be
accounted for by using Eq.\,\ref{eq:psi_eom_1} to
evolve $\orthrep{\boldsymbol{A}}^{z_{l}}(t)$ forward in time through
a time step $dt$. We may therefore rewrite this expression as
\begin{equation}
\begin{aligned}\boldsymbol{A}^{z_{l}}(t+dt)\boldsymbol{R}^{z_{l}}(t)=  \exp\Bigg[-\frac{{i}}{\hbar}(-dt)\mathcal{P}_{t+dt}^{z_{l}}\mathcal{H}_{t}^{z_{l}}\Bigg]\orthrep{\boldsymbol{A}}^{z_{l}}(t+dt)+\mathcal{\mathcal{O}}(dt^{2}).
\end{aligned}
\label{eq:linearised_eom_new_rep-1-1-1-1}
\end{equation}

The unitary dynamics arising from the projected Hamiltonian
leads to evolution of $\boldsymbol{R}^{z_{l}}(t+dt)$, according to
a linearization of Eq.\,\ref{eq:psi_eom_2}, backwards in time through
the time step $-dt$. 
Here we have not discussed how the $\boldsymbol{R}^{z_{l}}(t+dt)$
and $\boldsymbol{A}^{z_{l}}(t+dt)$ factors are constructed at the
intermediate step, or alternatively how the $\boldsymbol{A}^{z_{l}}(t+dt)$
factor is evaluated at the end. There are a number of different approaches
for doing so that have the same order of accuracy as the above integration
scheme. The PSI approach is one possible integration scheme for integrating
the linearized equations, and is discussed in more detail in Paper II.

%The PSI approach is one possible integration scheme that does
%not introduce any additional approximations over those discussed so
%far. In this approach, the updates at each node are performed in serial,
%with Hamiltonian updates following each of these steps. However, in
%general it is not necessary to use serial updating schemes. As was
%the case for the linearised form of the ML-MCTDH EOMs, it is possible
%to update the standard coefficient tensor at each node in parallel.
%We will leave discussion of the practical application of
%parallel integration schemes and their numerical efficiency for future
%work.

The process outlined here leads to integration schemes
that are accurate to first order. Sometimes this will not provide
satisfactory performance, and it may be beneficial to use higher order
schemes. Such approaches can be obtained through the use of higher
order linearization and higher order splitting techniques, e.g.
Stang splitting.\citep{LUBICH2015,KLOSS2017, BONFANTI2018252}
\pagebreak

\section{Parallel Integration Scheme for the PSI EOMs}
\subsection{A Parallel Algorithm}

The  projector splitting integrator (PSI) \citep{LUBICH2015,LUBICH2016,KLOSS2017,BONFANTI2018252,CERUTI2021} is an inherently serial algorithm, requiring sequential
updates of the coefficient tensors at each node. This is due to the
conversion between the orthonormality conditions that is discussed
in section II.C of Paper II. When updating the coefficient tensor
at node $z_{l}$, the PSI algorithm transforms the coefficient tensors
associated with all ancestors of node $z_{l}$, such that they satisfy
the orthonormality conditions given by Eq. 8 in Paper II. In updating
a node, the PSI updates the $R^{z_{l}}$ matrix before transferring
the orthonormality condition to another node. This updating of the
$R^{z_{l}}$ alters the coefficient tensors associated with the ancestors
of node $z_{l}$, and as such it is not possible to update multiple
nodes in parallel in a simple way. 

In order to parallelize the solution of the EOMs, note
that it is not necessary to enforce that the total tree structure
satisfies the orthonormality conditions given by Eq. 8. Instead, we
can construct all of the transformed coefficient tensors, $\widetilde{\boldsymbol{A}}^{z_{l}}(t)$,
associated with all nodes, and the corresponding $\boldsymbol{R}^{z_{l}}(t)$
matrices. Now as 
\begin{equation}
\widetilde{\boldsymbol{A}}^{z_{l}}=\boldsymbol{A}^{z_{l}}\boldsymbol{R}^{z_{l}},
\end{equation}
knowledge of these two matrices for each node in the ML-MCTDH wavefunction is, 
in principle, sufficient to reconstruct the standard ML-MCTDH representation. 
E.g. we can construct the standard ML-MCTDH coefficient tensors as
\begin{equation}
\boldsymbol{A}^{z_{l}}=\widetilde{\boldsymbol{A}}^{z_{l}}\Big(\boldsymbol{R}^{z_{l}}\Big)^{-1}.
\end{equation}

We can update the standard coefficient tensors at each node in
parallel, and as discussed in Appendix B of Paper I, we can likewise
do this using the $\widetilde{\boldsymbol{A}}^{z_{l}}(t)$ and $\boldsymbol{R}^{z_{l}}(t)$
matrices. For the root node we simply need to evaluate
\begin{equation}
\widetilde{\boldsymbol{A}}^{1}(t+dt)=\exp\left[-\frac{i}{\hbar}\boldsymbol{h}{}_{t}^{1}dt\right]\widetilde{\boldsymbol{A}}^{1}(t)+\mathcal{O}(dt^{2}),
\end{equation}
which is obtained as the solution to the linear ODE
\begin{equation}
\dot{\widetilde{\boldsymbol{A}}}{}^{1}(t)=-\frac{i}{\hbar}\boldsymbol{h}{}_{t}^{1}\widetilde{\boldsymbol{A}}^{1}(t).
\end{equation}
For all non-root nodes, $z_{l}$, we have
\begin{equation}
\begin{aligned}\boldsymbol{A}^{z_{l}}(t+dt)\boldsymbol{R}^{z_{l}}(t)= & \exp\Bigg[-\frac{{i}}{\hbar}(-dt)\mathcal{P}_{t+dt}^{z_{l}}\mathcal{H}_{t}^{z_{l}}\Bigg]\exp\Bigg[-\frac{{i}}{\hbar}dt\mathcal{H}_{t}^{z_{l}}\Bigg]\widetilde{\boldsymbol{A}}{}^{z_{l}}(t)+\mathcal{\mathcal{O}}(dt^{2}).\end{aligned}
\end{equation}
Treating each of the propagators independently we have that
\begin{equation}
\boldsymbol{A}^{z_{l}}(t+dt)\boldsymbol{R}^{z_{l}}(t+dt)=\widetilde{\boldsymbol{A}}^{z_{l}}(t+dt)=\exp\Bigg[-\frac{{i}}{\hbar}dt\mathcal{H}_{t}^{z_{l}}\Bigg]\widetilde{\boldsymbol{A}}{}^{z_{l}}(t)+\mathcal{\mathcal{O}}(dt^{2}),\label{eq:prop_1}
\end{equation}
and 
\begin{equation}
\boldsymbol{R}^{z_{l}}(t)=\exp\Bigg[-\frac{{i}}{\hbar}(-dt)\mathcal{P}_{t+dt}^{z_{l}}\mathcal{H}_{t}^{z_{l}}\Bigg]\boldsymbol{R}^{z_{l}}(t+dt).
\end{equation}

Within the PSI approach, the standard coefficient tensor $\boldsymbol{A}^{z_{l}}$ is updated by first
evolving the transformed coefficient tensor forward in time according
to the EOM 
\begin{equation}
\dot{\widetilde{\boldsymbol{A}}}{}^{z_{l}}(t)=-\frac{i}{\hbar}\mathcal{H}_{t}^{z_{l}}\widetilde{\boldsymbol{A}}^{z_{l}}(t),
\end{equation}
giving the solution found in Eq.\,\ref{eq:prop_1}. Following which we decompose the transformed coefficient tensor to give
\begin{equation}
\widetilde{\boldsymbol{A}}^{z_{l}}(t+dt)=\boldsymbol{A}^{z_{l}}(t+dt)\boldsymbol{R}^{z_{l}}(t+dt).\label{eq:ar_decomp}
\end{equation}
This decomposition is not unique and it is possible to insert
any arbitrary unitary factor and its inverse between these two terms.
In the PSI approach, this arbitrary unitary matrix does not matter
as we make use of both factors in this decomposition.  
We set our new standard coefficient tensors for node $z_l$ to $\boldsymbol{A}^{z_{l}}(t+dt)$, and transfer the $\boldsymbol{R}^{z_{l}}(t+dt)$ to other nodes in the tree.  To do this, we backwards time evolve $\boldsymbol{R}^{z_{l}}(t+dt)$ according to the EOM 
\begin{equation}
\dot{\boldsymbol{R}}^{z_{l}}(t)=-\frac{{i}}{\hbar}\mathcal{P}_{t+dt}^{z_{l}}\mathcal{H}_{t}^{z_{l}}\boldsymbol{R}^{z_{l}}(t)
\end{equation}
to obtain $\boldsymbol{R}^{z_{l}}(t)$ and transfer this factor to an adjacent node
in the tree. This transfer of the $\boldsymbol{R}^{z_{l}}(t)$
factor is what leads to the serial nature of the algorithm, and is
what needs to be changed to ensure a parallel algorithm. The factor of $\boldsymbol{R}^{z_{l}}(t)$ is simply included
in the definition of the transformed coefficient tensor of the parent
of node $z_{l}$ (see Eq. 13 and 14 of Paper II). As such, if we retain
this factor in the coefficient tensor of the parent of node $z_l$ we no longer
need to apply it back up the tree following the evolution of the coefficient
tensors at this node. Instead, all we need to do is remove this contribution
from the node $z_{l}$ to obtain the standard coefficient tensor.
In principle, this could be done by evolving the transformed coefficient
forward in time from $\widetilde{\boldsymbol{A}}{}^{z_{l}}(t)$
to $\widetilde{\boldsymbol{A}}{}^{z_{l}}(t+dt)$, and evolving $\boldsymbol{R}^{z_{l}}(t)$
forward in time to obtain $\boldsymbol{R}^{z_{l}}(t+dt)$, from which
we can compute the standard coefficient tensor as 
\begin{equation}
\boldsymbol{A}^{z_{l}}(t+dt)=\widetilde{\boldsymbol{A}}^{z_{l}}(t+dt)\Big(\boldsymbol{R}^{z_{l}}(t+dt)\Big)^{-1}.
\end{equation}

Applying this approach at all non-root nodes in parallel results in
updated standard coefficient tensors with leading order error
$\mathcal{O}(dt^2)$. However, whenever the ML-MCTDH wavefunction
is rank deficient the $\boldsymbol{R}^{z_{l}}(t+dt)$ matrix is singular,
and this operation is ill-defined. As such, care needs to be taken
in the construction of the standard coefficient tensors. As in the
standard PSI algorithm, we can construct a new set of coefficient
tensors that satisfy the orthonormality conditions required for the
standard ML-MCTDH EOMs using Eq.\,\ref{eq:ar_decomp}. However, this
decomposition is not unique. If we make use of both factors, this
non-uniqueness is not an issue, however, here we are attempting to
avoid transferring the $\boldsymbol{R}^{z_{l}}(t)$ matrix to other
nodes in the tree structure, and so we will not make use of it. As
such, when we apply this decomposition we obtain
\begin{equation}
\widetilde{\boldsymbol{A}}^{z_{l}}(t+dt)=\boldsymbol{A}^{z_{l}\prime}(t+dt)\boldsymbol{R}^{z_{l}\prime}(t+dt),\label{eq:ar_decomp-1}
\end{equation}
where the new coefficient tensors $\boldsymbol{A}^{z_{l}\prime}(t+dt)$
differ from the time-evolved coefficient tensors we want by an arbitrary
unitary matrix
\begin{equation}
\boldsymbol{A}^{z_{l}\prime}(t+dt)=\boldsymbol{A}^{z_{l}}(t+dt)\boldsymbol{U}(t+dt).
\end{equation}
If we can work out this unitary matrix, we can evaluate the needed time-evolved
coefficient tensor as
\begin{equation}
\boldsymbol{A}^{z_{l}}(t+dt)=\boldsymbol{A}^{z_{l}\prime}(t+dt)\boldsymbol{U}(t+dt)^{\dagger}.
\end{equation}
Now, using the semi-unitarity of $\boldsymbol{A}^{z_{l}}(t+dt)$, we
have 
\begin{equation}
\boldsymbol{A}^{z_{l}\dagger}(t+dt)\boldsymbol{A}^{z_{l}\prime}(t+dt)=\boldsymbol{U}(t+dt),
\end{equation}
but we do not know the value of $\boldsymbol{A}^{z_{l}}(t+dt).$ If
we instead Taylor series expand this term around the point $t$, we
obtain
\begin{equation}
\left[\boldsymbol{A}^{z_{l}\dagger}(t)+dt\dot{{\boldsymbol{A}}}^{z_{l}\dagger}(t)+dt^{2}\ddot{\boldsymbol{A}}^{z_{l}\dagger}(t)+\dots\right]\boldsymbol{A}^{z_{l}\prime}(t+dt)=\boldsymbol{U}(t+dt).
\end{equation}
Now, we can further simplify this to find
\begin{equation}
\boldsymbol{A}^{z_{l}\dagger}(t)\boldsymbol{A}^{z_{l}\prime}(t+dt)+dt\dot{{\boldsymbol{A}}}^{z_{l}\dagger}(t)\boldsymbol{A}^{z_{l}}(t)\boldsymbol{U}(t+dt)+\mathcal{O}(dt^{2})=\boldsymbol{U}(t+dt).
\end{equation}
For the standard PSI gauge this gives
\begin{equation}
\boldsymbol{U}(t+dt)=\boldsymbol{A}^{z_{l}\dagger}(t)\boldsymbol{A}^{z_{l}\prime}(t+dt)+\mathcal{O}(dt^{2}).\label{eq:Ut_1}
\end{equation}
For alternative choices of the dynamic gauge conditions there is an 
additional first order contribution to this expression.
%I'm not sure if I want to include this comment or not.  The invariant EOMs approach,
%at least as presented in Manthe's paper, is not actually correct.  It has a first order
%error at each time step and so has an uncontrolled overall error.  It wouldn't be too
%difficult to fix this.
In Ref. \onlinecite{Ceruti2021a} a parallel version of
the PSI integrator for the MCTDH wavefunction is presented that makes
use of this representation of the $\boldsymbol{U}(t+dt)$. In this
scheme, the transformed coefficient tensors at each child node are
time evolved, the standard coefficient tensors are formed, and the
conjugate transpose of this (approximate) unitary factor is absorbed
into the root node, following which the root node is evolved. This
is not the only possible scheme, and indeed in Ref. \onlinecite{WEIKE2021}
an integration scheme for the invariant EOMs is presented that allows
for all nodes to be updated in parallel, and does not require transfer
of factors between nodes in the tree structure. This scheme does not
apply the matrix $\boldsymbol{U}(t+dt)$ given by Eq.\,\ref{eq:Ut_1},
it instead notes that this matrix is not unitary, and application
of this matrix will lead to a reduction in the norm of the wavefunction,
and instead uses a modified form of this matrix in which the columns
have been normalized. This does not alter the overall order of the
approximation but may reduce the errors associated with the reduction
in the norm of the matrix. This change does not generally
lead to a unitary approximation to $\boldsymbol{U}(t+dt)$. However,
it is possible to construct a unitary approximation that is accurate
to the same error ($\mathcal{O}(dt^{2})$). To do this we note that
the matrix of singular values of the matrix given in Eq.\,\ref{eq:Ut_1}
differs from the identity matrix by a term that scales as $\mathcal{O}(dt^{2})$, 
so by performing a singular value decomposition of this matrix and
replacing the singular values matrix with the identity matrix, we obtain an
approximation to this matrix that is unitary, and has the
same $\mathcal{O}(dt^{2})$ error as the non-unitary approximations
given above. To demonstrate that such a scheme provides a way of integrating
the PSI EOMs, we will now consider an application of it to the 500
mode spin-boson model considered in section III.B of Paper II.

\subsection{Results}

We have applied the first order parallel integration scheme discussed
above to the integration of the PSI EOMs for a 500 mode spin-boson
model with $\alpha=2.0,\ \omega_{c}=25\Delta$. We have once again
looked at the convergence of the dynamics with respect to time step,
and have considered the same tree structure and integration time as
was considered in Fig. 4 of Paper II. In Table\,\ref{tab:System2_parallel_algorithm_convergence},
we present the convergence of the dynamics to a reference calculation
obtained using the serial PSI with $dt\Delta=6.25\times10^{\text{-4}}$.
Here we have considered three different first-order integration schemes.  
\begin{itemize}
    \item Parallel: A completely parallel integration scheme that avoids the backward in time propagation of the $\boldsymbol{R}^{z_{l}}$ matrices and allows for the coefficient tensors of each node to be evolved in parallel.
    \item Serial: The first order version of the PSI that only makes use of the forward step and sequentially updates the transformed coefficient tensors.
    \item No $\boldsymbol{R}^{z_{l}}$ evolution: A combination of the two integration schemes that uses the serial updating pattern of the first-order PSI approach, but avoids the backwards in time propagation of the $\boldsymbol{R}^{z_{l}}$ matrices by making use of the parallelizable approximation discussed above.
\end{itemize}

\begin{table}[h]
\begin{centering}
\begin{tabular}{c|>{\centering}p{2cm}>{\centering}p{2cm}>{\centering}p{2cm}>{\centering}p{2cm}>{\centering}p{2cm}>{\centering}p{2cm}>{\centering}p{2cm}}
 & & \multicolumn{2}{c}{Parallel} & \multicolumn{2}{c}{Serial} & \multicolumn{2}{c}{No $\boldsymbol{R}^{z_{l}}$ Evolution }\tabularnewline
\hline 
\hline 
$dt\Delta$ & $N_{MF}$ & $\Delta P$ & $\langle N_{H}\rangle$ & $\Delta P$ & $\langle N_{H}\rangle$ & $\Delta P$ & $\langle N_{H}\rangle$\tabularnewline
\hline 
$4\times10^{\text{-2}}$ & 25 & $2.0\times10^{\text{-}1}$ & $5.8\times10^{2}$ & $8.4\times10^{\text{-}3}$ & $6.4\times10^{2}$ & $2.2\times10^{\text{-}2}$ & $6.2\times10^{2}$\tabularnewline
$1\times10^{\text{-2}}$ & 85 & $3.9\times10^{\text{-}2}$ & $1.1\times10^{3}$ & $4.8\times10^{\text{-}5}$ & $1.1\times10^{3}$ & $2.7\times10^{\text{-}2}$ & $1.1\times10^{3}$\tabularnewline
$2.5\times10^{\text{-3}}$ & 325 & $1.2\times10^{\text{-}2}$ & $3.0\times10^{3}$ & $5.0\times10^{\text{-}6}$ & $2.9\times10^{3}$ & $1.6\times10^{\text{-}3}$ & $3.0\times10^{3}$\tabularnewline
$6.25\times10^{\text{-4}}$ & 1285 & $6.2\times10^{\text{-}3}$ & $9.6\times10^{3}$ & $2.6\times10^{\text{-}6}$ & $9.3\times10^{3}$ & $2.0\times10^{\text{-}4}$ & $9.3\times10^{3}$\tabularnewline
$1.5625\times10^{\text{-4}}$ & 5125 & $3.5\times10^{\text{-}4}$ & $3.1\times10^{4}$ & $8.4\times10^{\text{-}7}$ & $3.1\times10^{4}$ & $1.2\times10^{\text{-}5}$ & $3.1\times10^{4}$\tabularnewline
$3.90625\times10^{\text{-5}}$ & 20485 & $3.3\times10^{\text{-}5}$ & $1.0\times10^{5}$ & $1.4\times10^{\text{-}7}$ & $1.1\times10^{5}$ & $7.5\times10^{\text{-}7}$ & $1.0\times10^{5}$\tabularnewline
$9.765625\times10^{\text{-6}}$ & 81925 & $1.7\times10^{\text{-6}}$ & $4.1\times10^{5}$ & $9.9\times10^{\text{-}9}$ & $4.1\times10^{5}$ & $8.0\times10^{\text{-}8}$ & $4.1\times10^{5}$\tabularnewline
\hline 
\end{tabular}
\par\end{centering}
\caption{\label{tab:System2_parallel_algorithm_convergence} Convergence of
the relative deviations of the population dynamics for a spin-boson
model with $N=500$ bath modes and with $\alpha=2.0,\ \omega_{c}=25\Delta$
obtained using three first-order integration schemes.
Here the method labelled parallel is an entirely parallelizable variant of the PSI, 
the serial method corresponds to the standard serial
PSI approach but only uses the forward loop steps, while the ``No
$\boldsymbol{R}^{z_{l}}$ Evolution'' method uses the serial updating
scheme of the first order PSI but makes use of the parallelizable
approximation when evolving the coefficient tensors that avoid the backwards in time propagation of the
$\boldsymbol{R}^{z_{l}}$ matrices. For both methods, results obtained using the second-order, serial PSI with a time step of $dt\Delta=6.25\times10^{\text{{-}4}}$
were used as the reference calculation.}
\end{table}
These results demonstrate that the dynamics obtained using all three of these schemes converge towards the results obtained with the second order PSI, supporting the fact that the parallel
integration scheme provides a valid approach for solving the PSI EOMs.
However, the convergence of the parallel scheme occurs at a significantly slower rate than
was observed for the serial algorithm. Convergence of the deviation
to less than $10^{\text{{-}4}}$ is not obtained until $dt\lesssim4\times10^{\text{{-}}5}$, compared to $dt\lesssim1\times10^{\text{{-}}2}$ that was required
using both the first and second order serial schemes.  In fact, the parallel integration scheme requires comparable numbers of Hamiltonian
applications to the standard ML-MCTDH calculations presented in Ref. \onlinecite{MEYER2018149}. If tighter tolerances are required,
this approach can outperform standard ML-MCTDH, but it is still less
efficient than approaches that make use of improved regularisation
schemes,\citep{MEYER2018149} and roughly 2 orders of magnitude
less efficient than either the serial PSI approach. As a result, this approach
is not practical for large scale simulations. 
 The key limitation in the accuracy of the parallel
approach appears to arise from the additional approximations involved 
in constructing the standard ML-MCTDH representation after the coefficient tensors
have been propagated. This is further supported by considering the results obtained using
the ``No $\boldsymbol{R}^{z_{l}}$ evolution'' method, that makes use of the serial scheme
for updating the coefficient tensors but uses the parallelizable approximation to avoid
the backwards in time evolution of the $\boldsymbol{R}^{z_{l}}$ matrices.
We find once again that this scheme performs considerably worse than the fully serial 
PSI, although it does show slightly more rapid convergence than the fully 
parallelizable scheme. 

 The fully parallelizable scheme we have discussed here is closely related to
the approach presented in Ref. \onlinecite{WEIKE2021} for integrating the invariant EOMs.  
That approach makes use of a higher order scheme for evolving the transformed 
coefficient tensors at each node, as well as an adaptive time step controller, and so
may provide some improvements over the parallel schemes considered here.  However, 
the approach presented in Ref. \onlinecite{WEIKE2021}, makes use of a very similar scheme 
to avoid the backwards in time evolution of the $\boldsymbol{R}^{z_{l}}$ matrices, 
which we find to be the dominant sources of error.  Whether similar performance issues are
observed for the invariant EOMs approach, or whether the use of an alternative dynamic
gauge condition leads to improved accuracy, remains to be seen.

 While the scheme we have discussed here is fully parallelizable,
it is not a practical method. It exhibits
considerably poorer performance than the serial PSI, at least for this problem.
This poor performance appears to arise due to the strategy used for
constructing the standard ML-MCTDH coefficient tensors from the transformed
coefficient tensors. Whether alternative strategies exist that would
improve this performance is an open question, and will be left as
future work. 

\pagebreak
\section{Alternative Gauge Conditions for the Singularity-Free EOMs}
As in the case of the standard ML-MCTDH approach, it is possible to
make use of alternative gauge conditions for the singularity-free EOMs. 
One choice for standard ML-MCTDH, which provides numerical benefits in some situations,
involves partitioning the Hamiltonian into the contributions that are 
separable and non-separable at each node as
\begin{align}
\hat{{H}} & =\hat{{h}}_{sep}^{z_{l}}+\hat{{H}}_{sep}^{z_{l}}+\sum_{r\not\in sep}\hat{{h}}_{r}^{z_{l}}\hat{{H}}_{r}^{z_{l}},\\
 & =\hat{{h}}_{sep}^{z_{l}}+\hat{{H}}_{sep}^{z_{l}}+\hat{H}_{res}^{z_{l}}.
\end{align}
Here, $\hat{{h}}_{sep}^{z_{l}}$ contains all terms in that
Hamiltonian that act on the degrees of freedom associated with only
one of the children of node $z_{l}$, $\hat{{H}}_{sep}^{z_{l}}$
contains all terms that only act on the degrees of freedom accounted
for by the SHF, and $\hat{{H}}_{res}^{z_{l}}$ is the non-separable, residual terms.

Setting the elements of the constraint matrix to\citep{WANG2003,BECK20001,MANTHE2008}
\begin{equation}
\left[\boldsymbol{{x}}^{z_{l}}(t)\right]_{ij}=\frac{{1}}{\hbar}\bra{\phi_{i}^{z_{l}}(t)}\hat{h}_{sep}^{z_{l}}\ket{\phi_{j}^{z_{l}}(t)}\label{eq:hamiltonian_gauge}
\end{equation}
leads to an ML-MCTDH EOM for the root node which only depends
on the residual part of the Hamiltonian.\citep{BECK20001} This
can reduce the numerical effort required to solve the ML-MCTDH EOMs
whenever the separable contributions to the SPF Hamiltonian are dominant.
We can obtain a similar approach for the singularity-free EOMs, if we make
use of this constraint matrix for the SPFs and set the elements of
the transformed SHF constraint matrix to
\begin{equation}
\left[\boldsymbol{{y}}^{z_{l}}(t)\right]_{ij}=\frac{{1}}{\hbar}\bra{\Psi_{i}^{z_{l}}(t)}\hat{H}_{sep}^{z_{l}}\ket{\Psi_{j}^{z_{l}}(t)}.\label{eq:shf_hamiltonian_gauge}
\end{equation}
Inserting these gauge conditions into the singularity-free EOMs 
(Eqs.\ 44 and 45 in Paper I) gives
\begin{equation}
\begin{aligned}\dotorthrep{{\boldsymbol{{A}}}}^{z_{l}}(t)= & -\frac{{i}}{\hbar}\sum_{r\not\in sep}\boldsymbol{{h}}_{r}^{z_{l}}(t)\orthrep{{\boldsymbol{{A}}}}^{z_{l}}(t)\orthrep{{\boldsymbol{{H}}}}_{r}^{z_{l}}(t),\end{aligned}
\end{equation}
and
\begin{equation}
\dot{{\boldsymbol{{R}}}}{}^{z_{l}}(t)=  -\frac{{i}}{\hbar}\sum_{r\not\in sep}\boldsymbol{{A}}^{z_{l}\dagger}(t)\boldsymbol{{h}}_{r}^{z_{l}}(t)\boldsymbol{{A}}^{z_{l}}(t)\boldsymbol{{R}}^{z_{l}}(t)\orthrep{{\boldsymbol{{H}}}}_{r}^{z_{l}}(t).
\end{equation}
These equations only involve the non-separable
contribution to the Hamiltonian associated with node $z_{l}$. If
the separable terms evolve on a much more rapid timescale than the
non-separable terms, fewer Hamiltonian evaluations may be required
to evolve a single coefficient tensor through a step. 
%However, methods
%developed with this gauge choice and with the linearisation strategy
%described in Appendix B of paper I\ref{subsec:Linearisation-and-Evolution} 
%will likely show considerably poorer performance than the standard
%PSI approach (see Sec. II of the Supplementary Information). 

\begin{table}[h!]
\begin{centering}
\begin{tabular}{c|>{\centering}p{2cm}>{\centering}p{2cm}>{\centering}p{2cm}>{\centering}p{2cm}>{\centering}p{2cm}}
Gauge &  & \multicolumn{2}{c}{Alternative} & \multicolumn{2}{c}{Standard}\tabularnewline
\hline 
\hline 
$dt\Delta$ & $N_{MF}$ & $\Delta P$ & $\langle N_{H}\rangle$ & $\Delta P$ & $\langle N_{H}\rangle$\tabularnewline
\hline 
$4\times10^{\text{-2}}$ & 50 & $6.4\times10^{\text{-}2}$ & $4.9\times10^{2}$ & $1.6\times10^{\text{-}2}$ & $9.9\times10^{2}$\tabularnewline
$2\times10^{\text{-2}}$ & 90 & $9.9\times10^{\text{-}3}$ & $7.7\times10^{2}$ & $1.6\times10^{\text{-}4}$ & $1.4\times10^{3}$\tabularnewline
$1\times10^{\text{-2}}$ & 170 & $3.6\times10^{\text{-}4}$ & $1.3\times10^{3}$ & $1.2\times10^{\text{-5}}$ & $2.1\times10^{3}$\tabularnewline
$5\times10^{\text{-3}}$ & 330 & $2.8\times10^{\text{-}5}$ & $2.2\times10^{3}$ & $5.0\times10^{\text{-6}}$ & $3.2\times10^{3}$\tabularnewline
$2.5\times10^{\text{-3}}$ & 650 & $6.2\times10^{\text{-}6}$ & $3.9\times10^{3}$ & $3.4\times10^{\text{-6}}$ & $5.3\times10^{3}$\tabularnewline
$1.25\times10^{\text{-3}}$ & 1290 & $3.3\times10^{\text{-}6}$ & $7.2\times10^{3}$ & $1.8\times10^{\text{-7}}$ & $9.3\times10^{3}$\tabularnewline
$6.25\times10^{\text{-4}}$ & 2570 & $5.5\times10^{\text{-7}}$ & $1.3\times10^{4}$ & Reference & $1.7\times10^{4}$\tabularnewline
\hline 
\end{tabular}
\par\end{centering}
\caption{\label{tab:System2_parallel_algorithm_convergence-2}Convergence of
the relative deviations of the population dynamics obtained using
the second order PSI with both an alternative and the standard choice
for the gauge conditions for a spin-boson model with $N=500$ bath
modes and with $\alpha=2.0,\ \omega_{c}=25\Delta$. The population
dynamics obtained using the second-order serial PSI with the standard
choice of gauge conditions with $dt\Delta=6.25\times10^{\text{{-}4}}$
is used as the reference.}
\end{table}
In order to explore this, we now consider using the PSI algorithm to integrate these 
alternative EOMs for the 500 mode spin-boson model with $\alpha=2.0,\ \omega_{c}=25\Delta$, 
and with the same tree topology and evolution parameters as were considered
in Fig. 4 of Paper II.
%
%We next consider application of the PSI with the alternative gauge
%condition discussed in Appendix C of Paper I. Here we apply the standard
%PSI algorithm, however using the alternative EOMs for the evolution
%of the transformed coefficient tensors and $\boldsymbol{R}^{z_{l}}$ tensors
%that are given in Eqs. C2 and C3 of Paper I
%
%\begin{equation}
%\begin{aligned}\dot{\widetilde{\boldsymbol{A}}}{}^{z_{l}}(t)= & %-\frac{{i}}{\hbar}\sum_{r\not\in %sep}\boldsymbol{{h}}_{r}^{z_{l}}(t)\widetilde{\boldsymbol{A}}{}^{z_{l}}(t)\widetilde{\boldsymb%ol{H}}{}_{r}^{z_{l}}(t),\end{aligned}
%\end{equation}
%and
%
%\begin{equation}
%\begin{aligned}\dot{{\boldsymbol{{R}}}}{}^{z_{l}}(t)= & -\frac{{i}}{\hbar}\sum_{r\not\in %sep}\boldsymbol{{A}}^{z_{l}\dagger}(t)\boldsymbol{{h}}_{r}^{z_{l}}(t)\boldsymbol{{A}}^{z_{l}}(%t)\boldsymbol{{R}}^{z_{l}}(t)\widetilde{\boldsymbol{H}}{}_{r}^{z_{l}}(t).\end{aligned}
%\end{equation}
%
%We once again consider the 500 mode spin-boson model with $\alpha=2.0,\ \omega_{c}=25\Delta$
%and use the same tree topology and evolution parameters as were considered
%in Fig. 4 of Paper II. 
In Table\ \ref{tab:System2_parallel_algorithm_convergence-2} we present the convergence of the
dynamics towards reference calculations obtained using the PSI with
the standard gauge choice and a time step of $dt\Delta=6.25\times10^{\text{-4}}$,
and provide the results obtained using the standard gauge (and presented
in Table III of Paper II) for reference.

Here we observe that the results obtained using this new gauge converge
towards the reference, with a deviation of less than $10^{-4}$ observed
for all $dt\Delta\leq5\times10^{-3}$. This convergence is slower
than when using the standard gauge, where deviations of less than
$10^{-4}$ are observed for all $dt\Delta\leq1\times10^{-2}$. For
a given timestep the deviation obtained using this new gauge is roughly
one order of magnitude larger than with the standard gauge, and this
can likely be attributed to the larger leading error term associated
when applying the standard linearization strategy (or in this case
a second order variant of the strategy). In this process we assume
that the SPF and SHF basis functions are constant through the evolution.
The alternative dynamic gauge used here introduces explicit rotations
of these functions that are present even if the functions would otherwise
not evolve, and this evolution is not accounted for accurately. Similar
behavior has been observed for the CMF integration schemes applied
to standard ML-MCTDH.\citep{BECK20001} This may
possibly be improved through the use of alternative linearization
strategies, however, we will not explore this further here.
 At a given time step, the number of Hamiltonian applications required
to obtain the results is considerably smaller when using this new
gauge than compared to the standard PSI gauge. This can be attributed
to the differences in the forms of the EOMs. For the standard gauge
the EOMs involve contributions from all terms in the Hamiltonian.
For the new gauge, only those terms that are not separable at the
node $z_{l}$ enter into the EOMs for node $z_{l}$. As such, it is
not necessary to include the vast majority of terms that are present
in the Hamiltonian, and the resultant integration is considerably
easier. 

\pagebreak
\section{Implementation of the PSI: An Efficient Sum-of-Product Operator Representation} 
 Fast evaluation of the SPF and mean-field Hamiltonian
matrices is incredibly important for an efficient implementation of
the PSI. For general Hamiltonians, this is not always possible, and
the evaluation of these matrices may prevent efficient implementations.
However, there are a number of representations that allow for efficient
evaluation of these matrices. In this series of two papers (and in particular in
Eqs.\,16 and 18 of Paper II)
we have made use of the sum-of-product operator representation that
is commonly used in the standard ML-MCTDH approach. In this representation
we write the Hamiltonian in terms of a sum over $N_{prod}$ operators
each of which is expressed as a direct product of operators ($\hat{h}_{\alpha,r}$)
that each act on a single physical degree of freedom
\begin{equation}
\hat{{H}}=\sum_{r=1}^{N_{prod}}\bigotimes_{\alpha=1}^{D}\hat{h}_{\alpha,r},\label{eq:sop_ham}
\end{equation}
where $D$ is the total number of physical degrees of freedom
in the problem. At each node $z_{l}$, we can make use of this representation
to write the Hamiltonian operator as a sum over a set of $N_{prod}$
operators, each of which is expressed as a direct product of operators
acting on the physical degrees of SPFs and SHFs of this node
\begin{equation}
\hat{{H}}=\sum_{r=1}^{N_{prod}}\hat{h}_{r}^{z_{l}}\hat{H}_{r}^{z_{l}}.\label{eq:sop_node_naive}
\end{equation}
Here $\hat{h}_{r}^{z_{l}}$
and $\hat{H}_{r}^{z_{l}}$ are operators acting on the physical degrees
of freedom associated with the SPFs and SHFs of node $z_{l}$, respectively.
We will refer to these operators as the SPF and SHF Hamiltonian operators.
These operators are simply direct products of the operators $\hat{h}_{\alpha,r}$,
and can be constructed recursively as
\begin{equation}
\hat{h}_{r}^{z_{l}}=\bigotimes_{k}\hat{h}_{r}^{(z_{l},k)},\label{eq:spf_hamiltonian_op}
\end{equation}
 where $\hat{h}_{r}^{(z_{l},k)}$ is a SPF Hamiltonian operator
associated with the $k$-th child of node $z_{l}$, and 
\begin{equation}
\hat{H}_{r}^{(z_{l},k)}=\hat{H}_{r}^{z_{l}}\bigotimes_{j\neq k}\hat{h}_{r}^{(z_{l},j)},\label{eq:shf_hamiltonian_op}
\end{equation}
 For each term in this sum it is possible to obtain the SPF
and mean-field Hamiltonian matrices (which correspond to $\hat{h}_{r}^{z_{l}}$
and $\hat{H}_{r}^{z_{l}}$, respectively) independently using the
recursive expressions given in Eqs.\,16 and 18 of Paper II (or equivalently Eqs.\,\ref{eq:spf-update-ascending} and \ref{eq:mf-update-descending} below). However this approach is generally
not optimal, and there are a number of optimisations that can considerably
reduce the numerical effort required to evaluate the Hamiltonian matrices.

 If either of the operators $\hat{h}_{r}^{z_{l}}$ or $\hat{H}_{r}^{z_{l}}$
are the identity operator, then the evaluation of the SPF or mean-field
Hamiltonian matrix simplifies considerably. The recursive evaluation
of the SPF (Eq.\,18 of Paper II),
\begin{equation}
\begin{aligned}
h_{r;I^{z_l}J^{z_l}}^{z_{l}}=&\prod_{k=1}^{d^{z_{l}}}A_{I^{(z_l,k)}i_k}^{*(z_{l},k)}h_{r;I^{(z_l,k)}J^{(z_l,k)}}^{(z_{l},k)}A_{J^{(z_l,k)}
j_k}^{(z_{l},k)}\\
=&\prod_{k=1}^{d^{z_{l}}}M_{r;i_k j_k}^{(z_{l},k)},
\end{aligned}\label{eq:spf-update-ascending}
\end{equation}
and mean-field Hamiltonian matrices (Eq.\, 18 of Paper II),
\begin{equation}
\widetilde{H}_{r;ab}^{(z_{l},k)}=U_{I^{z_l}_{k;a} i}^{*z_{l}}\left(\prod_{\substack{k^{\prime}=1\\k^{\prime}\neq k}}^{d^{z_{l}}}M_{r;i_{k^\prime} j_{k^\prime}}^{(z_{l},k^\prime)}\right)\widetilde{H}_{r;i j}^{z_{l}}U_{I^{z_l}_{k; b} j}^{z_{l}}\label{eq:mf-update-descending}
\end{equation}
are simply efficient ways for evaluating
the matrix elements of the Hamiltonian operators with respect to the
orthonormal set of functions given by the SPFs and transformed SHFs,
respectively. As such, if these operators are the identity operator,
these matrix elements reduce to the overlap of these functions, and
therefore the matrix representation of any identity operators is simply
the identity matrix. Thus, we do not need to explicitly evaluate
the SPF and mean-field Hamiltonian matrices that arise from terms
where $\hat{h}_{r}^{z_{l}}=\hat{1}$ or $\hat{H}_{r}^{z_{l}}=\hat{1}$.
In many cases, and for the models considered here, this significantly reduces 
the number of operations that need to be performed. 

 Another significant improvement occurs if any two (or
more) of the operators $\hat{h}_{r}^{z_{l}}$ are identical or any
two (or more) operators $\hat{H}_{r}^{z_{l}}$ are identical, then it
is possible to further simplify Eq.\,\ref{eq:sop_node_naive}. In order to do this, we will start by introducing the notation $\hat{h}_{c;i}^{z_{l}}$
($\hat{H}_{c;i}^{z_{l}}$) to denote an operator acting on the SPF
(SHF) degrees of freedom that is common to the set of 
terms with indices $\boldsymbol{r}_{i}^{h;z_{l}}$($\boldsymbol{r}_{i}^{H;z_{l}})$ in the sum in Eq.\,\ref{eq:sop_node_naive}.  We will suppose that we have $I^{h;z_l}$ and $I^{H;z_l}$ common SPF and SHF operators, respectively.
We will also introduce the notation $\boldsymbol{r}^{ind;z_l}$ to denote
the indices of operators that do not contain any of these common operators.
Using this notation, we may rewrite Eq.\,\ref{eq:sop_node_naive} as
\begin{equation}
\begin{aligned}
\hat{H}=\sum_{r\in\boldsymbol{r}^{ind;z_{l}}}\hat{h}_{r}^{z_{l}}\hat{H}_{r}^{z_{l}}+\sum_{i=1}^{I^{h;z_l}}\hat{h}_{c;i}^{z_{l}}\sum_{r\in\boldsymbol{r}_{i}^{h;z_{l}}}\hat{H}_{r}^{z_{l}}+\sum_{i=1}^{I^{H;z_l}}\hat{H}_{c;i}^{z_{l}}\sum_{r\in\boldsymbol{r}_{i}^{H;z_{l}}}\hat{h}_{r}^{z_{l}}.\label{eq:sop_node_naive-1}
\end{aligned}
\end{equation}
 We next introduce the notation
\begin{equation}
\hat{h}_{s;i}^{z_{l}}=\sum_{r\in\boldsymbol{r}_{i}^{H;z_{l}}}\hat{h}_{r}^{z_{l}}\ \ \ \ \ \text{and}\ \ \ \ \ \hat{H}_{s;i}^{z_{l}}=\sum_{r\in\boldsymbol{r}_{i}^{h;z_{l}}}\hat{H}_{r}^{z_{l}}
\end{equation}
 for the sums of operators that act on the physical degrees
of freedom associated with the SPFs and SHFs, respectively. Now Eq.\,\ref{eq:sop_node_naive-1}
becomes
\begin{equation}
\hat{H}=\underbrace{\sum_{r\in\boldsymbol{r}^{ind;z_l}}\hat{h}_{r}^{z_{l}}\hat{H}_{r}^{z_{l}}}_{\text{(A)}}+\underbrace{\sum_{i=1}^{I^{h;z_l}}\hat{h}_{c;i}^{z_{l}}\hat{H}_{s;i}^{z_{l}}}_{\text{(B)}}+\underbrace{\sum_{i=1}^{I^{H;z_l}}\hat{h}_{s;i}^{z_{l}}\hat{H}_{c;i}^{z_{l}}}_{\text{(C)}}.\label{eq:sop_node}
\end{equation}

 This expression is still in the form of a sum-of-product
operator representation, now with a potentially reduced number
of terms in the sum. It is no longer possible to use the recursive expressions for the SPF and mean-field density matrices given by Eqs.\,\ref{eq:spf-update-ascending} and \ref{eq:mf-update-descending}.  It is necessary to make use of different strategies for evaluating these matrices depending on the terms from which they arise from.  In order to make use of this representation it is necessary to evaluate the $I^{h;z_l}$ sets of indices $\boldsymbol{r}_{i}^{h;z_{l}}$, the $I^{H;z_l}$ sets of indices $\boldsymbol{r}_{i}^{H;z_{l}}$, and the remaining set of indices  $\boldsymbol{r}^{ind;z_l}$.  We will now present a recursive approach for constructing these sets of indices.

\subsection{Partitioning of Indices} 
\subsubsection*{(B): Common SPF operator indices}
 We start by considering the leaf node of the ML-MCTDH wavefunction tree that corresponds to the physical degree of freedom $\alpha$, which we will label by $l_\alpha$.  Now we can express the sum-of-product Hamiltonian given by Eq.\,\ref{eq:sop_ham} in the form given by Eq.\,\ref{eq:sop_node_naive} as
\begin{equation}
\begin{aligned}
\hat{{H}}=\sum_{r=1}^{N_{prod}}\hat{h}_{\alpha,r}\bigotimes_{\beta\neq\alpha}^{D}\hat{h}_{\beta,r}  = \sum_{r=1}^{N_{prod}}\hat{h}_{\alpha,r}\hat{H}^{l_\alpha}_r=\sum_{r=1}^{N_{prod}}\hat{h}^{l_\alpha}_{r}\hat{H}^{l_\alpha}_r.
\end{aligned}
\end{equation}
 The SPF Hamiltonian operators acting on the leaf nodes are the primitive Hamiltonian operators associated with the physical degree of freedom $\alpha$, and so we can immediately construct the indices $\boldsymbol{r}_{i}^{h;z_{l}}$ for the common SPF Hamiltonian operator terms by inspecting these primitive Hamiltonian operators. For all non-leaf nodes, we need to use the recursive relation for the SPF Hamiltonian operators in order to construct these indices.  From the recursive definition of the SPF Hamiltonian operators of a node $z_l$, the operators $\hat{h}_{c;i}^{z_{l}}$ must satisfy
\begin{equation}
\hat{h}_{c;i}^{z_{l}}=\hat{h}_{r}^{z_{l}}=\bigotimes_{k}\hat{h}_{r}^{(z_{l},k)} =\bigotimes_{k}\hat{h}_{c;i_k}^{(z_{l},k)}\ \ \forall\ r\in\boldsymbol{r}_{i}^{h;z_{l}}.\label{eq:recursive_common_spf}
\end{equation}
 In order for $\hat{h}_{c;i}^{z_{l}}=\hat{h}_{r}^{z_{l}}\ \forall\ r\in\boldsymbol{r}_{i}^{h;z_{l}}$, it is necessary that the operators $\hat{h}_{r}^{z_{l},k}$ are also
the same for all $r\in\boldsymbol{r}_{i}^{h;z_{l}}$.  If we denote the set of indices for which the operator $\hat{h}_{r}^{(z_{l},k)} = \hat{h}_{c;i_k}^{(z_{l},k)}$ as $\boldsymbol{r}_{i_{k}}^{h;z_{l},k}$, we  have that $\boldsymbol{r}_{i}^{h;z_{l}}\subseteq\boldsymbol{r}_{i_{k}}^{h;(z_{l},k)}\ \ \forall$ children $(z_l,k)$ of node $z_l$.  Moreover, we can construct the indices $r\in\boldsymbol{r}_{i}^{h;z_{l}}$ as the intersection of these sets $\boldsymbol{r}_{i_{k}}^{h;z_{l},k}$, that is
\begin{equation}
    \boldsymbol{r}_{i}^{h;z_{l}} = \bigcap_{k} \boldsymbol{r}_{i_{k}}^{h;(z_{l},k)},\label{eq:rspf_set_intersection}
\end{equation}
where $\bigcap_k$ denotes the intersection of all the sets indexed by $k$. 
 In order to obtain all of the sets of indices $\boldsymbol{r}_{i}^{h;z_{l}}$ for all $i=1,\dots,I^{h;z_l}$, it is necessary to evaluate Eq.\,\ref{eq:rspf_set_intersection} for all of the values of $i_k=1,\dots,I^{h;(z_l, k)}$ for each of the child nodes $(z_l, k)$.  We only label those terms where $\boldsymbol{r}_{i}^{h;z_{l}}$ contains at least 2 indices as common SPF operator terms.  We can obtain the sets of indices $\boldsymbol{r}_{i}^{h;z_{l}}$ for all nodes $z_l$ by recursively applying this procedure starting from the leaf nodes, and continuing until we reach the root node.

\subsubsection*{(C): Common SHF operator indices}
 Having constructed the sets of indices for the common SPF operators, we are now in a positions where we can construct the sets of indices for the common SHF operators.  To do this we start by considering the root node.  For this node, the Hamiltonian can be written in the form
\begin{equation}
    \hat{H} = \left(\sum_{r=1}^{N_{prod}} \hat{h}^{0}_r\right) \times 1 =  \left(\sum_{r=1}^{N_{prod}} \hat{h}^{0}_r\right) \times H^{0}.
\end{equation}
 As such, for the root node all terms can be treated as having a common mean-field Hamiltonian.  For the root node all terms are of the form (B), and we have
\begin{equation}
    \boldsymbol{r}_1^{H;0}=\left\{1,2,\dots,N_{prod}\right\}.
\end{equation}
 From the recursive definition of the SHF Hamiltonian operators (Eq.\,\ref{eq:shf_hamiltonian_op}), we have
\begin{equation}
\hat{H}_{c;i_k}^{(z_l, k)} = \hat{H}_{r}^{(z_{l},k)}=\hat{H}_{r}^{z_{l}}\bigotimes_{j\neq k}\hat{h}_{r}^{(z_{l},j)} \ \forall\ r\in\boldsymbol{r}_{i_k}^{H;(z_l, k)}
=\hat{H}_{c;i}^{z_{l}}\bigotimes_{j\neq k}\hat{h}_{c;i_j}^{(z_{l},j)}.\label{eq:recursive_common_shf}
\end{equation}
 In order for $\hat{H}_{c;i_k}^{(z_l, k)} = \hat{H}_{r}^{(z_{l},k)} \ \ \forall\ \ r\in\boldsymbol{r}_{i_k}^{H;(z_l, k)}$, it is necessary that the operators $\hat{H}_{r}^{z_{l}}$ and $\hat{h}_{r}^{(z_{l},j)}\ \ \forall\ \ j\neq k$ are also common operators for their respective nodes.  As a consequence, for each of the children $(z_l,k)$ of node $z_l$ we find $\boldsymbol{r}_{i_k}^{H;(z_{l}, k)} \subseteq\boldsymbol{r}_i^{H;z_l}$ and $\boldsymbol{r}_{i_k}^{H;(z_{l}, k)} \subseteq\boldsymbol{r}_{i_j}^{h;(z_l,j)}\ \forall \ j\neq k$. Moreover, we can construct the indices $r\in\boldsymbol{r}_{i}^{H;(z_{l},k)}$ as the intersection of the SPF sets $\boldsymbol{r}_{i_{j}}^{h;(z_{l},j)}$ for the sibling nodes of $(z_{l},k)$ and the SHF set $\boldsymbol{r}_i^{H;z_l}$ of the parent node $z_l$. That is
\begin{equation}
    \boldsymbol{r}_{i_k}^{H;(z_{l},k)} = \boldsymbol{r}_i^{H;z_l}\bigcap_{j\neq k} \boldsymbol{r}_{i_{j}}^{h;(z_{l},j)}.\label{eq:rshf_set_intersection}
\end{equation}
 Now, as was the case for the common SPF Hamiltonian terms, it is necessary to evaluate these intersections for all allowed values of $i=1,\dots,I^{H;z_l}$, and $i_j = 1, \dots, I^{h;(z_l, j)}$ in order to construct all of the different $\boldsymbol{r}_{i_k}^{H;(z_{l},k)}$ for $i_k = 1, \dots, I^{H;(z_l, K)}$.  This expression can be applied recursively moving down the tree structure, provided the common SPF indices have been evaluated.  
 Once we have constructed all of the sets of indices corresponding to the common SPF and SHF Hamiltonian operators at each node, the remaining set $\boldsymbol{r}^{ind;z_l}$ is the set of all of the indices that have not already been assigned to a set.  
 Having given recursive expressions that enable us to construct the representation given by Eq.\,\ref{eq:sop_node} for each node, we will now discuss the process through which the Hamiltonian matrices corresponding to these terms can be evaluated.
\subsection{Evaluation of Hamiltonian Matrices}
\subsubsection*{(A): Standard Terms}
 The terms in sum (A) in Eq.\,\ref{eq:sop_node}, are exactly of the form given in the naive sum-of-product representation for the Hamiltonian.  As such, these terms can be treated in exactly the same manner as the standard terms.  In other words the Hamiltonian matrices can be constructed using Eqs.\,\ref{eq:spf-update-ascending} and \ref{eq:mf-update-descending}.  The Hamiltonian matrices of other nodes that are used in this expression will not necessarily correspond to terms in the sum (A) for these nodes.  In particular, for the evaluation of the SPF Hamiltonian matrices of node $z_l$, may involve SPF Hamiltonian matrices of node $(z_l,k)$ that are common to multiple indices $r$.  However, as we will see below, the SPF Hamiltonian matrices associated  with these common terms are evaluated using Eq.\,\ref{eq:spf-update-ascending}.  Similar arguments hold for the evaluation of the mean-field Hamiltonian matrices.  Importantly, the Hamiltonian matrices for node $z_l$ associated with these terms never depend on the matrices associated with operators $\hat{h}_{s;i}^{\alpha}$ or $\hat{H}_{s;i}^{\alpha}$ for any node $\alpha$. 

\subsubsection*{(B): Common SPF Hamiltonian Terms}
 The evaluation of the Hamiltonian matrices associated with the terms $\hat{h}_{c;i}^{z_{l}}\hat{H}_{s;i}^{z_{l}}$ that are present in sum (B) of Eq.\,\ref{eq:sop_node}, is more involved.  These are the terms where the SPF Hamiltonian operator is common to many values of $r\in\boldsymbol{r}_{i}^{h;z_{l}}$, and the SHF Hamiltonian operator is a sum over many different terms.  The SPF Hamiltonian terms can be obtained recursively according to Eq.\,\ref{eq:recursive_common_spf}.  As such, the SPF Hamiltonian matrix corresponding to the term $h_{c;i}^{z_l}$ can be obtained using an expression of the form given by Eq.\,\ref{eq:spf-update-ascending}, that now involves the matrices $\boldsymbol{h}_{c,{i_k}}^{(z_l,k)}$ that are the SPF Hamiltonian matrices associated with the operators $\hat{h}_{c;i_k}^{(z_{l}, k)}$.

 The evaluation of the mean-field Hamiltonian matrix associated with the terms in sum (B) of Eq.\,\ref{eq:sop_node} is more involved. Now as $\boldsymbol{r}_{i}^{h;z_{l}}\subseteq\boldsymbol{r}_{i_{k}}^{h;(z_{l},k)}$,  the sum in
\begin{equation}
\hat{H}_{i_{k},s}^{z_{l},k}=\sum_{r\in\boldsymbol{r}_{i_{k}}^{h;(z_{l},k)}}\hat{H}_{r}^{(z_{l},k)}
\end{equation}
 must contain all indices $r\in\boldsymbol{r}_{i}^{h;z_{l}}$ 
for all values of $i$ where $\boldsymbol{r}_{i}^{h;z_{l}}\cap\boldsymbol{r}_{i_{k}}^{h;(z_{l},k)}\neq\varnothing$. Grouping all of these terms together, We may rewrite this sum as
\begin{equation}
\hat{H}_{i_{k},s}^{(z_{l},k)}=\sum_{j\in\boldsymbol{J}^{h;z_l}}\sum_{r\in\boldsymbol{r}_{j}^{h;z_{l}}}\hat{H}_{r}^{(z_{l},k)}+\sum_{r\in\boldsymbol{r}_{\setminus\boldsymbol{J}^{h;z_l}}^{h;(z_{l},k)}}\hat{H}_{r}^{(z_{l},k)},
\end{equation}
 where we have introduced the set
\begin{equation}
    \boldsymbol{J}^{h;z_l}=\{j|\boldsymbol{r}_{j}^{h;z_{l}}\cap\boldsymbol{r}_{i_{k}}^{h;(z_{l},k)}\neq\varnothing\},
\end{equation}
and 
\begin{equation}
    \boldsymbol{r}_{\setminus\boldsymbol{J}^{h;z_l}}^{h;(z_{l},k)}=\boldsymbol{r}_{i_{k}}^{h;(z_{l},k)}\setminus\left(\bigcup_{j\in\boldsymbol{J}^{h;z_l}}\boldsymbol{r}_{j}^{h;z_{l}}\right).
\end{equation}
Inserting the recursive definition of the SHF Hamiltonian operators
(Eq.\,\ref{eq:shf_hamiltonian_op}), this becomes
\begin{equation}
\hat{H}_{i_{k},s}^{(z_{l},k)}=\sum_{j\in\boldsymbol{J}^{h;z_l}}\hat{H}_{s,j}^{z_{l}}\bigotimes_{l\neq k}\hat{h}_{c,j}^{(z_{l},l)}+\sum_{r\in\boldsymbol{r}_{\setminus\boldsymbol{J}^{h;z_l}}^{h;(z_{l},k)}}\hat{H}_{r}^{(z_{l},k)}\bigotimes_{j\neq k}\hat{h}_{r}^{(z_{l},j)}.
\end{equation}
From the above it is clear that we do not need the individual terms that define
the operators $\hat{H}_{s,j}^{z_{l}}$, we only need the overall operator.  The mean-field Hamiltonian matrix for the operator $\hat{H}_{i_{k},s}^{(z_{l},k)}$ can therefore be evaluated as
\begin{equation}
\hat{H}_{i_{k},s}^{z_{l},k}=\sum_{j\in\boldsymbol{J}^{h;z_l}}\hat{H}_{s,j}^{z_{l}}\bigotimes_{l\neq k}\hat{h}_{c,j}^{z_{l},l}+\sum_{r\in\boldsymbol{r}_{\setminus\boldsymbol{J}^{h;z_l}}^{h;z_{l},k}}\hat{H}_{r}^{z_{l},k}\bigotimes_{j\neq k}\hat{h}_{r}^{z_{l},j}.\label{eq:SHF_sum_simplification}
\end{equation}
In order to evaluate the matrix representation of this operator, it is necessary to construct matrix elements with respect to the transformed SHFs.  Doing so for each term we find that the matrix representation of this operator may be expressed as
\begin{equation}
\boldsymbol{H}_{i_{k},s}^{z_{l},k}=\sum_{j\in\boldsymbol{J}^{h;z_l}}\boldsymbol{M}_{s,j}^{z_{l}}+\sum_{r\in\boldsymbol{r}_{\setminus\boldsymbol{J}^{h;z_l}}^{h;z_{l},k}}\boldsymbol{M}_{r}^{z_{l},k},
\end{equation}
where each of the matrices $\boldsymbol{M}_{s,j}^{z_{l}}$ and $\boldsymbol{M}_{r}^{z_{l},k}$ can be calculated according to an equation with the same form as Eq.\,\ref{eq:mf-update-descending} but using the matrix representations of the operators present in Eq.\,\ref{eq:SHF_sum_simplification}.

\subsubsection*{(C): Common SHF Hamiltonian Terms}
 The evaluation of the Hamiltonian matrices associated with the terms $\hat{h}_{s;i}^{z_{l}}\hat{H}_{c;i}^{z_{l}}$ that are present in sum (C) of Eq.\,\ref{eq:sop_node} also requires an alternative approach.  As for the type (B) terms, the matrix representation of the common operator in this term ($\hat{H}_{c;i}^{z_{l}}$) can be evaluated in the standard way.  That is, the mean-field Hamiltonian matrix corresponding to $H_{c;i}^{z_l}$ can be obtained using a recursive expression of the form given by Eq.\,\ref{eq:mf-update-descending}, however now involving the matrix representations of the operators present in Eq.\,\ref{eq:recursive_common_shf}.

 As was the case for the type (B) terms, evaluation of the other term is more involved.  The SPF Hamiltonian operator is obtained as a sum over many terms 
\begin{equation}
\hat{h}_{s;i}^{z_{l}}=\sum_{r\in\boldsymbol{r}_{i}^{H;z_{l}}}\hat{h}_{r}^{z_{l}}.
\end{equation}
Now, as $\boldsymbol{r}_{i_k}^{H;(z_{l}, k)} \subseteq\boldsymbol{r}_{i_j}^{H;z_l}\ \forall\ k$, we may rewrite the sum over $r$ to give
\begin{equation}
\hat{h}_{s;i}^{z_{l}}=\sum_k \sum_{j\in\boldsymbol{J}^{H;(z_l,k)}}\sum_{r\in\boldsymbol{r}_{j}^{H;(z_{l},k)}}\hat{h}_{r}^{z_{l}}+\sum_{r\in\boldsymbol{r}_{\setminus\boldsymbol{J}^{H;z_l}}^{H;z_{l}}}\hat{h}_{r}^{z_{l}},
\end{equation}
 where the sum of $k$ runs over all children of this node, and we have introduced the sets associated with each of the children
\begin{equation}
    \boldsymbol{J}^{H;(z_l,k)}=\{j|\boldsymbol{r}_{i}^{H;z_{l}}\cap\boldsymbol{r}_{j}^{H;(z_{l},k)}\neq\varnothing\},
\end{equation}
and the remaining set of $r$ indices
\begin{equation}
    \boldsymbol{r}_{\setminus\boldsymbol{J}^{H;z_l}}^{H;z_{l}}=\boldsymbol{r}_{i}^{H;z_{l}}\setminus\left(\bigcup_k\bigcup_{j\in\boldsymbol{J}^{H;(z_l,k)}}\boldsymbol{r}_{j}^{H;(z_{l},k)}\right).
\end{equation}
Inserting the recursive definition of the SPF Hamiltonian operator (Eq.\,\ref{eq:spf_hamiltonian_op}) gives
\begin{equation}
\hat{h}_{s;i}^{z_{l}}=\sum_k \sum_{j\in\boldsymbol{J}^{H;(z_l,k)}}\sum_{r\in\boldsymbol{r}_{j}^{H;(z_{l},k)}} \bigotimes_{k^\prime} \hat{h}_{r}^{(z_{l},k^\prime)}+\sum_{r\in\boldsymbol{r}_{\setminus\boldsymbol{J}^{H;z_l}}^{H;z_{l}}}\bigotimes_{k^\prime} \hat{h}_{r}^{(z_{l},k^\prime)}.
\end{equation}
For each value of $j\in\boldsymbol{J}^{H;(z_l,k)}$, there is an index $i_{jk^\prime}$ associated with each of the sibling nodes to node $(z_l,k)$, namely the nodes $(z_l, k^\prime)$ for which $\boldsymbol{r}_{j}^{H;(z_{l},k)}\subset\boldsymbol{r}_{i_{jk^\prime}}^{h;(z_{l},k^\prime)}$.  All of the $r$ indices present in this term are accounted for by common SPF Hamiltonian operator terms in the sibling nodes.  As such, we may rewrite the above expression in the form
\begin{equation}
\begin{aligned}
\hat{h}_{s;i}^{z_{l}}=&\sum_k \sum_{j\in\boldsymbol{J}^{H;(z_l,k)}}\hat{h}_{s;j}^{(z_l, k)} \bigotimes_{k^\prime\neq k} \hat{h}_{c;i_{jk^\prime}}^{(z_l,k^\prime)}+\sum_{r\in\boldsymbol{r}_{\setminus\boldsymbol{J}^{H;z_l}}^{H;z_{l}}}\bigotimes_{k^\prime} \hat{h}_{r}^{(z_{l},k^\prime)}.
\end{aligned}
\end{equation}
The SPF Hamiltonian operator of node $z_l$ can therefore be expressed in terms of direct products between common and sum terms, or general terms.  For each term it is possible to construct the matrix representation using a recursive expression of the form given in Eq.\,\ref{eq:spf-update-ascending}, and the overall matrix representation of the operator $\hat{h}_{s;i}^{z_{l}}$ is simply the sum over all of these terms.

This strategy for treating the Hamiltonian can
provide considerable performance improvements over the naive sum-of-product
representation given in Eq.\,\ref{eq:sop_node_naive}, depending on
the form of the Hamiltonian. As an example, for the standard spin-boson
model Hamiltonian for a system with $N$ bath modes, the full
sum-of-product representation given in Eq.\,\ref{eq:sop_ham} contains
$N_{prod}=2N+1$ terms. Using this representation it would be
necessary to evaluate $\mathcal{O}(N)$ terms in order to construct
the Hamiltonian at each node. As the number of nodes scales as $\mathcal{O}(N\log(N))$,
the overall algorithm would scale as $\mathcal{O}(N^{2}\log(N))$.
When using a separation of the system and bath degrees of freedom
within the tree structure, this alternative representation requires
the evaluation of 3 distinct terms for each node where the SPFs represent
solely bath degrees of freedom, and 2 for the other nodes. Using this
alternative representation we take the scaling of the approach to
$\mathcal{O}(N\log(N))$. For the $10^{6}$ mode problems considered in Paper II
that required roughly 30 hours in order to obtain converged results,
the use of the naive representation would take $\mathcal{O}(N=10^{6})$
times longer, which would clearly be impractical.
Source code capable of constructing the partitioning of the Hamiltonian terms and evaluating the resultant terms will be made available upon request.
\pagebreak
\section{Further Benchmarks: Convergence With Respect To Krylov Subspace Integrator Parameters}

In applying the PSI it is necessary to solve the linear ODEs for the
transformed coefficient tensors and $\boldsymbol{R}^{z_{l}}$ matrices.
Here we have done this using an adaptive order, adaptive time step
Krylov subspace integration scheme that makes use of the error estimates
presented in Ref. \onlinecite{Sidje1998}. In all calculations discussed
in the main text we have used a fixed value for the Krylov subspace
integrator tolerance, $\epsilon_{kry}$, of $\epsilon_{kry}=10^{\text{-12}}$.
Here we now consider the effect this tolerance parameter has on the
convergence and efficiency of the PSI calculations. We once again
consider the 500 mode spin-boson moel with with $\alpha=2.0,\ \omega_{c}=25\Delta$
and use the same tree topology and evolution parameters as were considered
in Fig. 4 of Paper II. The convergence of the dynamics with respect
to time step for various choices of $\epsilon_{kry}$ is given in
Table\ \ref{tab:System1_convergence-2-1}.

\begin{table*}[h!]
\begin{centering}
\begin{tabular}{cc>{\centering}p{2cm}>{\centering}p{2cm}>{\centering}p{2cm}>{\centering}p{2cm}>{\centering}p{2cm}>{\centering}p{2cm}}
\multicolumn{1}{c|}{$\epsilon_{kry}$} & $ $ & \multicolumn{2}{c}{$10^{\text{-4}}$} & \multicolumn{2}{c}{$10^{\text{-6}}$} & \multicolumn{2}{c}{$10^{\text{-8}}$}\tabularnewline
\hline 
\hline 
\multicolumn{1}{c|}{$dt\Delta$} & $N_{MF}$ & $\Delta P$ & $\langle N_{H}\rangle$ & $\Delta P$ & $\langle N_{H}\rangle$ & $\Delta P$ & $\langle N_{H}\rangle$\tabularnewline
\hline 
\multicolumn{1}{c|}{$4\times10^{\text{-2}}$} & 50 & $1.2\times10^{\text{-}2}$ & $1.3\times10^{3}$ & $1.5\times10^{\text{-}2}$ & $1.3\times10^{2}$ & $1.5\times10^{\text{-}2}$ & $9.9\times10^{2}$\tabularnewline
\multicolumn{1}{c|}{$2\times10^{\text{-2}}$} & 90 & $1.1\times10^{\text{-}3}$ & $5.2\times10^{3}$ & $1.7\times10^{\text{-}4}$ & $1.7\times10^{3}$ & $6.7\times10^{\text{-}5}$ & $1.2\times10^{3}$\tabularnewline
\multicolumn{1}{c|}{$1\times10^{\text{-2}}$} & 170 & $6.8\times10^{\text{-3}}$ & $2.1\times10^{3}$ & $2.0\times10^{\text{-5}}$ & $1.4\times10^{3}$ & $1.8\times10^{\text{-5}}$ & $1.6\times10^{3}$\tabularnewline
\multicolumn{1}{c|}{$5\times10^{\text{-3}}$} & 330 & $4.4\times10^{\text{-3}}$ & $1.6\times10^{3}$ & $1.4\times10^{\text{-4}}$ & $2.0\times10^{3}$ & $4.4\times10^{\text{-6}}$ & $2.4\times10^{3}$\tabularnewline
\multicolumn{1}{c|}{$2.5\times10^{\text{-3}}$} & 650 & $3.5\times10^{\text{-2}}$ & $2.6\times10^{3}$ & $1.8\times10^{\text{-5}}$ & $3.4\times10^{3}$ & $3.3\times10^{\text{-6}}$ & $4.0\times10^{3}$\tabularnewline
\multicolumn{1}{c|}{$1.25\times10^{\text{-3}}$} & 1290 & $5.4\times10^{\text{-3}}$ & $5.1\times10^{3}$ & $6.1\times10^{\text{-5}}$ & $6.3\times10^{3}$ & $1.8\times10^{\text{-6}}$ & $7.3\times10^{3}$\tabularnewline
\multicolumn{1}{c|}{$6.25\times10^{\text{-4}}$} & 2570 & $4.1\times10^{\text{-4}}$ & $1.0\times10^{4}$ & $1.4\times10^{\text{-4}}$ & $1.1\times10^{4}$ & $8.0\times10^{\text{-8}}$ & $1.3\times10^{4}$\tabularnewline
\hline 
 &  &  &  &  &  &  & \tabularnewline
 & \multicolumn{1}{c|}{$\epsilon_{kry}$} & $ $ & \multicolumn{2}{c}{$10^{\text{-10}}$} & \multicolumn{2}{c}{$10^{\text{-12}}$} & \tabularnewline
\cline{2-7} \cline{3-7} \cline{4-7} \cline{5-7} \cline{6-7} \cline{7-7} 
 & \multicolumn{1}{c|}{$dt\Delta$} & $N_{MF}$ & $\Delta P$ & $\langle N_{H}\rangle$ & $\Delta P$ & $\langle N_{H}\rangle$ & \tabularnewline
\cline{2-7} \cline{3-7} \cline{4-7} \cline{5-7} \cline{6-7} \cline{7-7} 
 & \multicolumn{1}{c|}{$4\times10^{\text{-2}}$} & 50 & $1.3\times10^{\text{-}2}$ & $8.8\times10^{2}$ & $1.6\times10^{\text{-}2}$ & $9.9\times10^{2}$ & \tabularnewline
 & \multicolumn{1}{c|}{$2\times10^{\text{-2}}$} & 90 & $1.2\times10^{\text{-}4}$ & $1.3\times10^{3}$ & $1.6\times10^{\text{-}4}$ & $1.4\times10^{3}$ & \tabularnewline
 & \multicolumn{1}{c|}{$1\times10^{\text{-2}}$} & 170 & $1.7\times10^{\text{-5}}$ & $1.8\times10^{3}$ & $1.2\times10^{\text{-5}}$ & $2.1\times10^{3}$ & \tabularnewline
 & \multicolumn{1}{c|}{$5\times10^{\text{-3}}$} & 330 & $4.2\times10^{\text{-6}}$ & $2.8\times10^{3}$ & $5.0\times10^{\text{-6}}$ & $3.2\times10^{3}$ & \tabularnewline
 & \multicolumn{1}{c|}{$2.5\times10^{\text{-3}}$} & 650 & $3.1\times10^{\text{-6}}$ & $4.6\times10^{3}$ & $3.4\times10^{\text{-6}}$ & $5.3\times10^{3}$ & \tabularnewline
 & \multicolumn{1}{c|}{$1.25\times10^{\text{-3}}$} & 1290 & $1.7\times10^{\text{-6}}$ & $8.1\times10^{3}$ & $1.8\times10^{\text{-7}}$ & $9.3\times10^{3}$ & \tabularnewline
 & \multicolumn{1}{c|}{$6.25\times10^{\text{-4}}$} & 2570 & $1.1\times10^{\text{-7}}$ & $1.5\times10^{4}$ & Reference & $1.7\times10^{4}$ & \tabularnewline
\cline{2-7} \cline{3-7} \cline{4-7} \cline{5-7} \cline{6-7} \cline{7-7} 
\end{tabular}
\par\end{centering}
\caption{\label{tab:System1_convergence-2-1}The convergence with respect to
time step ($dt$) of the deviation in the population dynamics
of a spin-boson model obtained using the projector splitting integrator
for various values of the Krylov subspace integration tolerance, $\epsilon_{kry}$.
The spin-boson model considered has $\varepsilon/\Delta=0$, $\alpha=2.0$,
$\omega_{c}=25\Delta$, and with $N=500$ bath modes. These calculations
were performed using the same tree topology as was used in the main
text and deviations from the reference calculation (indicated in the
table) were calculated using dynamics obtained up to $t\Delta=0.8$.}
\end{table*}

For $\epsilon_{kry}\geq10^{\text{-6}}$ we observe rather large deviations
even at very small time steps. The results obtained are not converged to
within a tolerance of $10^{-4}$ for any of the time steps considered.
However, upon decreasing $\epsilon_{kry}$ to $\leq10^{-8}$, we find
that the deviations obtained become roughly independent of the value
$\epsilon_{kry}$ for a given time step, and so a Krylov subspace
tolerance of $\epsilon_{kry}\leq10^{\text{-8}}$ is sufficient to
converge the dynamics of this model. For all values of $\epsilon_{kry}\leq10^{\text{-8}}$,
we find that, unsurprisingly, as the Krylov subspace tolerance increases
the number of Hamiltonian applications required decreases. Using a
Krylov subspace tolerance of $\epsilon_{kry}\leq10^{\text{-8}}$ leads
to a $\sim25$\% decrease in the number of Hamiltonian applications
required compared to the calculations obtained with $\epsilon_{kry}\leq10^{\text{-12}}$
that are reported in the main text, without significant changes in
the accuracy of the calculations. For larger values of the Krylov
subspace integration tolerance the inaccurate evolution of the coefficient
tensors leads to significantly larger deviations.
\pagebreak
\section{Further Benchmarks: Wider Tree Structures}

In addition to the models considered in the main text, we now consider
a series of three spin-boson models that have previously been treated using
standard ML-MCTDH with improved regularisation schemes.\citep{MEYER2018149, WANG2021}
Here we will consider similar tree structures to those considered in Refs. \onlinecite{MEYER2018149} and \onlinecite{WANG2021}.  These correspond to considerably wider tree structures
(larger numbers of groups of SPFs at each node) than were considered in the main text. 

\subsection{Spin-Boson Model: $\alpha=0.5,\ \omega_{c}=25\Delta$}
We first consider a spin-boson model with parameters $\varepsilon/\Delta=0$,$\alpha=0.5$,$\omega_{c}=25\Delta$
and with either $N=250$ or $1000$ bath modes. Following Ref. \onlinecite{MEYER2018149}, a three
layer ML-MCTDH ansatz is used for $N=250$, and a four layer tree is used for $N=1000$. 
In each case, the top node has six children
(five for the bath and one for the system) with each node accounting for bath 
degrees of freedom using 8 SPFs and 2 SPFs are used for the
system node. 
For all layers below this, each node has up to 4 children
and uses 5 SPFs. In all calculations a Krylov subspace integrator tolerance of $\epsilon_{kry}=10^{-12}$ was used.

\subsubsection*{Comparison with Standard ML-MCTDH}

\begin{figure}[h]
\begin{centering}
\includegraphics[width=\textwidth]{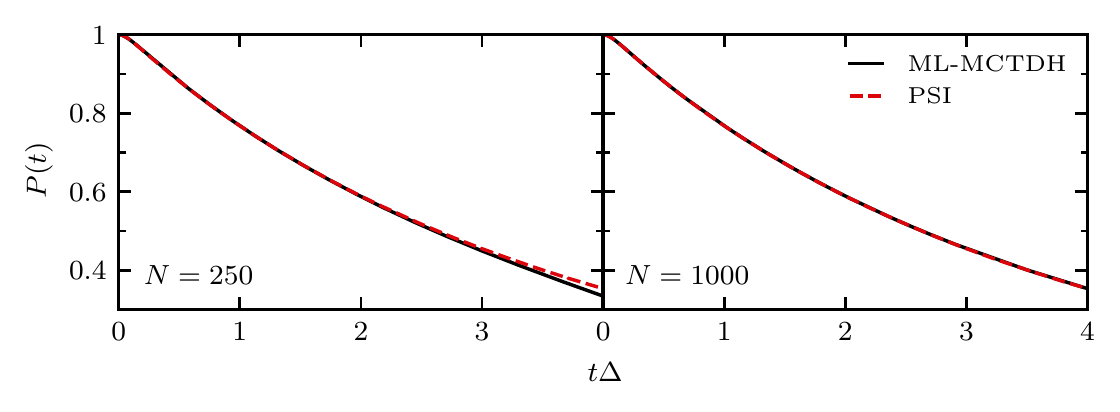}
\par\end{centering}
\caption{\label{fig:system_1_comparison} The time-dependent expectation value
$P(t)$ obtained for two spin-boson model with $\varepsilon/\Delta=0$,
$\alpha=0.5$, $\omega_{c}=25\Delta$, and with $N=250$ (left) or
$N=1000$ (right) bath modes. Here the results obtained using the
PSI are compared against the results presented in Ref. \onlinecite{MEYER2018149} that were obtained using the standard ML-MCTDH approach.}
\end{figure}

In Fig.\,\ref{fig:system_1_comparison}, the time-dependent popultion difference
value $P(t)=\langle\hat{\sigma}_{z}(t)\rangle$ obtained for these two
models using the PSI with a time step of $dt\Delta=0.000625$
are compared to the standard ML-MCTDH results that have previously
been reported in Ref. \onlinecite{MEYER2018149}.

\begin{figure}[h]
\begin{centering}
\includegraphics[width=0.5\textwidth]{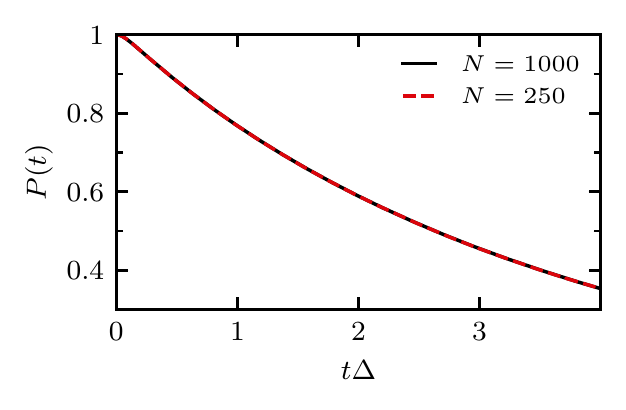}
\par\end{centering}
\caption{\label{fig:system_1_comparison_2} Comparison of the time-dependent expectation value $P(t)$ obtained using the PSI for a spin-boson model with $\varepsilon/\Delta=0$, $\alpha=0.5$, $\omega_{c}=25\Delta$, and with $N=250$ and $N=1000$ bath modes. }
\end{figure}

For $N=250$, the population dynamics obtained using the
PSI agree with the standard ML-MCTDH results for times $t\Delta<2$.
Past this point deviations are observed, with the PSI predicting a
larger value of $P(t)$ at longer times. Upon increasing the number
of bath modes to $N=1000$, both approaches agree over the entire
period considered ($t\Delta=4$). In Fig.\,\ref{fig:system_1_comparison_2}, 
we compare the results obtained with $N=250$ and $N=1000$ that were 
obtained using the PSI.  We see that the two PSI results agree closely over
the entire time period considered here (to within the thickness of the lines). 
As we have attempted to use the same bath discretisation strategy
as was used in Ref. \onlinecite{MEYER2018149}, it is unclear what leads 
to this difference in the convergence of dynamics with respect to the 
number of bath modes for these two approaches.

\subsubsection*{Convergence with respect to timestep}
In Table\,\ref{tab:System1_convergence}, we show the convergence of
the deviation with respect to the timestep used in the projector splitting
integrator. For both bath sizes, a deviation of less than $1\times10^{\text{-}4}$
is obtained for all timesteps less than $dt\Delta=0.02$. For the
$N=250$ case, this corresponds to 410 Hamiltonian evaluations and
$\sim$4800 Hamiltonian applications in order to obtain dynamics up
to $t\Delta=4$. This corresponds to a factor of $\sim$30 and $\sim3$
times fewer evaluations and applications, respectively, than were
used in previous ML-MCTDH calculations using both standard and improved regularisation strategies (see Table I of
Ref. \onlinecite{MEYER2018149}). For the $N=1000$ case, this corresponds to 810
Hamiltonian evaluations and $\sim$10,000 Hamiltonian applications
in order to obtain dynamics up to $t\Delta=8$. This again corresponds
to a factor of $\sim$30 and $\sim3$ times fewer evaluations and
applications, respectively, than were required by previous ML-MCTDH calculations
(see Table II of Ref. \onlinecite{MEYER2018149}).
\begin{table}[h]
\begin{centering}
\begin{tabular}{>{\centering}p{2cm}|>{\centering}p{2cm}>{\centering}p{2cm}>{\centering}p{2cm}>{\centering}p{2cm}>{\centering}p{2cm}>{\centering}p{2cm}}
$N$ &  & \multicolumn{2}{c}{250}  &  & \multicolumn{2}{c}{1000}\tabularnewline
\hline 
\hline 
$dt\Delta$ & $N_{MF}$ & $\Delta P$ & $\langle N_{H}\rangle$ & $N_{MF}$ & $\Delta P$ & $\langle N_{H}\rangle$\tabularnewline
\hline 
0.04 & 210 & $1.7\times10^{\text{-4}}$ & $3.4\times10^{\text{3}}$ & 410 & $5.0\times10^{\text{-}4}$ & $8.9\times10^{\text{3}}$\tabularnewline
0.02 & 410 & $1.3\times10^{\text{-}5}$ & $4.8\times10^{\text{3}}$ & 810 & $3.2\times10^{\text{-}5}$ & $1.0\times10^{\text{4}}$\tabularnewline
0.01 & 810 & $2.2\times10^{\text{-}6}$ & $7.4\times10^{3}$ & 1610 & $1.5\times10^{\text{-}6}$ & $1.5\times10^{4}$\tabularnewline
0.005 & 1610 & $1.2\times10^{\text{-}6}$ & $1.2\times10^{\text{4}}$ & 3210 & $4.4\times10^{\text{-}7}$ & $2.4\times10^{\text{4}}$\tabularnewline
0.0025 & 3210 & $2.1\times10^{\text{-}7}$ & $2.2\times10^{\text{4}}$ & 6410 & $2.0\times10^{\text{-}7}$ & $4.1\times10^{\text{4}}$\tabularnewline
0.00125 & 6410 & $2.3\times10^{\text{-}8}$ & $3.9\times10^{\text{4}}$ & 12810 & $1.9\times10^{\text{-}8}$ & $7.5\times10^{\text{4}}$\tabularnewline
0.000625 & 12810 & Reference & $7.1\times10^{\text{4}}$ & 25610 & Reference & $1.4\times10^{\text{5}}$\tabularnewline
\hline 
\end{tabular}
\par\end{centering}
\caption{\label{tab:System1_convergence}The convergence with respect to time step
($dt$) of the deviation in the population dynamics of
a spin-boson model obtained using the projector splitting integrator.
The spin-boson model considered has $\varepsilon/\Delta=0$, $\alpha=0.5$,
$\omega_{c}=25\Delta$, and with varying numbers of bath modes, $N=250$
or $1000$. For $N=250$ and $1000$, three and four level tree tensor
networks were used, respectively. To allow for a more direct comparison
with previously ML-MCTDH calculations\cite{MEYER2018149},
the deviations obtained for $N=250$ and $1000$ are evaluated to
times $t\Delta=4$ and $t\Delta=8$, respectively.}
\end{table}

As was found for the results shown in Table III of Paper II, there is no significant increase in the
deviations obtained for a given time step when increasing the size of the bath.
Similarly the number of Hamiltonian evaluations
and applications (after dividing by a factor of 2 two account for the longer evolution time
in the $N=1000$ case), does not significantly change when the number of bath modes is 
increased.  

\subsection{Spin-Boson Model: $\alpha=2.0,\ \omega_{c}=25\Delta$}
We next reconsider the spin-boson model considered in Paper II, however, with the
wider tree structures considered in Refs. \onlinecite{MEYER2018149} and \onlinecite{WANG2021}.  
Here we once again consider this model with varying numbers of bath modes ($N=500$, $2000$, $5000$, and $10^4$). Following Ref. \onlinecite{MEYER2018149}, four, five, and six layer ML-MCTDH wavefunctions were used for $N=500$, $2000$,
and $5000$, respectively, with the root node having five children
and each bath node using 8 SPFs. For the $N=10^4$
case, a seven layer ML-MCTDH wavefunction was used with the root node
having six children each of which have 8 SPFs, as was considered in
Ref. \onlinecite{WANG2021}. All other nodes had up to 4 children and used
5 SPFs. These correspond to considerably wider but less deeply nested tree structures than
were considered in the main text.

\begin{table*}[t]
\begin{centering}
\begin{tabular}{>{\centering}p{1.8cm}|>{\centering}p{1.8cm}>{\centering}p{1.8cm}>{\centering}p{1.8cm}>{\centering}p{1.8cm}>{\centering}p{1.8cm}>{\centering}p{1.8cm}>{\centering}p{1.8cm}>{\centering}p{1.8cm}}
$N$ & \multicolumn{2}{c}{500} & \multicolumn{2}{c}{2,000} & \multicolumn{2}{c}{5,000} & \multicolumn{2}{c}{10,000}\tabularnewline
\hline 
\hline 
$dt\Delta$ & $\Delta P$ & $\langle N_{H}\rangle$ & $\Delta P$ & $\langle N_{H}\rangle$ & $\Delta P$ & $\langle N_{H}\rangle$ & $\Delta P$ & $\langle N_{H}\rangle$\tabularnewline
\hline 
$4\times10^{\text{-2}}$ & $1.7\times10^{\text{-}2}$ & $1.2\times10^{3}$ & $8.8\times10^{\text{-}3}$ & $1.1\times10^{2}$ & $1.1\times10^{\text{-}2}$ & $1.1\times10^{2}$ & $1.4\times10^{\text{-}2}$ & $1.5\times10^{2}$\tabularnewline
$2\times10^{\text{-2}}$ & $2.2\times10^{\text{-}4}$ & $1.4\times10^{3}$ & $3.6\times10^{\text{-}4}$ & $1.3\times10^{3}$ & $2.5\times10^{\text{-}4}$ & $1.5\times10^{3}$ & $5.9\times10^{\text{-}4}$ & $1.7\times10^{3}$\tabularnewline
$1\times10^{\text{-2}}$ & $7.1\times10^{\text{-5}}$ & $2.0\times10^{3}$ & $5.3\times10^{\text{-5}}$ & $2.0\times10^{3}$ & $7.0\times10^{\text{-5}}$ & $2.0\times10^{3}$ & $7.7\times10^{\text{-5}}$ & $2.2\times10^{3}$\tabularnewline
$5\times10^{\text{-3}}$ & $1.8\times10^{\text{-5}}$ & $3.1\times10^{3}$ & $1.3\times10^{\text{-5}}$ & $3.1\times10^{3}$ & $2.0\times10^{\text{-6}}$ & $3.1\times10^{3}$ & $1.4\times10^{\text{-5}}$ & $3.3\times10^{3}$\tabularnewline
$2.5\times10^{\text{-3}}$ & $3.2\times10^{\text{-6}}$ & $5.0\times10^{3}$ & $1.1\times10^{\text{-5}}$ & $5.0\times10^{3}$ & $9.5\times10^{\text{-6}}$ & $5.0\times10^{3}$ & $4.7\times10^{\text{-6}}$ & $5.1\times10^{3}$\tabularnewline
$1.25\times10^{\text{-3}}$ & $5.0\times10^{\text{-7}}$ & $8.8\times10^{3}$ & $4.3\times10^{\text{-7}}$ & $8.8\times10^{3}$ & $9.1\times10^{\text{-7}}$ & $8.8\times10^{3}$ & $1.1\times10^{\text{-6}}$ & $8.8\times10^{3}$\tabularnewline
$6.25\times10^{\text{-4}}$ & Reference & $1.6\times10^{4}$ & Reference & $1.6\times10^{4}$ & Reference & $1.6\times10^{4}$ & Reference & $1.6\times10^{4}$\tabularnewline
\hline 
\end{tabular}
\par\end{centering}
\caption{\label{tab:The-convergence-with}The convergence with respect to time step
($dt$) of the deviation in the population dynamics of
a spin-boson model obtained using the projector splitting integrator.
The spin-boson model considered has $\varepsilon/\Delta=0$, $\alpha=2$,
$\omega_{c}=25\Delta$, and with varying numbers of bath modes (specified
in the tables). The deviations were evaluated using reference calculations
obtained with $dt\Delta=6.25\times10^{\text{-}4}$ and were computed
from dynamics up to $t\Delta=0.8$.}
\end{table*}

In Table\,\ref{tab:The-convergence-with}, we show the convergence
with respect to time-step of the deviations in the population dynamics
obtained for this spin-boson model with varying numbers of bath modes.
As was the case for the tree structures considered in the main text,
the population dynamics is converged to within a deviation of $10^{\text{-}4}$
for all simulations using a time step of $dt\Delta\leq0.01$
regardless of the number of bath modes. Furthermore, the convergence behavior
does not change significantly compared to the tree topology considered
in the main text.

The number of Hamiltonian applications required is essentially
independent to the number of bath modes, which supports the conclusion made in the main
text that it is the physics of the problem that limits the time step that
can be used with the PSI approach.

\subsection{Polaron-Transformed Spin-Boson Model: $\alpha=2.0,\ \omega_{c}=2.5\Delta$}
We now consider the final model considered in Ref. \onlinecite{MEYER2018149}, a polaron-transformed spin-boson model with $\alpha=2.0,\ \omega_{c}=2.5\Delta$. This spin-boson
model corresponds to the 1 TLS case of the multi-spin-boson model that was considered in
Paper II.  It has previously been found that the standard ML-MCTDH approach can
fail to provide converged results \citep{MEYER2018149} for this model.  It is necessary
to use a very small regularisation parameter when treating this polaron-transformed
spin-boson model,\citep{MEYER2018148,MEYER2018149} which leads to very stiff EOMs that
can become infeasible to solve.  In Ref. \onlinecite{MEYER2018149} converged results were
obtained for this model by using an alternative regularisation scheme for the ML-MCTDH
EOMs.  
\begin{table}[h!]
\begin{centering}
\begin{tabular}{c|c>{\centering}p{2cm}>{\centering}p{2cm}>{\centering}p{2cm}>{\centering}p{2cm}}
 &  & \multicolumn{2}{c}{Standard} & \multicolumn{2}{c}{Polaron}\tabularnewline
\hline 
\hline 
$dt\Delta$ & $N_{MF}$ & $\Delta P$ & $\langle N_{H}\rangle$ & $\Delta P$ & $\langle N_{H}\rangle$\tabularnewline
\hline 
$3.2\times10^{\text{-1}}$ & 60 & $2.0\times10^{\text{-}4}$ & $9.5\times10^{2}$ & $2.0\times10^{\text{-}3}$ & $1.2\times10^{3}$\tabularnewline
$1.6\times10^{\text{-1}}$ & 110 & $4.9\times10^{\text{-}5}$ & $1.3\times10^{3}$ & $3.2\times10^{\text{-}4}$ & $1.4\times10^{3}$\tabularnewline
$8\times10^{\text{-2}}$ & 210 & $1.4\times10^{\text{-}5}$ & $1.9\times10^{3}$ & $5.0\times10^{\text{-}5}$ & $2.1\times10^{3}$\tabularnewline
$4\times10^{\text{-2}}$ & 410 & $3.4\times10^{\text{-}6}$ & $3.1\times10^{3}$ & $4.0\times10^{\text{-}6}$ & $3.4\times10^{3}$\tabularnewline
$2\times10^{\text{-2}}$ & 810 & $8.2\times10^{\text{-}7}$ & $5.6\times10^{3}$ & $1.2\times10^{\text{-}6}$ & $5.8\times10^{3}$\tabularnewline
$1\times10^{\text{-2}}$ & 1610 & $1.6\times10^{\text{-}7}$ & $1.0\times10^{4}$ & $2.9\times10^{\text{-}7}$ & $1.0\times10^{4}$\tabularnewline
$5\times10^{\text{-3}}$ & 3210 & Reference & $1.9\times10^{4}$ & Reference & $1.9\times10^{4}$\tabularnewline
\hline 
\end{tabular}
\par\end{centering}
\caption{\label{tab:The-convergence-with-1}The convergence with respect to
time step ($dt$) of the deviation in the population dynamics
of a spin-boson model obtained using the projector splitting integrator.
Here we consider convergence of the dynamics obtained using the standard
and the polaron-transformed Hamiltonian for a spin-boson model with
$\varepsilon/\Delta=0$, $\alpha=2$, $\omega_{c}=2.5\Delta$, and
with $N=512$ bath modes. In all cases, the dynamics was obtained
using the alternative integration scheme described above and the deviation
was evaluated for dynamics obtained to times $t\Delta=8$.}
\end{table}

Following Ref. \onlinecite{MEYER2018149}, we consider this model with $N=512$ bath
modes.  We have used a four layer ML-MCTDH wavefunction, where the root node
had six children (five for the bath and one for the system) and each bath node used
8 SPFs while the system node used 2 SPFs. For all other (non-leaf) layers, each node had
up to 4 children and used 5 SPFs. In Table\,\ref{tab:The-convergence-with-1}, 
we present the the standard and polaron-transformed dynamics obtained using the PSI for 
this model.

We find that for large time steps slightly fewer function evaluations are required when using the
standard spin-boson Hamiltonian compared to the polaron-transformed
Hamiltonian and smaller deviations are found. However, upon decreasing the time step the number of Hamiltonian
evaluations and the deviations observed from their references become
similar. For the standard representation, the dynamics obtained
is converged to within a deviation of $10^{\text{-}4}$ by a time step
of $dt\Delta=0.16$, which corresponds to 110 Hamiltonian evaluations
and an average of $\sim1300$ Hamiltonian evaluations per node. For
the polaron-transformed Hamiltonian, convergence is obtained for all
time steps less than or equal to $dt\Delta=0.08$, which required 210
Hamiltonian evaluations and an average of $\sim2100$ Hamiltonian
evaluations per node. While this problem has previously been treated
using standard ML-MCTDH based approaches with improved regularisation
schemes, it is difficult to provide a direct comparison between the
two sets of results as the timescale over which dynamics has been
obtained has not been presented in Ref. \onlinecite{MEYER2018149}. Here we have
considered the population dynamics up to times $t\Delta=8$. 

By comparing the results for this model to those obtained with $\omega_c = 25\Delta$ (Table
\ref{tab:The-convergence-with}), we see that comparable numerical effort, as measured via
the number of Hamiltonian evaluations and applications, is required to obtain a results with 
comparable accuracy up to $t\Delta = 8$ for this model as is required to obtain results to $t\Delta =0.8$ 
for the previous model. This supports the conclusion that the numerical effort associated with the
PSI is strongly tied to the physics of the problem.  When we reduce the time scale of evolution 
of the bath by a factor of $10$, we are able to increase the time step by a factor of $10$ and 
obtain comparable accuracy. 

\pagebreak
\section{Further Benchmarks: Multi-Spin-Boson Model}
Finally we will reconsider the multi-spin-boson model that was considerend in Sec. III.C of Paper II.  We will use the same set of multi-spin-boson model parameters (given in Table IV in Paper II), and only examine
the cases where $M=1$ or $2$.

\subsection{Convergence With Respect to Bath Size}
We will first investigate the convergence of the population dynamics with respect to the number of frequencies
that have been used to discretize the bath.  For all calculations in the main text we used $N_b = 512$ modes.  Here we will look at considerably larger numbers of bath modes.  In all calculations we use the same strategy for constructing the ML-MCTDH wavefunction as discussed in the main text for this model, and will 
use $N_{spf}=24$ and $48$ for the $M=1$ and $2$ calculations, respectively. For both cases this is sufficient to obtain deviations of less than $10^{\text{-}4}$ (see Table V of Paper II).  We present the deviation as a
function of the bath size in Table\,\ref{tab:ptmsb_bathconv}.
\begin{table*}[h]
\begin{centering}
\begin{tabular}{r|>{\centering}p{2cm}>{\centering}p{2cm}>{\centering}p{2cm}>{\centering}p{2cm}}
$M$ & \multicolumn{2}{c}{1}  & \multicolumn{2}{c}{2}\tabularnewline
\hline 
\hline 
$N$ & $\langle\Delta P\rangle$ & $\langle N_{H}\rangle$ & $\langle\Delta P\rangle$ & $\langle N_{H}\rangle$ \tabularnewline
\hline 
128 & $1.3\times10^{\text{-}3}$ & $2.7\times10^{3}$ & $1.2\times10^{\text{-}2}$ & $2.6\times10^{3}$ \tabularnewline
256 & $6.0\times10^{\text{-}4}$ & $2.6\times10^{3}$ & $5.2\times10^{\text{-}3}$ & $2.5\times10^{3}$ \tabularnewline
512 & $2.6\times10^{\text{-}4}$ & $2.6\times10^{3}$ & $1.8\times10^{\text{-}3}$ & $2.5\times10^{3}$\tabularnewline
1024 & $1.1\times10^{\text{-}4}$ & $2.6\times10^{3}$ & $1.4\times10^{\text{-}3}$ & $2.5\times10^{3}$ \tabularnewline
2048 & $4.8\times10^{\text{-}5}$ & $2.6\times10^{3}$ & $4.4\times10^{\text{-}4}$ & $2.5\times10^{3}$ \tabularnewline
4096 & $1.4\times10^{\text{-}5}$ & $2.5\times10^{3}$ & $5.4\times10^{\text{-}4}$ & $2.5\times10^{3}$ \tabularnewline
8192 & Reference & $2.5\times10^{3}$ & Reference & $2.4\times10^{3}$\tabularnewline
%16384 & Reference & $2.5\times10^3$ & &\tabularnewline
\hline 
\end{tabular}
\par\end{centering}
\caption{\label{tab:ptmsb_bathconv}Convergence of the population
dynamics obtained for the polaron-transformed MSB models with varying numbers
of TLSs, $M$, as a function of the number of frequencies used in the discretized bath, $N$. Convergence
is measured through the average relative deviation of the population
dynamics from the reference calculations indicated in the Table. We also present the average number of Hamiltonian evaluations
per node required to obtain this dynamics out to a time $t\Delta=6$.}
\end{table*}
\begin{comment}
I am currently running 16,384 mode simulation for $M=2$, and already have them for $M=1$.  The $M=2$ results are still a few days off, and I don't think they will really add much (although based on the the time-period I do have results for, the deviations between $N=8192$ and $N=16,384$ should be less than $10^{\text{-}4}$.  I'm not sure it is worth waiting for them.
\end{comment}
For $M=1$ convergence of the deviation to less than $10^{\text{-}4}$ is not obtained until 
there are $N = 2048$ distinct frequencies in the bath.
For $M=2$, the deviations between the results obtained with $N=4096$ and $N=8192$ distinct frequencies are larger than $10^{\text{-}4}$.  These correspond to calculations with $8,192$ and $16,384$ bath modes, respectively, which demonstrates that there are models for which very large number of bath modes are required to obtain accurate results.  The significant increase in the number of bath modes required when moving from $M=1$ to $M=2$ can be understood in terms of the increased complexity in the bath correlation functions present. 
In order to accurately describe the effect of the bath on the system dynamics it is necessary to have sufficiently many bath modes so that the resultant Fourier series approximation of each distinct bath correlation function is accurate.  As we increase the number of bath correlation functions and the allow for different functional forms, which occurs as we increase the number of TLSs $M$, we need more bath modes to obtain the same accuracy.  

These results also demonstrate that the calculations presented in the main text
are not converged to the continuum bath limit. However, for the fixed $N=512$ model they 
have been converged with respect to all other parameters.  For the $M=4$ and $M=6$ cases we would expect
similar results, that is the dynamics is not converged to the continuum bath limit and large numbers of bath modes would be required to do so.  %, however, here we 
%do not present such calculations.  In order to obtain sufficiently accurate results for these two cases
%it is necessary to use large values of $N_{spf}$, and consequently the simulations take a large amount
%of time to complete.  Increasing the number of frequencies to $N = 2048$, would increase
%both the memory and computer time requirements for simulations of these systems by over a factor of 4.  
%While there is no reason that such calculations couldn't be performed, With the serial implementation 
%of the PSI, the resource requirements make such calculations impractical (the $M=6$ calculations with $N=2048$ would require $\gtrsim 130$ GB of memory and take $\gtrsim 2,000$ hours to complete).

\subsection{Polaron-Transformed Multi-Spin-Boson Model}
As a final application of the PSI approach we will consider the polaron-transformed
form of the multi-spin-boson model Hamiltonian that was considered in Sec. III.C of 
Paper II.  Using a transformation
\begin{equation}
\hat{{T}}=\exp\left[-\sum_{i=1}^{M}\hat{{\sigma}}_{i,z}\sum_{k}\frac{{g_{ik}}}{\omega_{k}}\left(\hat{{a}}_{k}^{\dagger}-\hat{{a}}_{k}\right)\right]
\end{equation}
that displaces the bath modes depending on the states of each of the
TLSs, and applying this transform to the Hamiltonian in Eq. 39 of Paper II gives
\begin{equation}
\begin{aligned}\hat{{H}}_{pol}=  & \hat{{T}}^{\dagger}\hat{{H}}\hat{{T}}\\
=& \sum_{i=1}^{M} \varepsilon_{i}\hat{{\sigma}}_{i,z}-\sum_{i,j=1}^{M}J_{ij}\hat{\sigma}_{i,z}\hat{\sigma}_{j,z}+\sum_{k}\omega_{k}\hat{{a}}_{k}^{\dagger}\hat{{a}}+
\sum_{i=1}^{M}\Delta_{i}\left[\hat{{\sigma}}_{i,+}\exp\left[-2\sum_{k}\frac{{g_{ik}}}{\omega_{k}}\left(\hat{{a}}_{k}^{\dagger}-\hat{{a}}_{k}\right)\right]+\mathrm{{h.c.}}\right],
\end{aligned}\label{eq:polaron_transformed_hamiltonian}
\end{equation}
where the TLS-TLS coupling constant is given by
\begin{equation}
J_{ij}=\frac{{1}}{\hbar}\sum_{k}\frac{g_{ik}g_{jk}}{\omega_{k}}=\frac{1}{\pi\hbar}\int_{0}^{\infty}\frac{J_{ij}(\omega)}{\omega}\mathrm{{d}}\omega.
\end{equation}
The polaron transform introduces a direct coupling between the TLSs in this
model, as well as a set of terms that directly couple all bath modes to each other in a multiplicative fashion and to each of the TLSs. 

We now present the population dynamics for polaron-transformed MSB models
with one or two TLSs that have been obtained using the PSI with ML-MCTDH wavefunctions
that make use of varying numbers of SPFs.  These results are shown in Table\,\ref{tab:ptmsb}. 
We have used the same tree topologies as were used in Sec. III.C of Paper II for 
the standard representation of the MSB Hamiltonian, and all calculations presented 
in this section were obtained with a time step of $dt\Delta=0.04$.  When evaluating
the deviations in the population dynamics, we have used the same references as were
used in the main text.  Namely, we have used the results obtained using the standard 
Hamiltonian and $N_{spf} = 48$ and $80$ when $M=1$ and $2$, respectively.  These results
are entirely analogous to those presented in Table V of Paper II, however, now using 
the polaron-transformed representation for the Hamiltonian.

\begin{table*}[h]
\begin{centering}
\begin{tabular}{r|>{\centering}p{2cm}>{\centering}p{2cm}>{\centering}p{2cm}>{\centering}p{2cm}}
$M$ & \multicolumn{2}{c}{1}  & \multicolumn{2}{c}{2}\tabularnewline
\hline 
\hline 
$N_{spf}$ & $\langle\Delta P\rangle$ & $\langle N_{H}\rangle$ & $\langle\Delta P\rangle$ & $\langle N_{H}\rangle$ \tabularnewline
\hline 
4 & $1.6\times10^{\text{-}2}$ & $2.3\times10^{3}$ & $2.5\times10^{\text{-}2}$ & $2.2\times10^{3}$ \tabularnewline
8 & $5.7\times10^{\text{-}4}$ & $2.6\times10^{3}$ & $1.1\times10^{\text{-}1}$ & $2.5\times10^{3}$ \tabularnewline
12 & $4.5\times10^{\text{-}5}$ & $2.9\times10^{3}$ & $6.4\times10^{\text{-}3}$ & $2.7\times10^{3}$\tabularnewline
16 & $1.5\times10^{\text{-}5}$ & $3.1\times10^{3}$ & $3.0\times10^{\text{-}3}$ & $2.8\times10^{3}$ \tabularnewline
24 & $2.1\times10^{\text{-}6}$ & $3.1\times10^{3}$ & $2.0\times10^{\text{-}3}$ & $3.0\times10^{3}$ \tabularnewline
32 & $7.3\times10^{\text{-}7}$ & $3.4\times10^{3}$ & $7.7\times10^{\text{-}4}$ & $3.2\times10^{3}$ \tabularnewline
48 & $4.7\times10^{\text{-}7}$ & $4.0\times10^{3}$ & $4.3\times10^{\text{-}4}$ & $3.8\times10^{3}$\tabularnewline
64 &  &  & $5.5\times10^{\text{-}5}$ & $4.0\times10^{3}$ \tabularnewline
\hline 
\end{tabular}
\par\end{centering}
\caption{\label{tab:ptmsb}Convergence of the population
dynamics obtained for the polaron-transformed MSB models with varying numbers
of TLSs, $M$, as a function of the number of SPFs, $N_{spf}$. Convergence
is measured through the average relative deviation of the population
dynamics from the reference calculations indicated in Table V of Paper II that were obtained using the standard representation for this Hamiltonian. We also present the average number of Hamiltonian evaluations
per node required to obtain this dynamics out to a time $t\Delta=6$.}
\end{table*}
For both $M=1$ and $M=2$ the deviations from the reference calculations,
which were obtained using the standard representation of the MSB model,
decrease with increasing numbers of SPFs.  As such, the dynamics we obtain
using the PSI converge towards the same result regardless of the 
representation used, as expected.  This supports the notion that we are converging
towards the correct dynamics. For $M=1$ we find deviations of less than
$10^{\text{-}4}$ for all $N_{spf} \geq 12$, while for $M=2$ we find that
$N_{spf}=64$ is required. For the $M=1$ case, the deviations obtained for 
a given $N_{spf}$ are comparable to those obtained using the standard
representation (Table V of Paper II), however, typically more Hamiltonian
evaluations are required.  For $M=2$, we find slightly larger deviations 
at a given $N_{spf}$ when using the polaron-transformed model, as well as
larger numbers of Hamiltonian applications being required.  For this 
problem, the polaron-transform increases the cost of the calculations.

\bibliography{si}